\documentclass[twocolumn]{aastex631}

\usepackage{savesym}
\savesymbol{tablenum}
\usepackage{siunitx}
\restoresymbol{SIX}{tablenum}

\usepackage[version=4]{mhchem}
\usepackage{graphicx,epsf}
\usepackage{subfigure}
\usepackage{graphicx}	
\usepackage{amsmath}	
\usepackage{amssymb}	
\usepackage{units}
\usepackage{newtxtext,newtxmath}
\usepackage{siunitx}
\usepackage{footnote}

\newcommand{\Msun}{{\rm M}_\odot}
\newcommand{\Rsun}{{\rm R}_\odot}

\newcommand{\kms}{\textrm{km}\,\textrm{s}^{-1}}

\def\arcsec{\hbox{$^{\prime\prime}$}}

\def\ca{{CaST}}
\def\cas{{CaSTs}}
\def\sng{{SN~2021gno}}
\def\sni{{SN~2021inl}}

\newcommand{\Einstein}{NASA Einstein Fellow}

\DeclareRobustCommand{\ion}[2]{\relax\ifmmode\ifx\testbx\f@series{\mathbf{#1\,\mathsc{#2}}}\else{\mathrm{#1\,\mathsc{#2}}}\fi\else\textup{#1\,{\mdseries\textsc{#2}}}\fi}

\shorttitle{Calcium-Strong Supernovae 2021gno and 2021inl}
\shortauthors{Jacobson-Gal\'an et al.}

\graphicspath{{./}{figures/}}

\begin{document}

\title{The Circumstellar Environments of Double-Peaked, Calcium-strong Supernovae 2021gno and 2021inl}

\correspondingauthor{Wynn Jacobson-Gal\'{a}n (he, him, his)}
\email{wynnjg@berkeley.edu}

\author[0000-0002-3934-2644]{W.~V.~Jacobson-Gal\'{a}n}
\affil{Department of Astronomy and Astrophysics, University of California, Berkeley, CA 94720, USA}

\author[0000-0001-8638-2780]{P.~Venkatraman}
\affil{Department of Astronomy and Astrophysics, University of California, Berkeley, CA 94720, USA}

\author[0000-0003-4768-7586]{R.~Margutti}
\affil{Department of Astronomy and Astrophysics, University of California, Berkeley, CA 94720, USA}

\author[0000-0003-4307-0589]{D.~Khatami}
\affil{Department of Astronomy and Theoretical Astrophysics Center, University of California, Berkeley, CA 94720, USA}

\author[0000-0003-0794-5982]{G.~Terreran}
\affil{Las Cumbres Observatory, 6740 Cortona Dr, Suite 102, Goleta, CA 93117-5575, USA}

\author[0000-0002-2445-5275]{R.~J.~Foley}
\affiliation{Department of Astronomy and Astrophysics, University of California, Santa Cruz, CA 95064, USA}

\author{R.~Angulo}
\affiliation{Department of Astronomy and Astrophysics, University of California, Santa Cruz, CA 95064, USA}

\author[0000-0002-4269-7999]{C.~R.~Angus}
\affiliation{DARK, Niels Bohr Institute, University of Copenhagen, Jagtvej 128, 2200 Copenhagen, Denmark}

\author[0000-0002-4449-9152]{K.~Auchettl}
\affiliation{Department of Astronomy and Astrophysics, University of California, Santa Cruz, CA 95064, USA}
\affiliation{School of Physics, The University of Melbourne, VIC 3010, Australia}
\affiliation{ARC Centre of Excellence for All Sky Astrophysics in 3 Dimensions (ASTRO 3D)}

\author[0000-0003-0526-2248]{P.~K.~Blanchard}
\affiliation{Center for Interdisciplinary Exploration and Research in Astrophysics (CIERA), and Department of Physics and Astronomy, Northwestern University, Evanston, IL 60208, USA}

\author[0000-0002-4674-0704]{A.~Bobrick}
\affil{Technion - Israel Institute of Technology, Physics department, Haifa Israel 3200002}

\author[0000-0002-7735-5796]{J.~S.~Bright}
\affil{Department of Astronomy and Astrophysics, University of California, Berkeley, CA 94720, USA}

\author{C.~D.~Couch}
\affiliation{Department of Astronomy and Astrophysics, University of California, Santa Cruz, CA 95064, USA}

\author[0000-0003-4263-2228]{D.~A.~Coulter}
\affil{Department of Astronomy and Astrophysics, University of California, Santa Cruz, CA 95064,
USA}

\author{K.~Clever}
\affiliation{Department of Astronomy and Astrophysics, University of California, Santa Cruz, CA 95064, USA}

\author{K.~W.~Davis}
\affil{Department of Astronomy and Astrophysics, University of California, Santa Cruz, CA 95064,
USA}

\author[0000-0001-5486-2747]{T.~J.~L.~de Boer}
\affil{Institute for Astronomy, University of Hawaii, 2680 Woodlawn Drive, Honolulu, HI 96822, USA}

\author[0000-0003-4587-2366]{L.~DeMarchi}
\affiliation{Center for Interdisciplinary Exploration and Research in Astrophysics (CIERA), and Department of Physics and Astronomy, Northwestern University, Evanston, IL 60208, USA}

\author[0000-0002-3696-8035]{S.~A.~Dodd}
\affil{Department of Astronomy and Astrophysics, University of California, Santa Cruz, CA 95064,
USA}

\author[0000-0002-6230-0151]{D.~O.~Jones}
\affiliation{Department of Astronomy and Astrophysics, University of California, Santa Cruz, CA 95064, USA}
\affiliation{\Einstein}

\author{J.~Johnson}
\affiliation{Department of Astronomy and Astrophysics, University of California, Santa Cruz, CA 95064, USA}

\author[0000-0002-5740-7747]{C.~D.~Kilpatrick}
\affil{Center for Interdisciplinary Exploration and Research in Astrophysics (CIERA), and Department of Physics and Astronomy, Northwestern University, Evanston, IL 60208, USA}

\author[0000-0003-2720-8904]{N.~Khetan}
\affiliation{DARK, Niels Bohr Institute, University of Copenhagen, Jagtvej 128, 2200 Copenhagen, Denmark}

\author{Z.~Lai}
\affiliation{Department of Astronomy and Astrophysics, University of California, Santa Cruz, CA 95064, USA}

\author[0000-0001-5710-8395]{D.~Langeroodi}
\affil{DARK, Niels Bohr Institute, University of Copenhagen, Jagtvej 128, 2200 Copenhagen, Denmark}

\author[0000-0002-7272-5129]{C.-C.~Lin}
\affil{Institute for Astronomy, University of Hawaii, 2680 Woodlawn Drive, Honolulu, HI 96822, USA}

\author[0000-0002-7965-2815]{E.~A.~Magnier}
\affil{Institute for Astronomy, University of Hawaii, 2680 Woodlawn Drive, Honolulu, HI 96822, USA}

\author[0000-0002-0763-3885]{D. Milisavljevic}
\affil{Department of Physics and Astronomy, Purdue University, 525 Northwestern Avenue, West Lafayette, IN 47907, USA}

\author[0000-0002-5004-199X] {H.~B.~Perets}
\affil{Technion - Israel Institute of Technology, Physics department, Haifa Israel 3200002}

\author[0000-0002-2361-7201]{J.~D.~R.~Pierel}
\affil{Space Telescope Science Institute, Baltimore, MD 21218}

\author[0000-0002-7868-1622]{J.~Raymond}
\affil{Center for Astrophysics \textbar{} Harvard \& Smithsonian, 60 Garden Street, Cambridge, MA 02138, USA}

\author[0000-0002-3825-0553]{S.~Rest}
\affiliation{Department of Physics and Astronomy, The Johns Hopkins University, Baltimore, MD 21218, USA}

\author[0000-0002-4410-5387]{A.~Rest}
\affil{Space Telescope Science Institute, Baltimore, MD 21218}
\affiliation{Department of Physics and Astronomy, The Johns Hopkins University, Baltimore, MD 21218, USA}

\author[0000-0003-1724-2885]{R.~Ridden-Harper}
\affiliation{School of Physical and Chemical Sciences | Te Kura Mat\={u}, University of Canterbury, Private Bag 4800, Christchurch 8140, New Zealand}

\author[0000-0002-9632-6106]{K.~J.~Shen}
\affil{Department of Astronomy and Theoretical Astrophysics Center, University of California, Berkeley, CA 94720, USA}

\author[0000-0003-2445-3891]{M.~R.~Siebert}
\affiliation{Department of Astronomy and Astrophysics, University of California, Santa Cruz, CA 95064, USA}

\author{C.~Smith}
\affiliation{Department of Astronomy and Astrophysics, University of California, Santa Cruz, CA 95064, USA}

\author[0000-0002-5748-4558]{K.~Taggart}
\affil{Department of Astronomy and Astrophysics, University of California, Santa Cruz, CA 95064, USA}

\author[0000-0002-1481-4676]{S.~Tinyanont}
\affil{Department of Astronomy and Astrophysics, University of California, Santa Cruz, CA 95064, USA}

\author[0000-0001-5567-1301]{F.~Valdes}
\affiliation{NSF's National Optical-Infrared Research Laboratory, 950}

\author[0000-0002-5814-4061]{V.~A.~Villar}
\affiliation{Department of Astronomy and Astrophysics, Pennsylvania State University, 525 Davey Laboratory, University Park, PA 16802, USA}
\affiliation{Institute for Computational \& Data Sciences, The Pennsylvania State University, University Park, PA, USA}
\affiliation{Institute for Gravitation and the Cosmos, The Pennsylvania State University, University Park, PA 16802, USA}

\author[0000-0001-5233-6989]{Q.~Wang}
\affiliation{Department of Physics and Astronomy, The Johns Hopkins University, Baltimore, MD 21218, USA}

\author[0000-0002-0840-6940]{S.~K.~Yadavalli}
\affiliation{Department of Astronomy and Astrophysics, Pennsylvania State University, 525 Davey Laboratory, University Park, PA 16802, USA}

\author[0000-0002-0632-8897]{Y.~Zenati}
\affiliation{Department of Physics and Astronomy, The Johns Hopkins University, Baltimore, MD 21218, USA}

\author[0000-0001-6455-9135]{A.~Zenteno}
\affiliation{Cerro Tololo Inter-American Observatory, NSF’ s National Optical- Infrared Astronomy Research Laboratory, Casilla 603, La Serena, Chile}

\begin{abstract}

We present panchromatic observations and modeling of calcium-strong supernovae (SNe) 2021gno in the star-forming host galaxy NGC 4165 and 2021inl in the outskirts of elliptical galaxy NGC 4923, both monitored through the Young Supernova Experiment (YSE) transient survey. The light curves of both SNe show two peaks, the former peak being derived from shock cooling emission (SCE) and/or shock interaction with circumstellar material (CSM). The primary peak in SN 2021gno is coincident with luminous, rapidly decaying X-ray emission ($L_x = 5 \times 10^{41}$ erg s$^{-1}$) detected by Swift-XRT at $\delta t = 1$ day after explosion, this observation being the second ever detection of X-rays from a calcium-strong transient. We interpret the X-ray emission in the context of shock interaction with CSM that extends to $r < 3 \times 10^{14}$ cm. Based on X-ray modeling, we calculate a CSM mass $M_{\rm CSM} = (0.3 - 1.6) \times 10^{-3}$ M$_{\odot}$ and density $n = (1-4) \times 10^{10}$ cm$^{-3}$. Radio non-detections indicate a low-density environment at larger radii ($r > 10^{16}$ cm) and mass loss rate of $\dot{M} < 10^{-4}$ M$_{\odot}$ yr$^{-1}$. SCE modeling of both primary light curve peaks indicates an extended progenitor envelope mass $M_e = 0.02 - 0.05$ M$_{\odot}$ and radius $R_e = 30 - 230$ R$_{\odot}$. The explosion properties suggest progenitor systems containing either a low-mass massive star or a white dwarf (WD), the former being unlikely given the lack of local star formation. Furthermore, the environments of both SNe are consistent with low-mass hybrid He/C/O WD + C/O WD mergers.

\end{abstract}


\section{Introduction} \label{sec:intro}

Calcium-rich (Ca-rich) transients are a peculiar class of stellar explosions whose progenitor system remains ambiguous \citep{filippenko03, perets10, kasliwal12}. These SNe are defined primarily based on an observed integrated emission line flux ratio of [\ion{Ca}{ii}]/[\ion{O}{i}] > 2 when the explosion has reached its nebular phase \citep{milisavljevic17} and the current sample consists of $N \gtrsim 25$ confirmed objects. Consequently, these SNe are labeled as ``Ca-rich'' compared to other transients when observed in their optically thin regime. However, because modeling of these SNe has indicated that they do not in fact produce more Ca in abundance relative to O \citep{milisavljevic17, wjg20, wjg21}, but rather are simply ``rich" in \ion{Ca}{ii} emission, we choose to adopt an alternative naming convention and refer to these as ``Ca-strong transients'' (\cas{}) from this point forward \citep{shen19}. 

Beyond their prominent \ion{Ca}{ii} emission, \cas{} have other observational characteristics that make them a well-defined explosion class. Firstly, these SNe are typically low-luminosity explosions ($M_{\rm peak} > -16.5$~mag) that have a fast photometric evolution (e.g., rise-times $\lesssim 15$~days) \citep{taubenberger17}. Physically, \cas{} are typically low energy explosions ($E_k \approx 10^{50}$~erg) that produce small amounts of ejecta ($\lesssim 0.6~\Msun$) and ${}^{56}\textrm{Ni}$ ($\lesssim 0.1~\Msun$); the latter being the dominant radioactive isotope that dictates their peak light curve luminosities. Spectroscopically, most \cas{} exhibit type I spectra with prominent \ion{He}{i} transitions at early-times and then experience an expedited transition to nebular phases where [\ion{Ca}{ii}] dominates. Lastly, the explosion environments of early samples of \cas{} showed a strong preference towards the outskirts of early-type galaxies where no star formation was detected, indicating a compact progenitor star e.g., a white dwarf (WD) system \citep{perets10, perets11, kasliwal12, lyman14, foley15,lunnan17, Dong22}. However, as the sample of confirmed \cas{} has grown, there has been increased diversity in the host environments of new objects. For example, \cas{} such as iPTF15eqv \citep{milisavljevic17}, iPTF16hgs \citep{de18}, SN~2016hnk \citep{galbany19, wjg19} and SN~2019ehk \citep{wjg20, nakaoka21} were all discovered in star-forming host galaxies, while a number of \cas{} reported in a recent sample by \cite{de20_carich} continued to show a preference towards early-type hosts. 

Many of these SNe were found at relatively large offsets from their host galaxy nuclei \citep[][and references therein]{per+21}, showing a different offset distribution than type Ia SNe \citep{kasliwal12}, which prompts suggestions of the progenitors residing in globular clusters or ejected at high velocities from their original formation closer to the host galaxy nuclei \citep{perets10,foley15,shen19}. However, a more detailed study \citep{per+21} showed that the large offsets originate from the SNe in early-type galaxies (also consistent with the two new SNe we discuss here), where a large fraction of the CaST SNe are found. In such galaxies the old stellar population extend to large distances, and the overall large offset distribution is consistent with the distribution of the old stellar populations in such galaxies, further supporting an old stellar progenitors for likely the majority of the CaST SNe \citep{per+14,per+21}. 

While the heterogeneous environments of \cas{} make it difficult to constrain a single progenitor channel, there have been significant constraints made to the parameter space of viable \ca{} progenitor systems. Firstly, the discovery of multiple \cas{} with double-peaked light curves (e.g., iPTF16hgs, SN~2018lqo, SN~2019ehk; \citealt{de18, de20_carich, wjg20}) has indicated that the progenitors of at least some of these transients must arise from stars surrounded by either extended envelopes or confined CSM. Another major breakthrough in the study of these objects came from the discovery of the closest \ca{} to date, SN~2019ehk, which exploded in the spiral host galaxy M100 at $D \approx 16.2$~Mpc \citep{wjg20, nakaoka21, de21}. SN~2019ehk was detected within $\sim$10 hours of explosion and produced luminous X-ray emission, coincident with shock-ionized spectral emission lines and a double-peaked light curve (\citealt{wjg20}, hereafter WJG20a). The combination of these observations (X-ray to radio) revealed that the SN~2019ehk progenitor star exploded into a confined shell of H- and He-rich CSM with mass of $\sim${}$7 \times 10^{-3}~\Msun$. Furthermore, SN~2019ehk is the first \ca{} with pre-explosion \textit{HST} imaging, which revealed no detectable progenitor, but did constrain the possible progenitor channels to a low-mass massive star ($<${}$10~\Msun$) or a WD system. Lastly, given its close proximity, SN~2019ehk was imaged out to $\sim$400~days post-explosion, which allowed for the tightest constraints to date to be made on the total amount of synthesized radioactive decay isotopes ${}^{56}\textrm{Ni}$ and ${}^{57}\textrm{Ni}$ in a \ca{}; the isotope mass ratio suggesting a progenitor channel involving the merger of low-mass WDs \citep{wjg21}. 

In this paper, we present, analyze and model multi-wavelength observations (X-ray to radio) of two new \cas{}, SNe~2021gno and 2021inl, both with double-peaked optical light curves. \sng{} was discovered with an apparent magnitude of 18.2 mag by the Zwicky Transient Facility (ZTF) on 20 March 2021 (MJD 59293.2) and is located at $\alpha = 12^{\textrm{h}}12^{\textrm{m}}10.29^{\textrm{s}}$, $\delta = +13^{\circ}14'57.04^{\prime \prime}$ \citep{Bruch21tns}. While \sng{} was originally classified as both a type II and type IIb SN \citep{hung21}, the spectral time series, coupled with its light curve evolution, indicates that it belongs in the \ca{} class. \sni{} was discovered by ZTF on 07 April 2021 (MJD 59311.2) with a detection magnitude of 19.5~mag and is located at $\alpha = 13^{\textrm{h}}01^{\textrm{m}}33.24^{\textrm{s}}$, $\delta = +27^{\circ}49'55.10^{\prime \prime}$ \citep{ztf21inl}. \sni{} was classified as a type Ib-peculiar and was noted to be spectroscopically consistent with the ``Ca-rich'' transient class \citep{taggart21}.

\begin{table}[ht]
\begin{center}
\caption{Main parameters of SN\,2021gno and its host galaxy \label{tbl:params21gno}}
\begin{tabular}{lcccccc}
\hline
\hline
Host Galaxy &  &  & &  & &   NGC~4165 \\ 
Galaxy Type &  &  & &  & & SAB(r)a \\
Galactic Offset &  &  & &  & &  $24.3\arcsec (3.6 ~ \rm kpc)$ \\
Redshift &  &  & &  & &  $0.0062 \pm 0.0002$\\  
Distance &  &  & &  & &  $30.5 \pm 5.6$~Mpc\footnote{\cite{Theureau07}}\\ 
Distance Modulus, $\mu$ &  &  & &  & & $32.4 \pm 0.4$~mag\\ 
$\textrm{RA}_{\textrm{SN}}$ &  &  & &  & &  $12^{\textrm{h}}12^{\textrm{m}}10.29^{\textrm{s}}$\\
$\textrm{Dec}_{\textrm{SN}}$ &  &  & &  & & $+13^{\circ}14'57.04^{\prime \prime}$\\
Time of Explosion (MJD) &  &  & &  & & 59292.7 $\pm$ 0.6\\ 
$E(B-V)_{\textrm{MW}}$ &  &  & &  & & 0.030 $\pm$ 0.001~mag\\
$E(B-V)_{\textrm{host}}$ &  &  & &  & & 0.0\footnote{No host reddening detected at explosion site.}\\
$m_{g}^{\mathrm{peak}}$ &  &  & &  & & $17.50 \pm 0.03$~mag\\
$M_{g}^{\mathrm{peak}}$ &  &  & &  & & $-14.9 \pm 0.1$~mag\footnote{Extinction correction applied.}\footnote{Relative to second $g$-band light curve peak}\\
\hline
\end{tabular}
\end{center}
\label{table:Observations}
\tablecomments{No extinction corrections have been applied to the presented apparent magnitudes.}
\end{table}

\begin{table}[ht]
\begin{center}
\caption{Main parameters of SN\,2021inl and its host galaxy \label{tbl:params21inl}}
\vskip0.1in
\begin{tabular}{lcccccc}
\hline
\hline
Host Galaxy &  &  & &  & & NGC~4923 \\ 
Galaxy Type &  &  & &  & & E/S0 \\
Galactic Offset &  &  & &  & & $60.0\arcsec (23.3 ~ \rm kpc)$ \\
Redshift &  &  & &  & & $0.0182 \pm 0.0001$\footnote{\cite{Albareti17}}\\  
Distance &  &  & &  & & $79.9 \pm 0.4$~Mpc\\ 
Distance Modulus, $\mu$ &  &  & &  & & $34.50 \pm 0.01$~mag\\ 
$\textrm{RA}_{\textrm{SN}}$ &  &  & &  & & $13^{\textrm{h}}01^{\textrm{m}}33.24^{\textrm{s}}$\\
$\textrm{Dec}_{\textrm{SN}}$ &  &  & &  & & $+27^{\circ}49'55.10^{\prime \prime}$\\
Time of Explosion (MJD) &  &  & &  & & 59309.4 $\pm$ 0.1\\ 
$E(B-V)_{\textrm{MW}}$ &  &  & &  & & 0.008 $\pm$ 0.001~mag\\
$E(B-V)_{\textrm{host}}$ &  &  & &  & & 0.0\footnote{No host reddening detected at explosion site.}\\
$m_{g}^{\mathrm{peak}}$ &  &  & &  & & $20.2 \pm 0.1$~mag\\
$M_{g}^{\mathrm{peak}}$ &  &  & &  & & $-14.3 \pm 0.2$~mag\footnote{Extinction correction applied.}\footnote{Relative to second $g$-band light curve peak}\\
\hline
\end{tabular}
\end{center}
\label{table:Observations}
\tablecomments{No extinction corrections have been applied to the presented apparent magnitudes.}
\end{table}

\begin{figure*}
\centering
\subfigure[]{\includegraphics[width=0.49\textwidth]{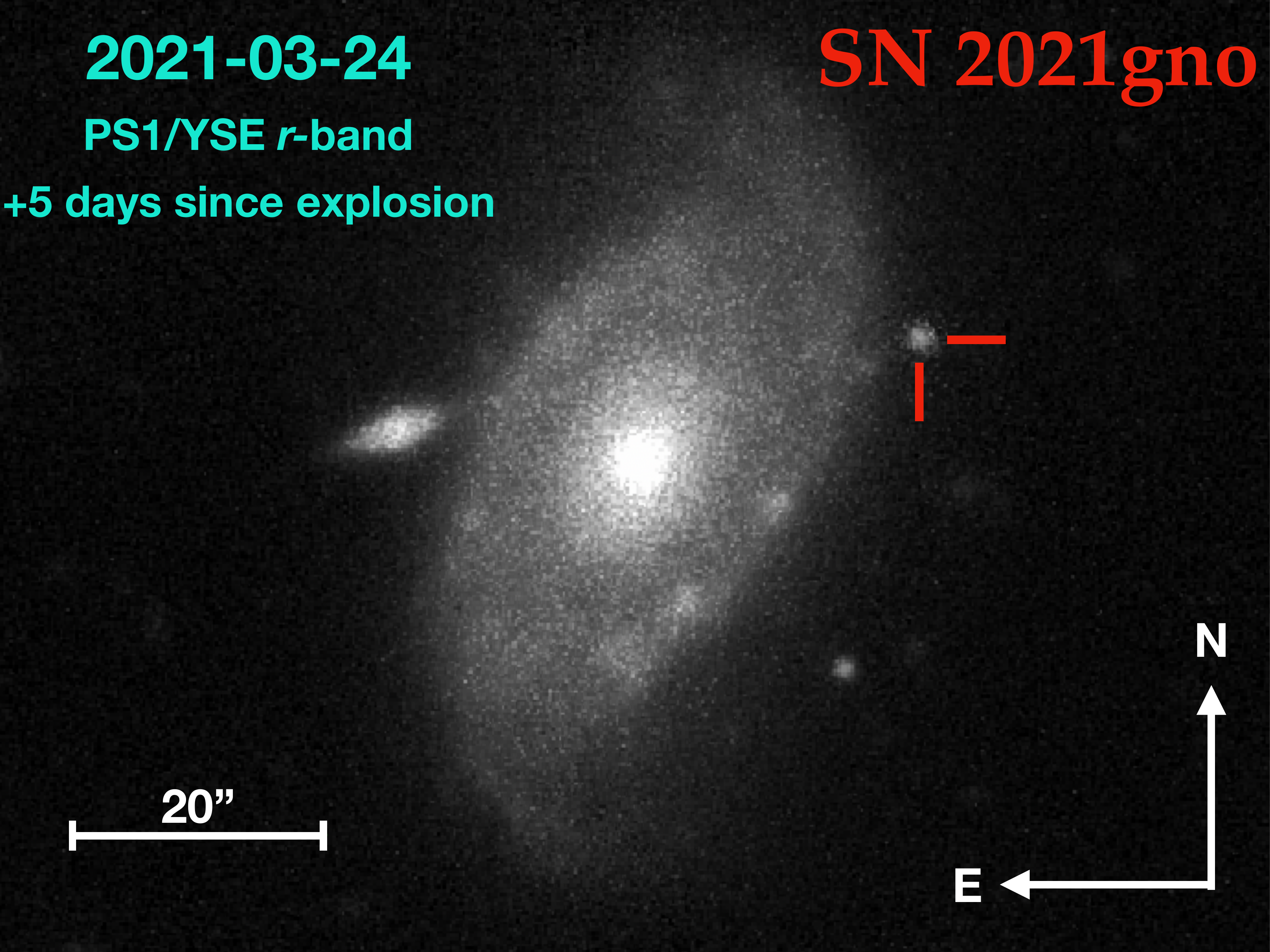}}
\subfigure[]{\includegraphics[width=0.49\textwidth]{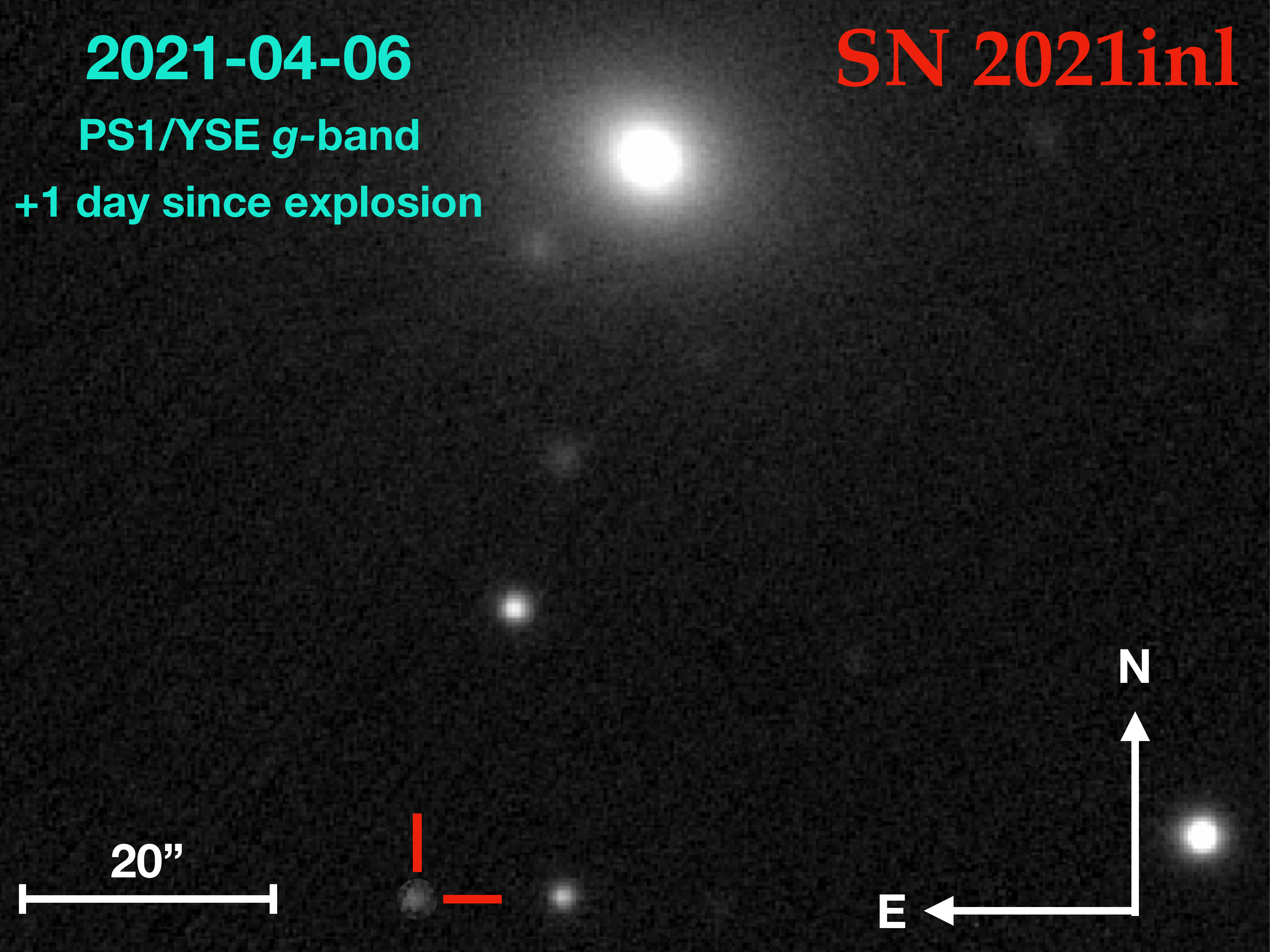}}
\caption{(a) PS1/YSE $r-$band explosion image of Ca-strong SN~2021gno in host galaxy NGC~4165. (b) PS1/YSE $g-$band explosion image of Ca-strong SN~2021inl, offset from host galaxy NGC~4923 by 60\arcsec.  \label{fig:sn_image} }
\end{figure*}

\sng{} is located 23.3\arcsec\:NW of the nucleus of the SABa galaxy NGC~4165. For \sng{}, we use the redshift-independent distance of $30.5 \pm 5.6$~Mpc, which was calculated using the Tully-Fisher relation \citep{Theureau07}. For \sni{}, we use a redshift $z = 0.0182 \pm 0.0001$ \citep{Albareti17}, which corresponds to a distance of $79.9 \pm 0.4$~Mpc for standard $\Lambda$CDM cosmology ($H_{0}$ = 70 km s$^{-1}$ Mpc$^{-1}$, $\Omega_M = 0.27$, $\Omega_{\Lambda} = 0.73$); unfortunately no redshift-independent distance is available. Possible uncertainties on the \sni{} distance could be the choice of $H_{0}$ and/or peculiar velocities of the host galaxy, the uncertainty on the former can, for example, contribute to $\lesssim 5$\% uncertainty of the SN luminosity. For each SN, we define the time of explosion as the mean phase between last non-detection and first detection. This results in a time of explosion of MJD $59292.7 \pm 0.6$~days (19 March 2021) for \sng{} and MJD $59309.4 \pm 0.1$~days (05 April 2021) for \sni{}. The main parameters of SNe~2021gno and 2021inl and their host-galaxies are displayed in Tables \ref{tbl:params21gno} and \ref{tbl:params21inl}, respectively.

\section{Observations} \label{sec:obs}

\subsection{UV/Optical/NIR Photometry}\label{SubSec:Phot}

\begin{figure*}[t]
\centering
\includegraphics[width=\textwidth]{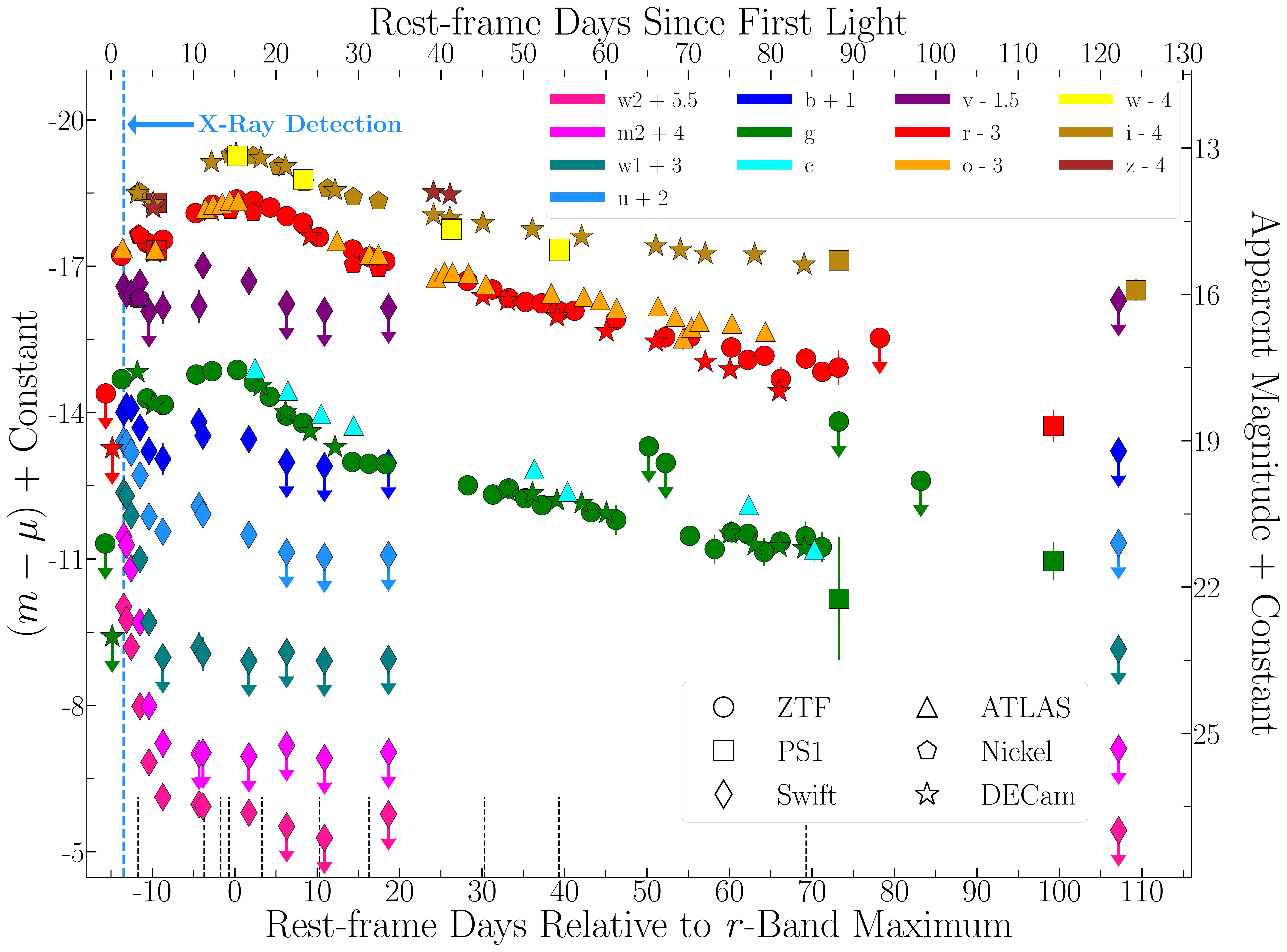}
\caption{UV/Optical/NIR light curve of \sng{} with respect to second $r$-band maximum ($\delta t\approx 15$~days). Peak of primary light curve peak occurs at phase $\delta t \approx 2$~days. Observed photometry (absolute and apparent magnitudes) is presented in the AB magnitude system. ATLAS data/3$\sigma$ upper limits are presented as triangles, PS1/YSE as squares, {\it Swift} as diamonds, ZTF as circles, Nickel as polygons, and DECam as stars. The epochs of our spectroscopic observations are marked by black dashed lines. Blue vertical dashed line mark the time of the X-ray detection in \sng. \label{fig:optical_LC} }
\end{figure*}

\sng{} was observed with the Ultraviolet Optical Telescope (UVOT; \citealt{Roming05}) onboard the Neil Gehrels \emph{Swift} Observatory \citep{Gehrels04} from 20 March 2021 until 21 April 2021 ($\delta t=$ 0.84 -- 33.0 days since explosion). We performed aperture photometry with a 5$\arcsec$ region with \texttt{uvotsource} within HEAsoft v6.26\footnote{We used the calibration database (CALDB) version 20201008.}, following the standard guidelines from \cite{Brown14}. In order to remove contamination from the host galaxy, we employed images acquired at $t\approx122$~days after explosion, assuming that the SN contribution is negligible at this phase. This is supported by visual inspection in which we found no flux associated with \sng{}. We subtracted the measured count rate at the location of the SN from the count rates in the SN images following the prescriptions of \cite{Brown14}. Consequently, we detect bright UV emission from the SN directly after explosion (Figure \ref{fig:optical_LC}) until maximum light. Subsequent non-detections in $w1, m2, w2$ bands indicate significant cooling of the photosphere and/or Fe-group line blanketing.

\begin{figure*}
\centering
\subfigure[]{\includegraphics[width=0.52\textwidth]{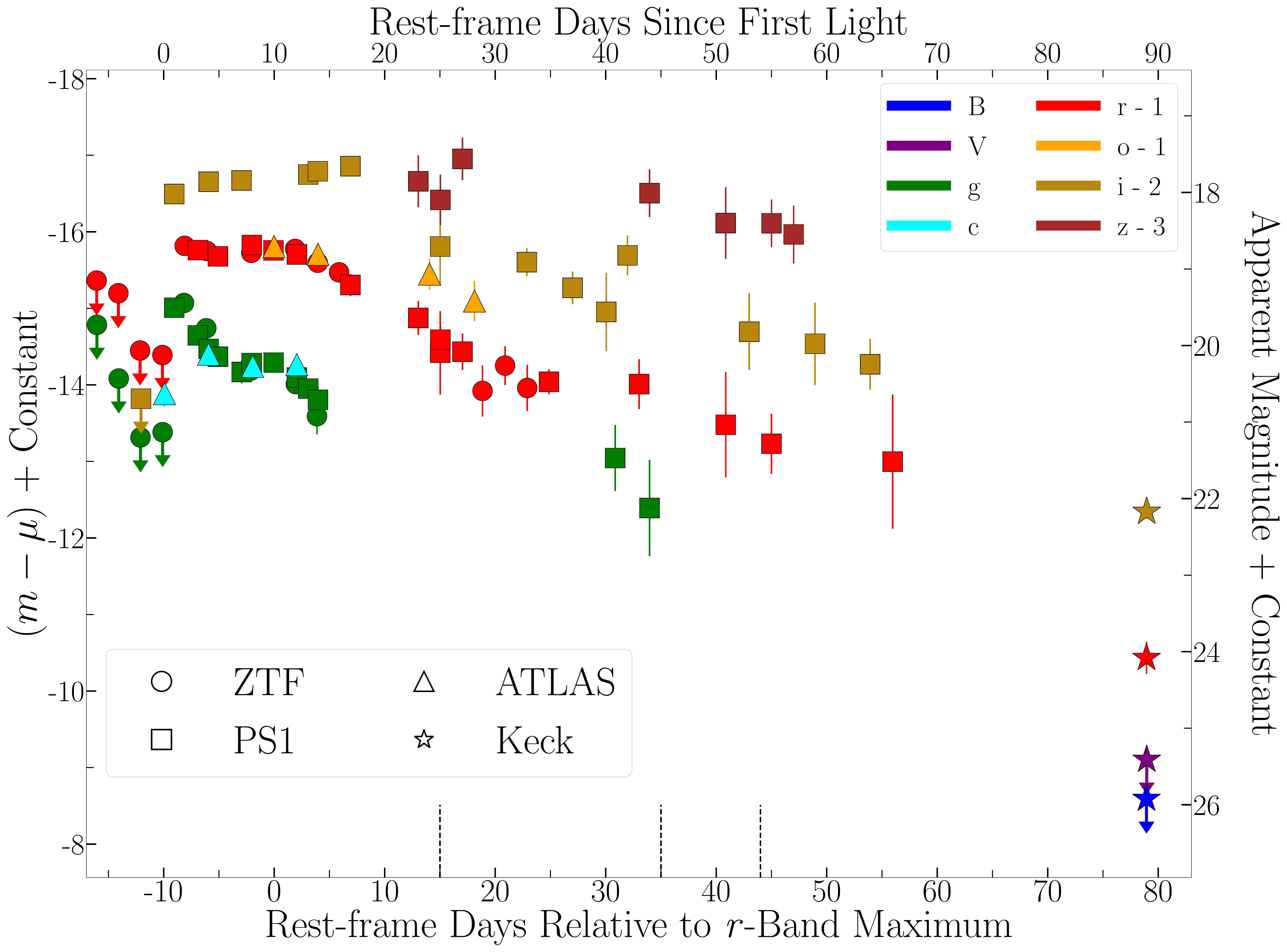}}
\subfigure[]{\includegraphics[width=0.46\textwidth]{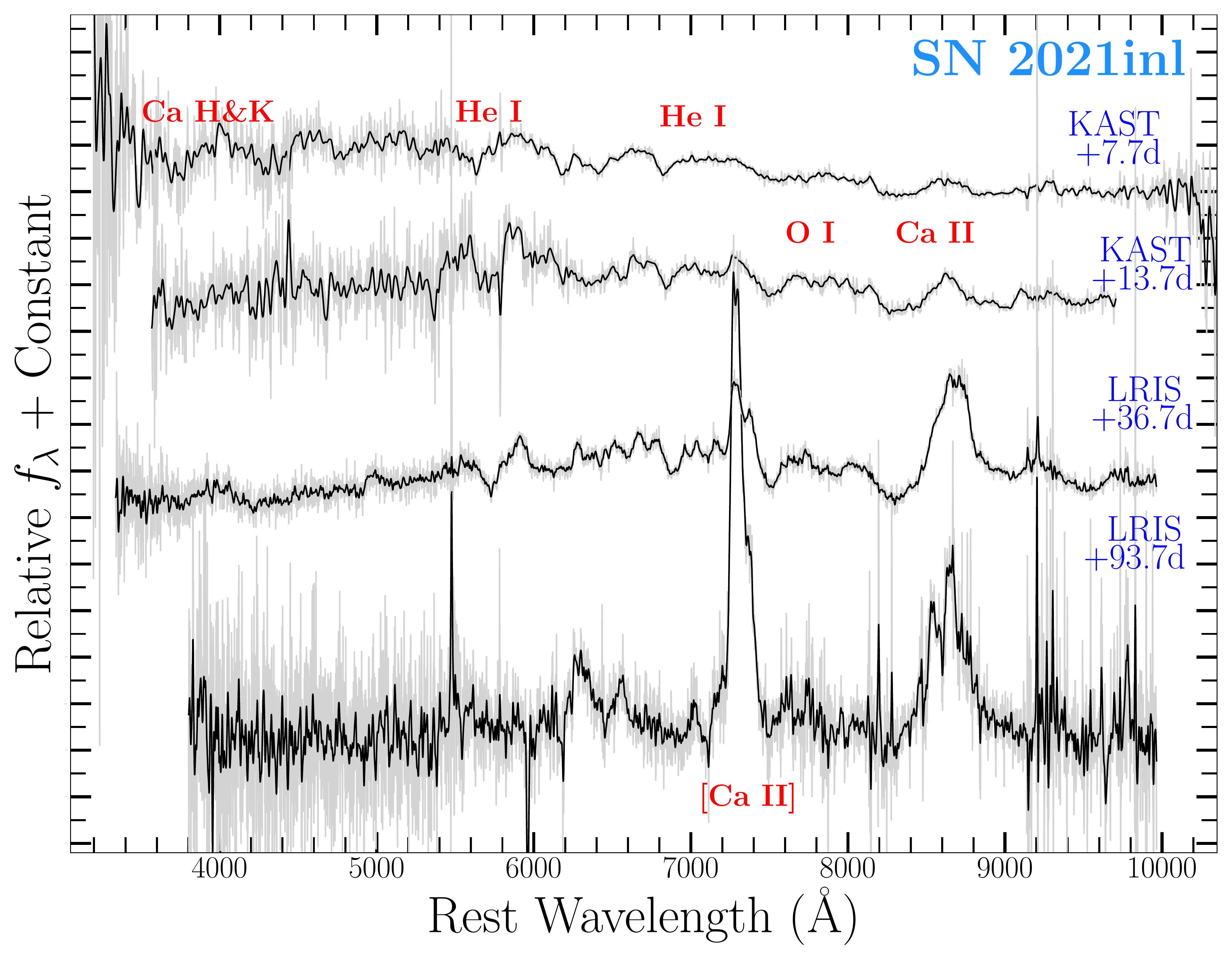}}
\caption{(a) Optical/NIR light curve of \sni{} with respect to second $r$-band maximum ($\delta t\approx 10$~days). Peak of primary light curve peak occurs at phase $\delta t \approx 2$~days. Observed photometry presented in AB magnitude system. ATLAS data/3$\sigma$ upper limits are presented as triangles, PS1/YSE as squares, {\it Swift} as diamonds, ZTF as circles, Nickel as polygons, and Keck LRIS as stars. The epochs of our spectroscopic observations are marked by black dashed lines. (b) Spectral observations of \sni{} with phases (blue) marked with respect to explosion. Raw spectra are shown in gray, and smoothed spectra with black lines. \label{fig:21inl_spec_LC} }
\end{figure*}

Additional $griz$-band imaging of \sng{} and \sni{} was obtained through the Young Supernova Experiment (YSE) \citep{Jones2021} with the Pan-STARRS telescope \citep[PS1;][]{Kaiser2002} between 24 March 2021 and 21 July 2021 ($\delta t= 4.80-123.5$~days since explosion) and 06 April 2021 and 10 June 2021 ($\delta t= 0.97-66.0$~days since explosion), respectively. PS1 images of SNe~2021gno and 2021inl are presented in Figure \ref{fig:sn_image}. Furthermore, \sng{} was observed with the DECam Extension survey of YSE between 22 March 2021 and 09 January 2022 ($\delta t = 1.49 - 294.7$~days) on the Cerro Tololo Inter-American Observatory Blanco 4-m telescope \citep{Rest22}. The YSE photometric pipeline is based on {\tt photpipe} \citep{Rest+05}. Each image template was taken from stacked PS1 exposures, with most of the input data from the PS1 3$\pi$ survey. All images and templates are resampled and astrometrically aligned to match a skycell in the PS1 sky tessellation. An image zero-point is determined by comparing PSF photometry of the stars to updated stellar catalogs of PS1 observations \citep{Chambers2017}. The PS1 templates are convolved with a three-Gaussian kernel to match the PSF of the nightly images, and the convolved templates are subtracted from the nightly images with {\tt HOTPANTS} \citep{becker15}. Finally, a flux-weighted centroid is found for each SN position and PSF photometry is performed using ``forced photometry": the centroid of the PSF is forced to be at the SN position. The nightly zero-point is applied to the photometry to determine the brightness of the SN for that epoch.

Both SNe~2021gno and 2021inl were observed with ATLAS ($\delta t = 0.70 - 84.6$ and $\delta t = 0.10 - 28.1$~days since explosion, respectively), a twin 0.5m telescope system installed on Haleakala and Mauna Loa in the Hawai'ian islands that robotically surveys the sky in cyan (\textit{c}) and orange (\textit{o}) filters \citep[][]{2018PASP..130f4505T}. The survey images are processed as described in \cite{2018PASP..130f4505T} and photometrically and astrometrically calibrated immediately \citep[using the RefCat2 catalogue;][]{2018ApJ...867..105T}. Template generation, image subtraction procedures and identification of transient objects are described
in \cite{Smith20}. Point-spread-function photometry is carried out on the difference images and all sources greater than 5$\sigma$ are recorded and all sources go through an automatic validation process that removes spurious objects \citep{Smith20}. Photometry on the difference images (both forced and non-forced) is from automated point-spread-function fitting as documented in \cite{2018PASP..130f4505T}. The photometry presented here are weighted averages of the nightly individual 30\,sec exposures, carried out with forced photometry at the position of the SNe.

The complete light curves of SNe~2021gno and 2021inl are presented in Figures \ref{fig:optical_LC} and \ref{fig:21inl_spec_LC}, respectively, and all photometric observations are listed in Appendix Table \ref{tbl:phot_table_s}. In addition to our observations, we include $g/r-$band photometry of SNe~2021gno and 2021inl from the Zwicky Transient Facility (ZTF; \citealt{bellm19, graham19}) forced-photometry service \citep{Masci19}, which span from 20 March 2021 to 15 June 2021 ($\delta t= 0.54-87.5$~days since explosion) and 07 April 2021 to 08 May 2021 ($\delta t= 1.85-32.9$~days since explosion).

The Milky Way (MW) $V$-band extinction and color excess along the \sng{} line of sight is $A_{V} = 0.093$~mag and \textit{E(B-V)} = 0.03~mag \citep{schlegel98, schlafly11}, respectively, which we correct for using a standard \cite{fitzpatrick99} reddening law (\textit{$R_V$} = 3.1). Additionally, the MW $V$-band extinction and color excess along the \sni{} line of sight is $A_{V} = 0.025$~mag and \textit{E(B-V)} = 0.008~mag \citep{schlegel98, schlafly11}. For both SNe, we do not correct for host-galaxy contamination given the absence of Na I D absorption in all spectra at the host redshift.

\subsection{Optical/NIR spectroscopy}\label{SubSec:Spec}

\begin{figure*}[t!]
\centering
\includegraphics[width=\textwidth]{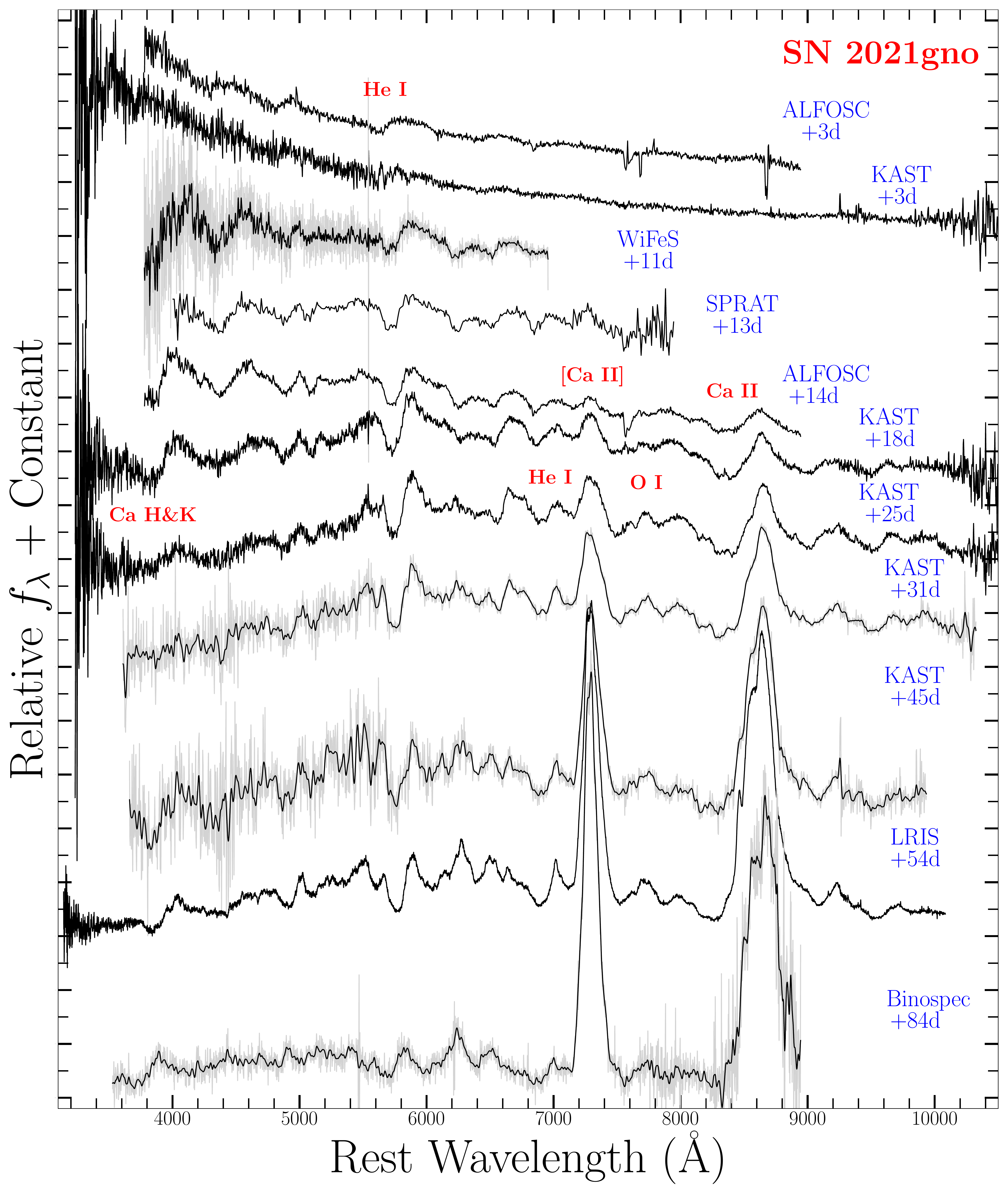}
\caption{Spectral observations of \sng{} with phases (blue) marked with respect to explosion. Raw spectra are shown in gray,
and smoothed spectra with black lines. \label{fig:spectral_series}  }
\end{figure*}

In Figures \ref{fig:spectral_series} and \ref{fig:21inl_spec_LC}(b), we present the complete series of optical spectroscopic observations of \sng{} and \sni{} from $\delta t= 3-84$~days and $\delta t= 25-111$~days relative to explosion, respectively. A full log of spectroscopic observations is presented in Appendix Tables \ref{tab:spec_table1} and \ref{tab:spec_table2}.

SNe~2021gno and 2021inl were observed with Shane/Kast \citep{KAST} and Keck/LRIS \citep{oke95} between $\delta t= 3-54$~days and $\delta t= 25-111$~days relative to explosion, respectively. For all these spectroscopic observations, standard CCD processing and spectrum extraction were accomplished with \textsc{IRAF}\footnote{https://github.com/msiebert1/UCSC\_spectral\_pipeline}. The data were extracted using the optimal algorithm of \citet{1986PASP...98..609H}.  Low-order polynomial fits to calibration-lamp spectra were used to establish the wavelength scale and small adjustments derived from night-sky lines in the object frames were applied. We employed custom IDL routines to flux calibrate the data and remove telluric lines using the well-exposed continua of the spectrophotometric standard stars \citep{1988ApJ...324..411W, 2003PASP..115.1220F}. Details of these spectroscopic reduction techniques are described in \citet{2012MNRAS.425.1789S}.

Spectra of \sng{} were also obtained with the Alhambra Faint Object Spectrograph (ALFOSC) on The Nordic Optical Telescope (NOT), as well as Binospec on MMT, and SpeX at the NASA Infrared Telescope Facility (IRTF). All of the spectra were reduced using standard techniques, which included correction for bias, overscan, and flat-field. Spectra of comparison lamps and standard stars acquired during the same night and with the same instrumental setting have been used for the wavelength and flux calibrations, respectively. When possible, we further removed the telluric bands using standard stars. Given the various instruments employed, the data-reduction steps described above have been applied using several instrument-specific routines. We used standard \textsc{IRAF} commands to extract all spectra.

\begin{figure*}
\centering
\subfigure[]{\includegraphics[width=0.49\textwidth]{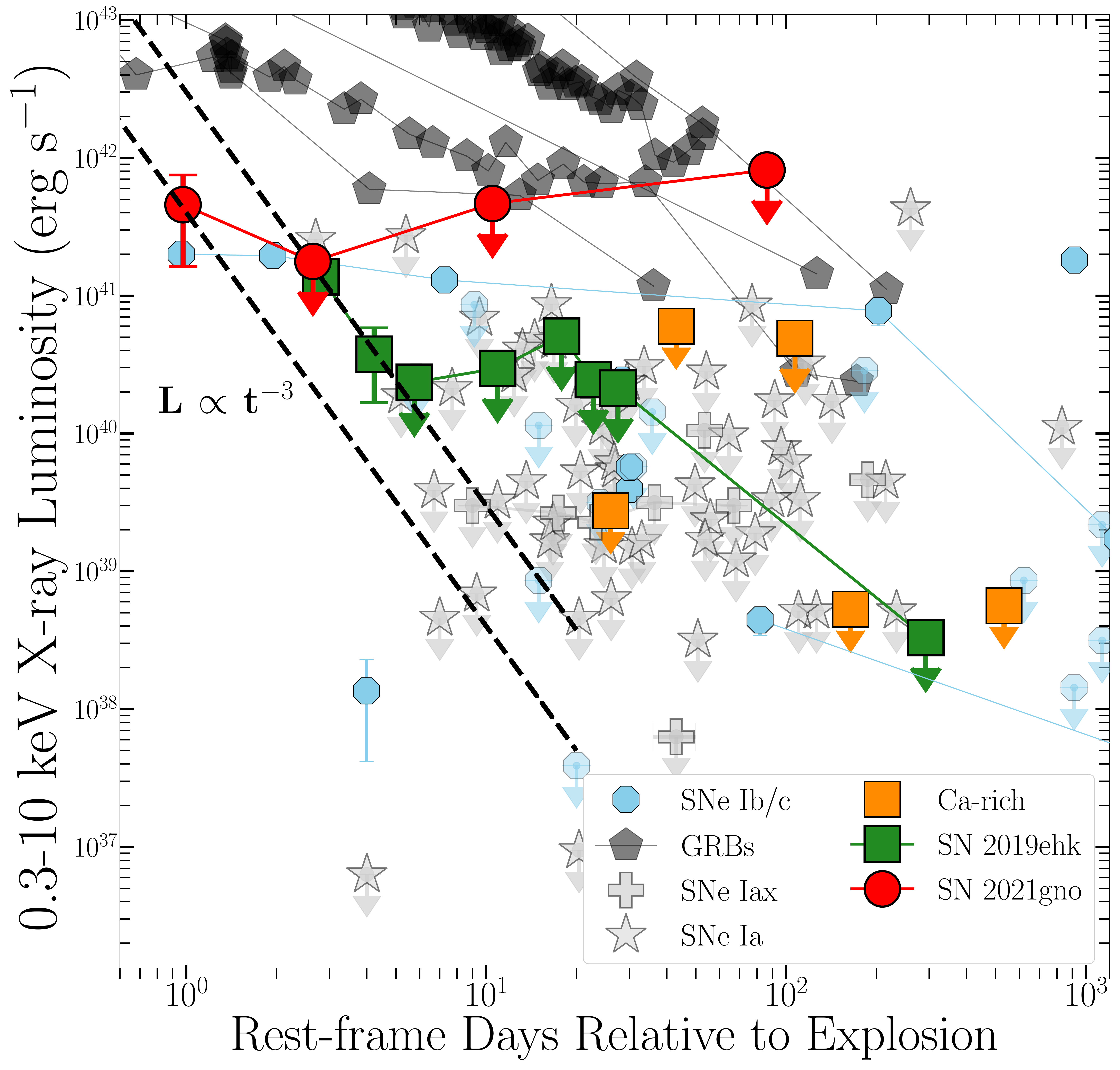}}
\subfigure[]{\includegraphics[width=0.49\textwidth]{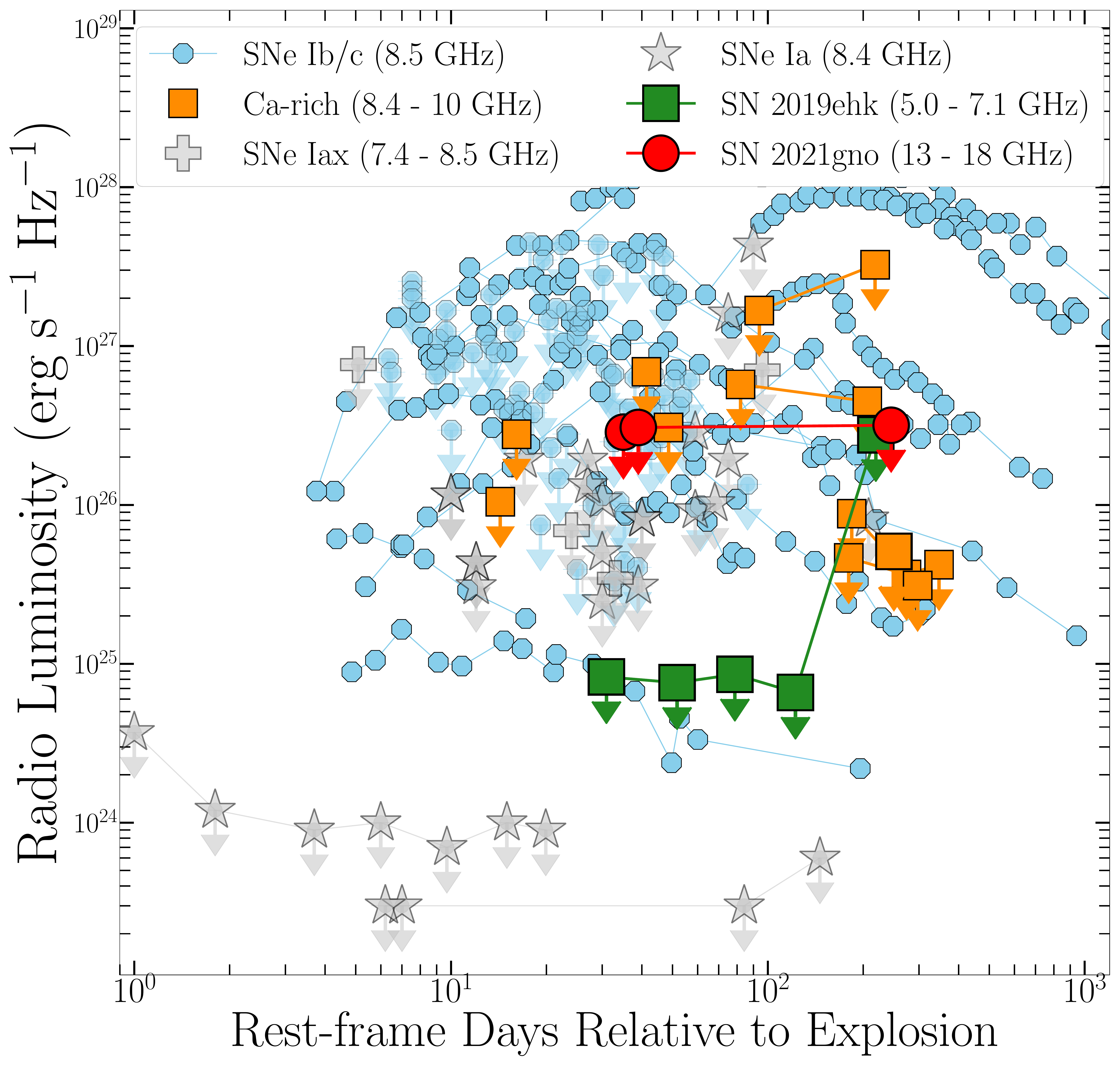}}
\caption{(a) X-Ray light curve of \sng{} (red circles) and other thermonuclear transients e.g., SNe~Iax (grey plus signs), SNe Ia (grey stars) and \cas \ (orange squares). Core-collapse SNe~Ib/c are shown as light blue octagons and GRBs are displayed as black polygons. The decline rate of SN~2019ehk's X-ray emission (green squares; $L_x\propto t^{-3}$) is shown as a black dashed line, which is also consistent with SN~2021gno's decline rate (black dashed line). (b) Radio non-detections of \sng{} (red circles) compared to non-detection limits of thermonuclear SNe and SNe~Ib/c.  \label{fig:xray_radio_LC} }
\end{figure*}

\subsection{X-ray observations with \emph{Swift}-XRT}\label{SubSec:XRT}

The X-Ray Telescope (XRT, \citealt{burrows05}) on board the \emph{Swift} spacecraft \citep{Gehrels04} started observing the field of \sng{} from 20 March 2021 to 06 Nov. 2021 ($\delta t = 0.81-233.6$~days since explosion with a total exposure time of 28.8 ks, IDs 14199 and 14214). We analyzed the data using HEAsoft v6.26 and followed the prescriptions detailed in \cite{margutti13}, applying standard filtering and screening. A bright source of X-ray emission is clearly detected in each individual observation with significance of $>3\sigma$ against the background in the first two epochs ($\delta t = 0.81 - 1.14$~days; total exposure time of 4.73 ks) and count rates of $(3.8\pm1.6) \times 10^{-3}$ and $(2.3\pm1.1) \times 10^{-3}$~c~s$^{-1}$, respectively. Given how close in time the first XRT observations are to one another, we chose to merge the two event files and use the combined epoch for analysis of the X-ray spectrum. 

To test the validity of the X-ray emission observed in \sng{}, we first employ a binomial test to understand the likelihood that the fading X-ray emission was a chance coincidence. In this test, we compared the observed counts in the combined early-time epoch to a late-time, template XRT image (3.6 ks) of the explosion site at $\sim 234$~days. We find a probability of chance fading X-ray emission of only $\sim 0.34\%$, further indicating that the observed X-ray photons were in fact derived from the SN at early-times. Furthermore, in the template image, no X-ray emission is detected above the background level at later phases. The complete X-ray light curve is presented in Figure \ref{fig:xray_radio_LC}(a).



From the merged event file at $t\le 2.1$~days, we extracted a spectrum using a 15$\arcsec$ region centered at the location of \sng{}. We find that the X-ray spectrum of the SN emission has a best-fitting photon index $\Gamma=0.7\pm0.5$ ($1\sigma$ error) corresponding to an unabsorbed 0.3-10 keV flux of $F_x= (4.1 \pm 2.2) \times 10^{-12}\,\rm{erg\,s^{-1}cm^{-2}}$. No evidence for intrinsic neutral hydrogen absorption is found ($\rm{NH_{MW}} < 2.2\times 10^{18}\,\rm{cm^{-2}}$). The Galactic neutral hydrogen column density along our line of sight is $\rm{NH_{MW}}=2.4\times 10^{20}\,\rm{cm^{-2}}$ \citep{Kalberla05}, which is used to account for the contribution of the host galaxy in modeling the X-ray excess. We then use the best-fitting spectral parameters inferred from the merged observations to flux-calibrate the count-rate upper limits derived for the following epochs (Table \ref{tab:xray_obs}). Given the distance to \sng{}, these measurements indicate a steeply decaying, large X-ray luminosity of $L_x\le 4.6\times 10^{41}\,\rm{erg\,s^{-1}}$ at $t\le2.1$~days (Figure \ref{fig:xray_radio_LC}), rivaling even the gamma-ray burst SN, 1998bw \citep{Kouveliotou04}. This very early-time observation represents only the second X-ray detection in a \ca{}, the first being SN~2019ehk, which showed luminous rapidly fading X-ray emission at $t\le4.2$~days since explosion (WJG20a). Furthermore, SN~2021gno's decay in X-ray luminosity is consistent with the steep light curve slope of $L_x\propto t^{-3}$ found in SN~2019ehk. 

The hard 0.3-10~keV X-ray spectrum of SN\,2019ehk is suggestive of thermal bremsstrahlung emission with temperature $T>10$~keV. Consequently, we fit the \sng{} contribution with a bremsstrahlung spectral model with $T=20$ keV and find an inferred emission measure of $EM=\int n_e n_I dV$ is $EM=(1.8 \pm 0.7) \times 10^{64}\,\rm{cm^{-3}}$ (at $\delta t \approx 1.0$ d), where $n_e$ and $n_I$ are the number densities of electrons and ions, respectively. Furthermore, in \S\ref{Sec:Radio_Xray_Modeling}, we apply the estimated $EM$ from the XRT detections to derive parameters of the CSM surrounding the progenitor system of \sng{} .

\subsection{Radio observations}\label{SubSec:radio}

We observed \sng{} with the Arcminute Microkelvin Imager Large Array (AMI-LA; \citealt{Zwart08, Hickish18}) on 21, 25 April and 19 Nov. 2021 ($\delta t = 35, \ 39, \ \& \ 245$~days since explosion) and found no evidence for radio emission from the SN. These data were all taken at a central frequency of 15.5 GHz across a 5 GHz bandwidth consisting of 4096 channels, which we average down to 8 for imaging. Radio frequency interference (RFI) flagging and bandpass and phase reference calibration were performed using a custom reduction pipeline \citep{Perrott13}. Additional flagging and imaging was performed in the Common Astronomy Software Applications (CASA; \citealt{McMullin07}) package. For imaging we use natural weighting with a clean gain of 0.1. To measure the source flux density we use the CASA task IMFIT. The resolution of the AMI-LA (characteristic beam dimensions 400 x 3000) when observing at the declination of J1820 means that the source is unresolved in all epochs. Details of each observation are presented in Appendix Table \ref{tab:radio} and the derived radio luminosity limits for \sng{} are plotted in Figure \ref{fig:xray_radio_LC}(b).

\section{Host Galaxy and Explosion Site}\label{sec:host}


\sng{} is located 3.6~kpc in projection from the nucleus in the outer arm of its SBa type host galaxy NGC~4165 (Fig. \ref{fig:sn_image}a). We determine the host galaxy oxygen abundance 12 + log(O/H) by using an SDSS spectroscopic observation taken on 20 April 2004 of the galactic core; given the SN location, the metallicity at the explosion site is likely lower. Using a combination of line flux ratios ([\ion{O}{iii}] / H$\beta$ and [\ion{N}{ii}]/H$\alpha$) in Equations 1 \& 3 of \cite{pettini04}, we determine a range of host metallicities of 12 + log(O/H) = $8.94 - 9.15$~dex ($1.03 - 1.06$~Z$_{\odot}$). Using the same spectrum, we find an H$\alpha$ emission line luminosity of $L_{\textrm{H$\alpha$}} = 8.7 \times 10^{39}$~erg~s$^{-1}$, which corresponds to a star formation rate of SFR = $0.07 \ \Msun$ yr$^{-1}$ \citep{Kennicutt98}. 

In order to understand the SFR and metallicity at the exact location of \sng{}, we acquired an additional host spectrum at the explosion site using the Goodman spectrograph on SOAR on 27 January 2021, when the SN emission is not expected to be detected given its brightness at this phase and the S/N of the spectrum. We find no detectable host galaxy emission lines at the SN location and perform a manual, optimal Gaussian extraction with a $6\sigma$ region, $3\sigma$ on each side of the SN location, which translates to distance of 0.22~kpc. We then derive a limit on the H$\alpha$ emission line luminosity by simulating a marginal detection as a Gaussian profile (FWHM = 100~$\kms$) with a peak flux of three times the spectrum’s root-mean-square (RMS)
flux. We then calculate $L_{\textrm{H$\alpha$}} < 4.3 \times 10^{36}$~erg~s$^{-1}$ and a local SFR of $< 3.4 \times 10^{-5} \ \Msun$ yr$^{-1}$. This estimate is consistent with the low SFR of $\sim 9.2 \times 10^{-5} \ \Msun$ yr$^{-1}$ inferred from the explosion site of SN~2019ehk (WJG20a) and suggests that \sng{} is more likely to have originated from an older progenitor system (e.g., low-mass massive star or WD binary). Furthermore, the H$\alpha$ luminosity at the explosion site of SN 2019ehk is only consistent with the \ion{H}{ii} region luminosity at the location of $\sim20\%$ of H-stripped SNe \citep{galbany18, Kuncarayakti18}. 


Similar to many other \cas{}, \sni{} is located at a large projected offset ($\sim $23~kpc) from early-type, E/S0 host galaxy NGC~4923 (Fig. \ref{fig:sn_image}b). While the explosion site indicates no star formation at the SN location, we also use an SDSS spectroscopic observation taken on 22 Feb. 2007 of the galactic core to infer properties of NGC~4923. To derive properties of the host galaxy, we model the SDSS spectrum as well as \textit{GALEX} All-Sky Survey Source Catalog (GASC; \citealt{Seibert2012}) UV, SDSS \textit{ugriz}, and NIR Two Micron All Sky Survey (2MASS; \citealt{Jarrett2000}) \textit{JHK$_s$} photometry with the Fitting and Assessment of Synthetic Templates code (FAST; \citealt{Kriek2009}). Our model grid includes stellar initial mass functions by \cite{Salpeter1955} and \cite{Chabrier2003}, star-formation history that is exponentially decreasing and a delayed function, and stellar population libraries presented by \cite{Bruzual2003} and \cite{Conroy2009}. For models without host galaxy dust reddening, we find a total stellar mass of $M_{\star} \approx (4.6 - 7.6) \times 10^{10} \ \Msun$, metallicity of $Z \approx Z_{\odot}$ and SFR $\lesssim 10^{-5} \ \Msun$~yr$^{-1}$. We also find consistent $M_{\star}$ and $Z$ measurements within a grid of models that included dust ($A_V = 0.6$), but all models found an SFR = 0. Overall, these models indicate that \sni{}, given its large offset from a host with no apparent star formation, is \emph{not} from a massive star progenitor. 

\section{Optical Light Curve Analysis}\label{sec:LC_analysis}

\subsection{Photometric Properties}\label{subsec:phot_properties}

SNe~2021gno and 2021inl are the fourth and fifth confirmed \cas{} with clearly defined double-peaked light curves as shown in Figure \ref{fig:optical_LC} \& \ref{fig:21inl_spec_LC}(a), respectively. The other double-peaked objects in the present \ca{} sample are iPTF16hgs \citep{de18}, SN~2018lqo \citep{de20}, and SN~2019ehk \citep{wjg20, nakaoka21, de21}. Similar to other double-peaked SNe, we define the phase of these SNe relative to both the secondary, ``Nickel-powered'' peak and to explosion as defined at the end of \S\ref{sec:intro}. For both \cas{}, we calculate the time of maximum by fitting a third-order polynomial to $g-$ and $r-$band photometry. For \sng{}, we find best-fit $g$- and $r$-band peak absolute magnitudes of $M_g = -14.90 \pm 0.03$~mag at MJD $59305.2\pm0.6$ and $M_r = -15.3 \pm 0.2$~mag at MJD $59307.6\pm0.6$, respectively, resulting in an $r$-band rise-time of $t_r = 15.3 \pm 0.6$~days. For \sni{}, we find best-fit $g$- and $r$-band peak absolute magnitudes of $M_g = -14.3 \pm 0.1$~mag at MJD $59318.6\pm0.1$ and $M_r = -14.8 \pm 0.2$~mag at MJD $59317.4\pm0.1$, respectively, resulting in an $r-$band rise-time of $t_r = 8.04 \pm 0.10$~days.

\begin{figure}[t!]
\centering
\includegraphics[width=0.45\textwidth]{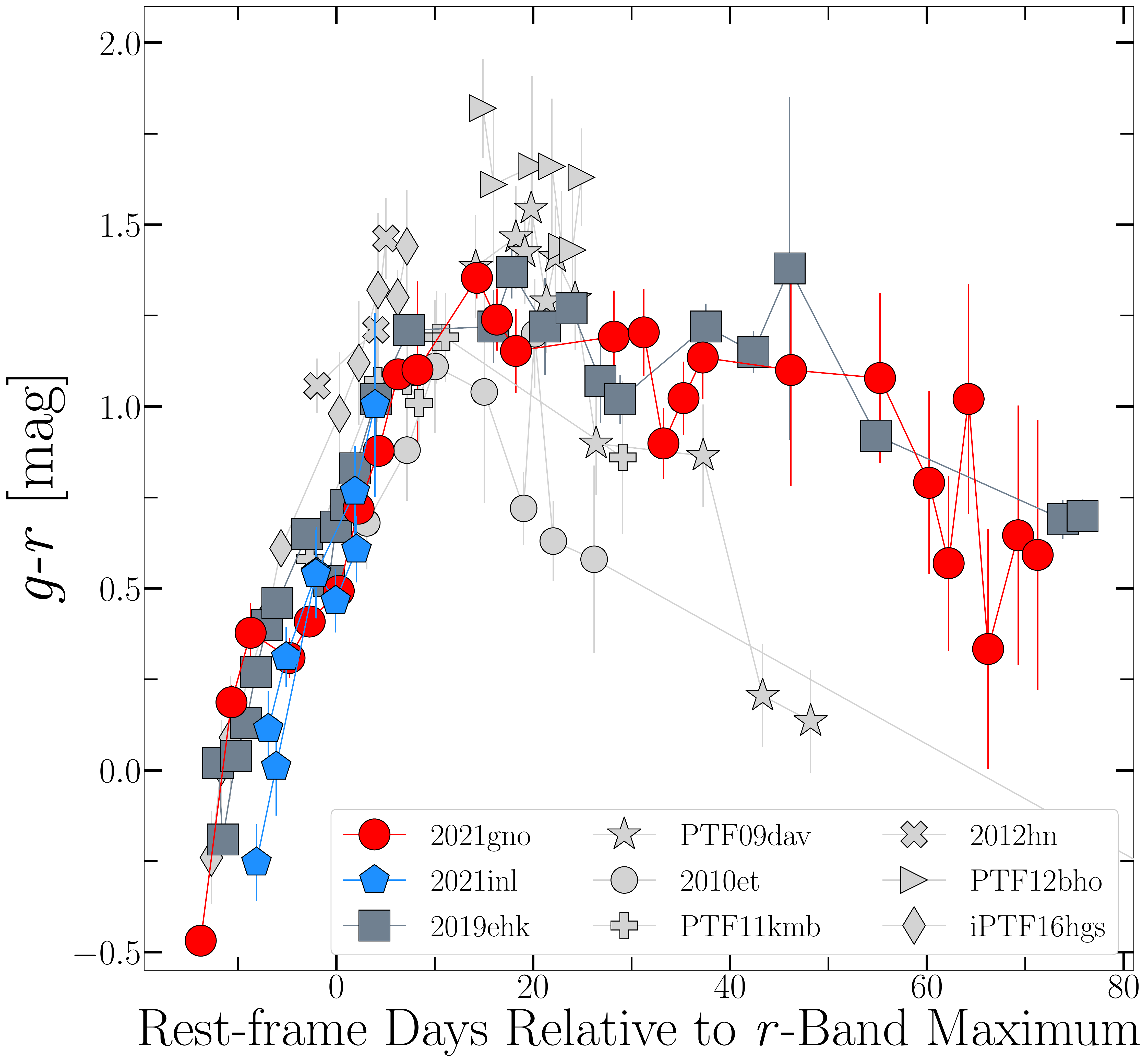}
\caption{\textit{g-r} color comparison of \sng\ (red circles), \sni\ (blue polygons), and current sample of \cas. All photometry has been extinction corrected.  \label{fig:colors}}
\end{figure}

\begin{figure*}
\centering
\subfigure[]{\includegraphics[width=0.49\textwidth]{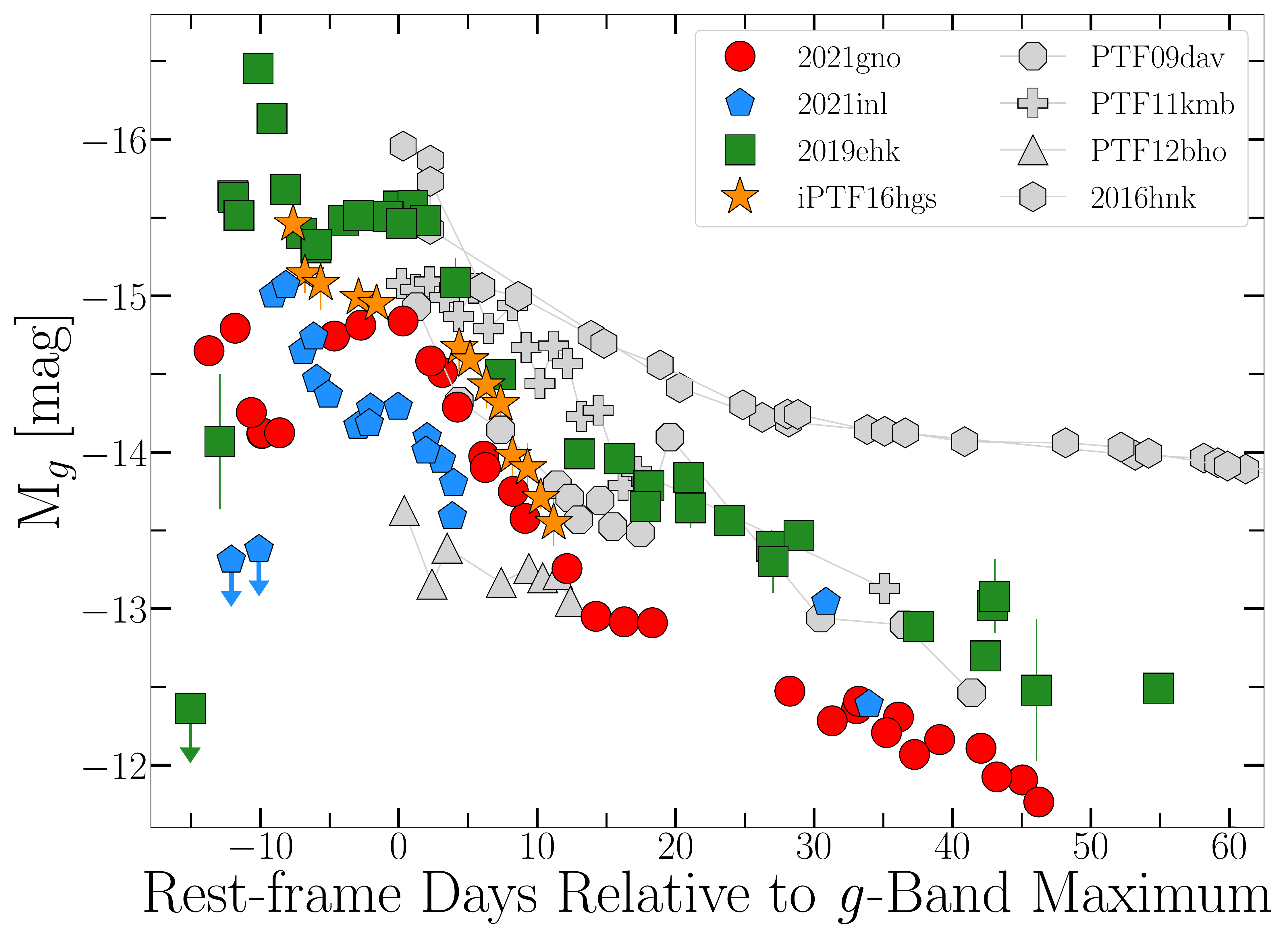}}
\subfigure[]{\includegraphics[width=0.49\textwidth]{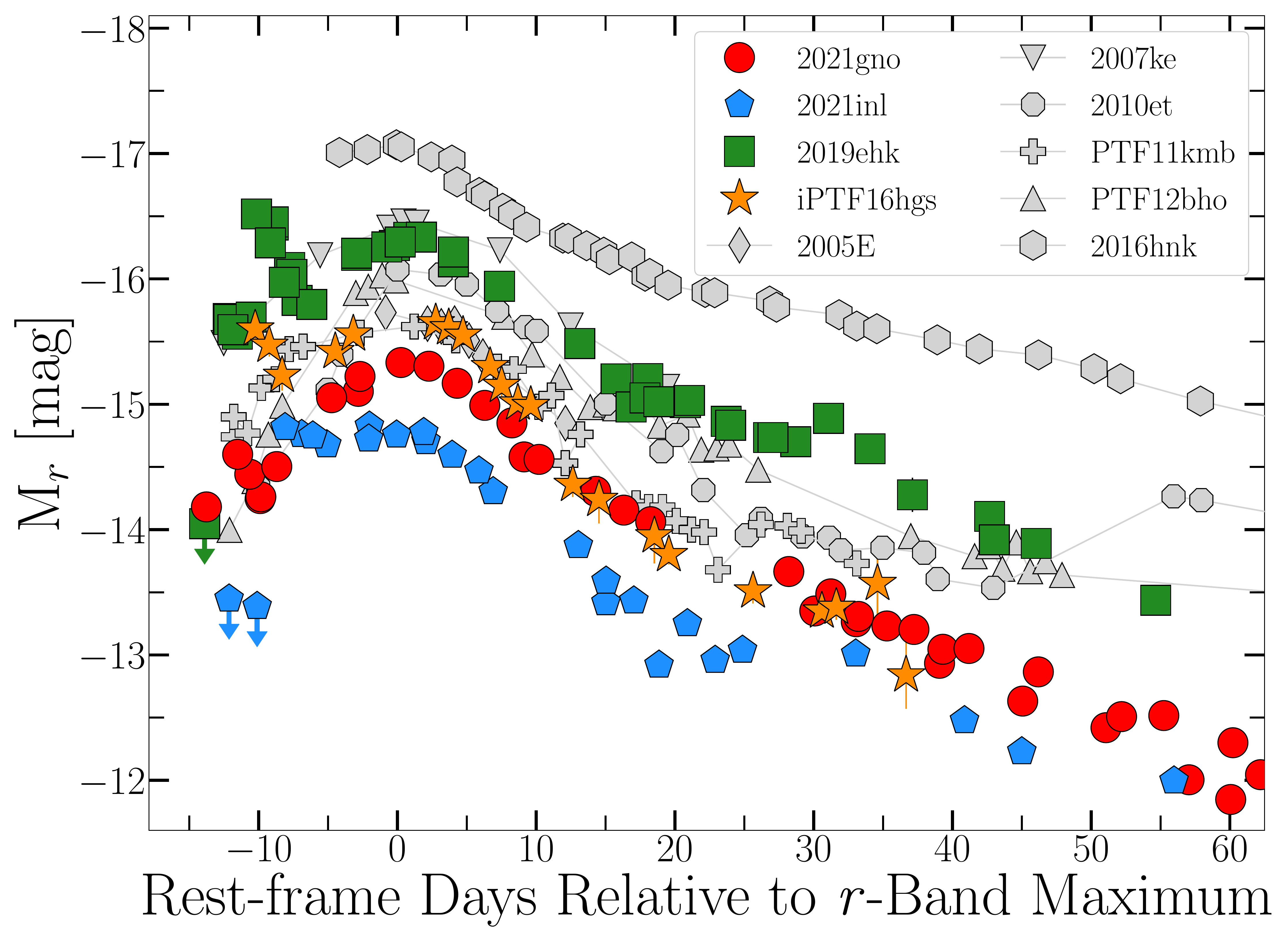}}
\caption{(a) Early-time $g-$band comparison of \sng\ (red circles), \sni\ (blue polygons), and classified \cas{} \citep{sullivan11, lunnan17, de18, wjg20,wjg19}. The peculiar, ``calcium-strong'' SN~2016hnk also presented for reference (gray hexagons). SNe~2021gno and 2021inl are the fourth and fifth objects in this class to show a double-peaked light curve, iPTF16hgs (orange stars), SN~2019ehk (green squares), and SN~2018lqo \citep{de20_carich} being the first three confirmed cases. (b) $r-$band comparison of \sng\ (red circles), \sni\ (blue polygons), and classified \cas.  \label{fig:carich_comp_LC} }
\end{figure*}

In Figure \ref{fig:carich_comp_LC}, we present $r-$ and $g-$band light curve comparisons of SNe~2021gno and 2021inl to a sample of confirmed \cas{}. Overall, both objects have a consistent light curve evolution to other \cas{} e.g., $t_r \lesssim 15$~days, $M_{\rm peak} > -16.5$~mag, and fast-decaying post-maximum photometry. Both SNe are amongst the lowest luminosity events compared to other \cas{}, with \sni{} being $\sim$1~mag fainter than SNe~2005E \citep{perets10} and 2019ehk \citep{wjg20, nakaoka21} and $\sim$2~mag less luminous than the peculiar "Calcium-strong" SN~2016hnk \citep{galbany19, wjg19}. Despite being intrinsically fainter, \sng{}'s overall photometric evolution is most similar to SN~2019ehk and iPTF16hgs \citep{de18}; all three objects contain double-peaked light curves, as well as consistent rise-times and post-peak decline rates in both $g-$ and $r-$bands. \sni{}'s post-maximum decline is also consistent with SNe~2019ehk, 2021gno and iPTF16hgs, with all objects exhibiting a relatively rapid decay in $g-$band flux following the Ni-powered SN peak. Additionally, we compare the $g-r$ colors of SNe~2021gno and 2021inl to a \ca{} sample in Figure \ref{fig:colors}. Same as the photometric evolution, the overall $g-r$ color evolution of these two objects at $\delta t < 70$~days post-peak is quite consistent with the colors typically observed in other \cas{}. Similar to other objects observed early and with high cadence observations (e.g., iPTF16hgs, SN~2019ehk), SNe~2021gno and 2021inl display blue colors at the start of their evolution ($g-r < 0$~mag), but quickly transform into instrinsically red explosions ($g-r >1$~mag) following SN peak. 

\subsection{Bolometric Light Curve}\label{subsec:bol_LC}

For \sng{}, we construct a bolometric light curve by fitting the ZTF, PS1, Nickel, ATLAS and \textit{Swift} photometry with a blackbody model that is dependent on radius and temperature. The extremely blue UV colors and early-time color evolution of \sng{} during its first light peak impose non-negligible deviations from the standard \textit{Swift}-UVOT count-to-flux conversion factors. We account for this effect following the prescriptions by \cite{brown10}. Each spectral energy distribution (SED) of \sng{} was generated from the combination of multi-color UV/optical/NIR photometry in the $w2$, $m2$, $w1$, $u$, $b$, $v$, $g$, $c$, $o$, $r$, $i$, and $z$ bands (1500--10000\,\AA). Similarly, for \sni{}, we construct a bolometric light curve by fitting the ZTF, PS1, Nickel, and ATLAS photometry with the same blackbody model to multi-band $g$, $c$, $o$, $r$, $i$, and $z$ bands (3000--10000\,\AA). For both SNe, we extrapolated between light curve data points using a low-order polynomial spline in regions without complete color information. All uncertainties on blackbody radii and temperature were calculated using the co-variance matrix generated by the SED fits. For the secondary, Nickel-powered light curve peak, we find peak bolometric luminosities of $(4.12 \pm \: 1.57) \times 10^{41} \: \mathrm{erg\:s^{-1}}$ and $(2.37 \pm \: 0.05) \times 10^{41} \: \mathrm{erg\:s^{-1}}$ for \sng{} and \sni{}, respectively.

\begin{figure*}
\centering
\subfigure[]{\includegraphics[width=0.49\textwidth]{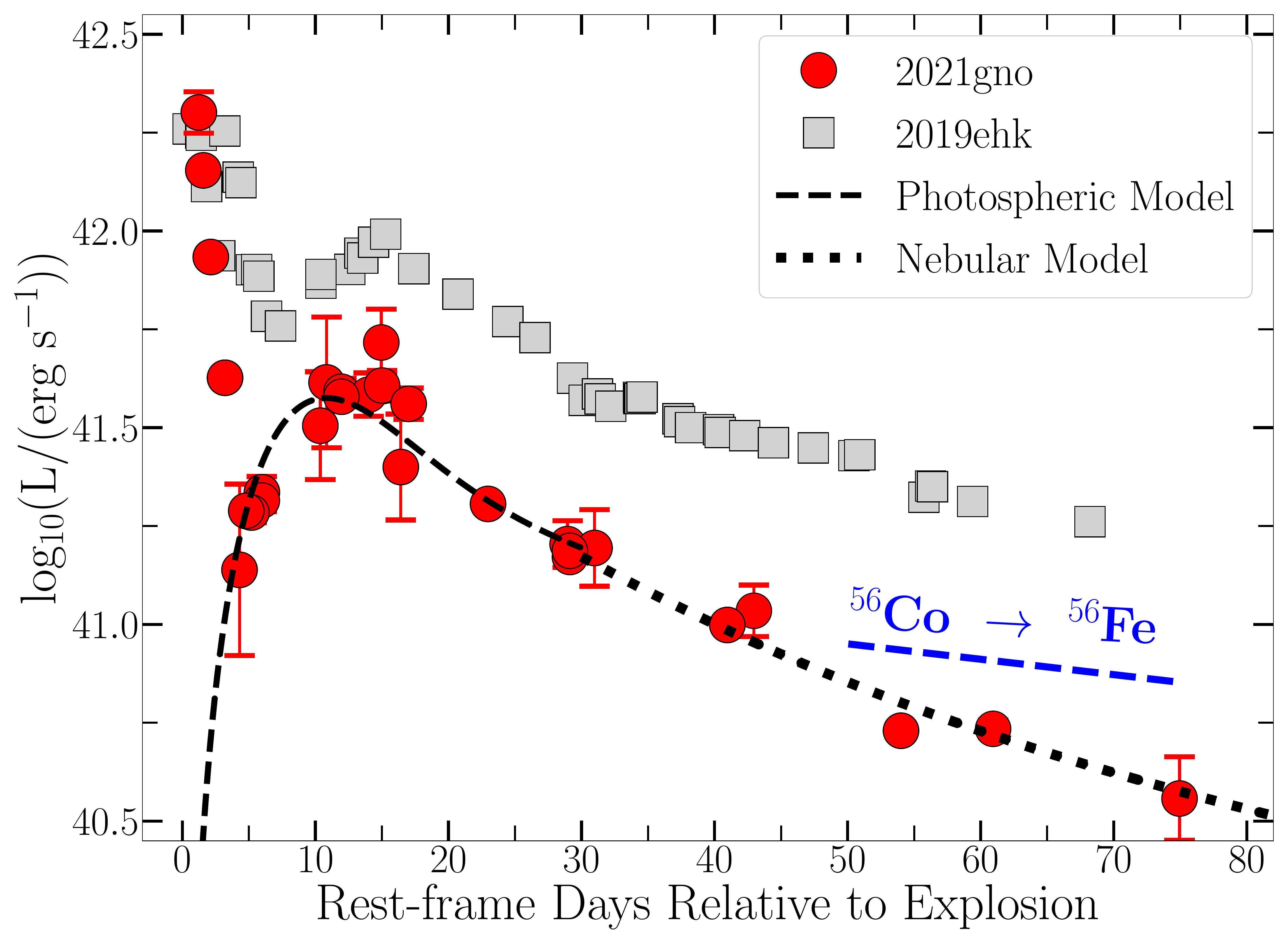}}
\subfigure[]{\includegraphics[width=0.49\textwidth]{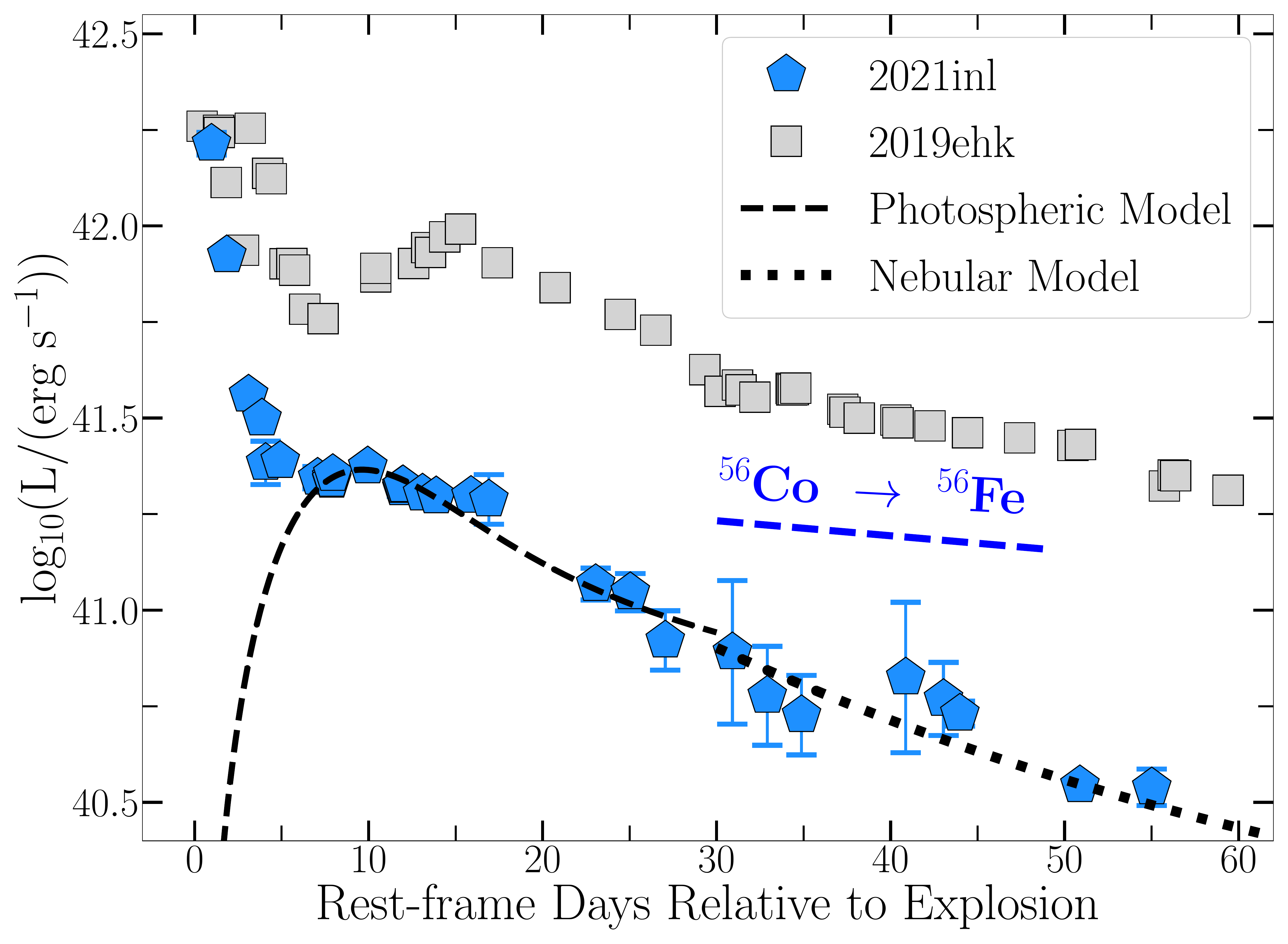}}
\caption{(a) Bolometric light curves of SNe~2021gno (red circles) and 2019ehk (gray squares). Secondary, ${}^{56}\textrm{Ni}$-powered peak in \sng{} is at a phase of $\delta t \approx 10$~days, while the primary peak from shock cooling emission or CSM interaction is during phases $\delta t < 5$~days. Photospheric light curve model for the early-time light curve of SN~2021gno (\S\ref{subsec:bol_LC}) is plotted as dashed black line. Modeling of the nebular phase data plotted as dotted black line. Blue dashed line shows the luminosity decline rate for a radioactive decay powered light curve with complete $\gamma-$ray trapping. (b) Bolometric light curves of SNe~2021inl (blue polygons) and 2019ehk (gray squares). Secondary, ${}^{56}\textrm{Ni}$-powered peak in \sni{} is at a phase of $\delta t \approx 10$~days. \label{fig:bol_LC} }
\end{figure*}

\begin{figure}[t!]
\centering
\includegraphics[width=0.45\textwidth]{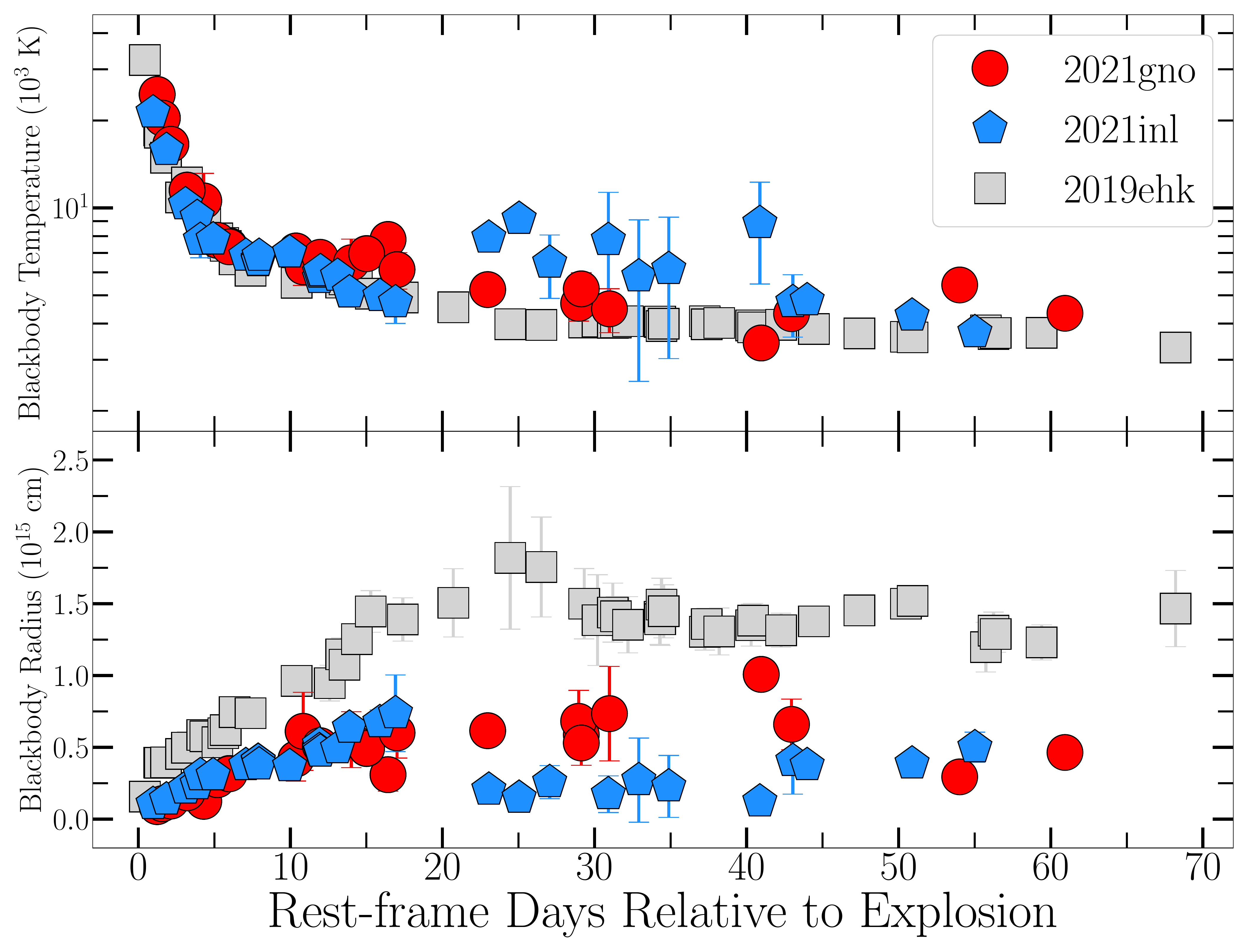}
\caption{Blackbody radii (bottom panel) and temperatures (top panel) derived from SED modeling of all multi-color optical photometry from SNe~2021gno (red circles), 2021inl (blue polygons), and 2019ehk (gray squares). \label{fig:BB_RT}}
\end{figure}

In Figures \ref{fig:bol_LC}(a)/(b), we present the bolometric light curves of SNe~2021gno and 2021inl, in addition to their blackbody radius and temperature evolution in Figure \ref{fig:BB_RT}. In both figures, we also present the bolometric luminosities, blackbody radii and temperatures of \ca{} SN~2019ehk (WJG20a). Overall, both SNe are less luminous than SN~2019ehk throughout all of their bolometric evolution except the very first data point of the first light curve peak where the luminosities are comparable. However, the post-peak bolometric decline in SNe~2021gno and 2021inl is consistent with the rate observed in SN~2019ehk; all of these objects declining faster than the typical decay of ${}^{56}\textrm{Co}$ $\rightarrow$ ${}^{56}\textrm{Fe}$ that assumes complete trapping of $\gamma$-rays. Furthermore, as shown in Figure \ref{fig:BB_RT}, the blackbody temperature of SNe~2019ehk, 2021gno, and 2021inl are all nearly identical throughout the early-time evolution, $\delta t < 70$~days since explosion. However, the blackbody radius of SN~2019ehk is larger than both SNe throughout their evolution, while \sng{} and \sni{} are consistent with one another for most early-time epochs. Additionally, it should be noted that the blackbody approximation may not be appropriate when emission lines (e.g., \ion{Ca}{ii}) begin to dominate the spectrum of SNe~2021gno and 2021inl, which occurs  $t > 40$ days after explosion. Consequently, a blackbody assumption for these objects in those phases is most likely an over-simplification and could result in additional uncertainty on the presented bolometric luminosities.

For \sng{}, the earliest inferred blackbody radius is $\sim9 \times 10^{13}$~cm ($\sim$1300~$\Rsun$) at $\delta t = 0.84$~days since explosion. This suggests a compact progenitor star with radius $R_{\star} \lesssim 10-100~\Rsun$, which allows for the first detected blackbody radius to be reached given a shock velocity of $v_s \approx 1.2 \times 10^4~\kms$. Similarly, the first blackbody radius of $\sim 10^{14}$~cm ($\sim$1400~$\Rsun$) in \sni{} at $\delta t = 0.97$~days also allows for a compact progenitor radius of $R_{\star} \lesssim 10-100~\Rsun$ for $v_s \approx 1.2 \times 10^{4}~\kms$. Similar inferences we made for SN~2019ehk whose initial blackbody radius at $\delta t \approx 0.4$~days after explosion rules out an extended progenitor. Furthermore, in all three SNe, WD progenitors are still permitted given the time it would take the SN shock to reach the first blackbody radii from a much smaller initial stellar radius. 

To determine physical parameters of both SNe such as ejecta mass ($M_{\rm ej}$), kinetic energy ($E_{\rm k}$), and ${}^{56}\textrm{Ni}$ mass ($M_{\textrm{Ni}}$), we model both bolometric light curves with the analytic expressions presented in Appendix A of \cite{valenti08} and in \cite{wheeler15}. Same as in SN~2019ehk, we exclude the first light curve peak and model two distinct phases of the light curve: photospheric ($\delta t < 30$~days; \citealt{arnett82}) and nebular ($\delta t > 40$~days; \citealt{sutherland84,Cappellaro97}). The analytic formalism applied in this modeling self-consistently implements the possibility of incomplete $\gamma$-ray trapping and a typical opacity of $\kappa = 0.1$ cm$^2$ g$^{-1}$ is applied in each model. Furthermore, we correct for the known degeneracy between kinetic energy and ejecta mass (e.g., see Eqn. 1 in WJG20a) by applying photospheric velocities of $v_{\rm ph} \approx 6000~\kms$ for \sng{} and $v_{\rm ph} \approx 7500~\kms$ for \sni{}, both of which are derived from \ion{Si}{ii} absorption features in the SN spectra. For \sng{}, we find an ejecta mass of $M_{\rm ej} = 0.60 \pm 0.01~\Msun$, kinetic energy of $E_{\rm k} = (1.3 \pm 0.2) \times 10^{50}$~erg, and ${}^{56}\textrm{Ni}$ mass of $M_{\textrm{Ni}} = (1.20 \pm 0.02) \times 10^{-2}~\Msun$. For \sni{}, we calculate an ejecta mass of $M_{\rm ej} = 0.29 \pm 0.01~\Msun$, kinetic energy of $E_{\rm k} = (9.6 \pm 0.4) \times 10^{49}$~erg, and ${}^{56}\textrm{Ni}$ mass of $M_{\textrm{Ni}} = (6.90 \pm 0.06) \times 10^{-3}~\Msun$. In both SNe, the photospheric and nebular model fits (shown in Fig. \ref{fig:bol_LC}) return consistent parameter values. Overall, the explosion parameters in \sng{} are very consistent with those derived for SN~2019ehk (WJG20a) despite a slightly lower $M_{\textrm{Ni}}$, which explains the larger luminosities observed in SN~2019ehk. However, \sni{}'s explosion parameters are all lower than that observed in SNe~2019ehk and 2021gno, but are consistent with the values generally found in the \ca{} class \citep{de20}.

\section{Optical/NIR Spectral Analysis}\label{sec:spectro_analysis}

\subsection{Spectroscopic Properties}\label{subsec:spec_analysis}
The complete spectral series of SNe~2021gno and 2021inl are presented in Figures \ref{fig:spectral_series} and \ref{fig:21inl_spec_LC}(b), both of which include obvious ion identifications for both SNe. During their photospheric phase, both SNe display prominent \ion{He}{i}, \ion{O}{i}, \ion{Ca}{ii}, \ion{Si}{ii} and Fe-group element transitions; neither SNe showing evidence for detectable \ion{H}{i}. In the first spectrum of \sng{} at +3~days post-explosion, we find that all the broad features can be identified as fast-moving \ion{He}{i} $\lambda\lambda$~4471, 5016, 5876, 6678 profiles and find a consistent expansion velocity of $\sim$$1.3\times 10^{4}~\kms$ from the minimum of the absorption profile. Based on the absorption profiles in the \sng{} maximum light spectrum, we find characteristic ejecta velocities of $\sim$$7000-8000\,\kms$ for \ion{He}{i}, $\sim$$6500\,\kms$ for \ion{Si}{ii}, and $\sim$$7000\,\kms$ for \ion{Ca}{ii}. We find similar expansion velocities in \sni{}, such as $\sim$$(1-1.2)\times 10^4~\kms$ for \ion{He}{i}, $\sim$$7500~\kms$ for \ion{Si}{ii}, and $\sim$$8000\,\kms$ for \ion{Ca}{ii}. Overall, the ejecta velocities estimated for both SNe are consistent with ion velocities found for SN~2019ehk (WJG20a) and other \cas{} \citep{kasliwal12, lunnan17,  de20_carich}. 

\begin{figure}[t!]
\centering
\includegraphics[width=0.45\textwidth]{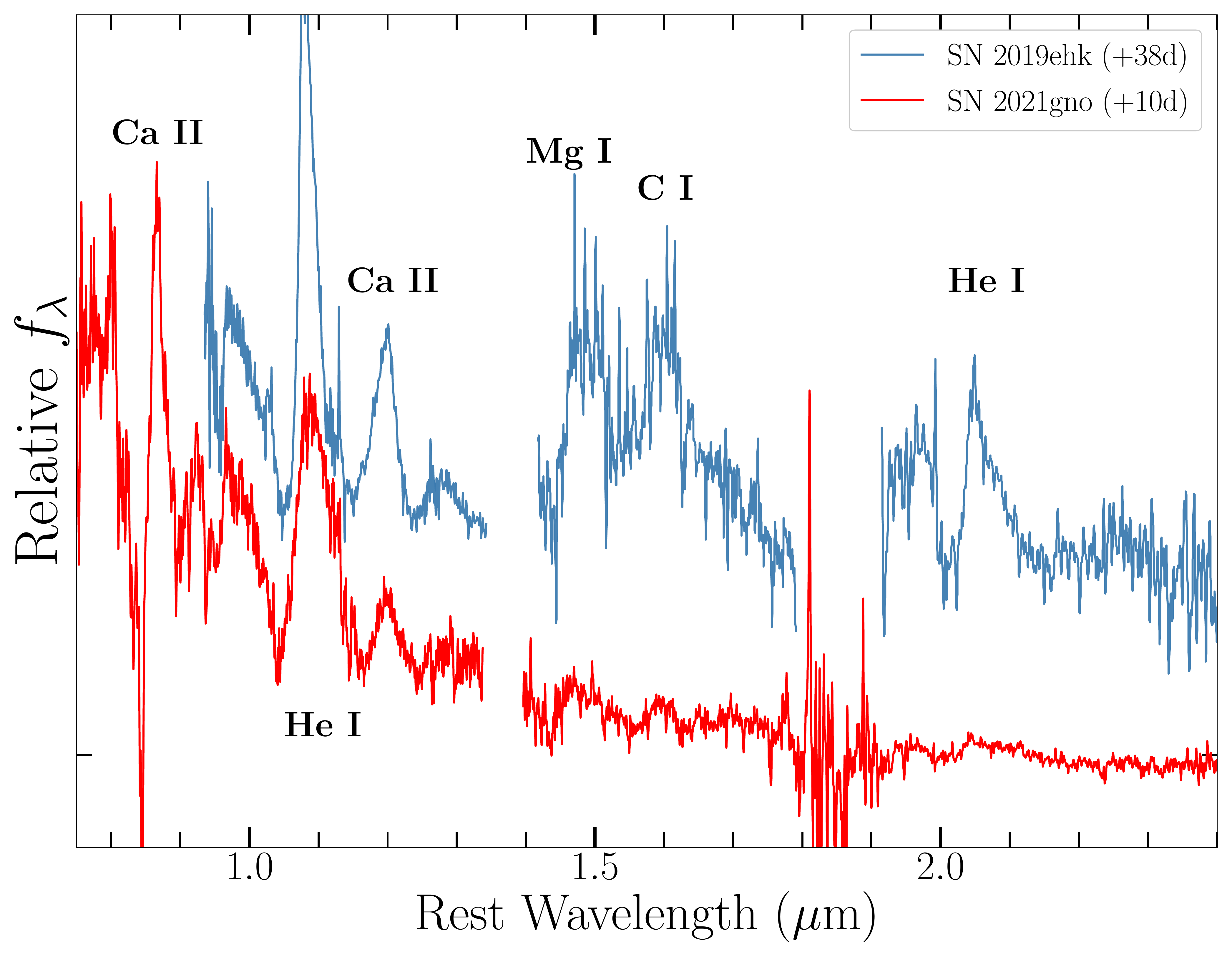}
\caption{SPEX NIR spectrum (red) of \sng{} at +10~days relative to second $B$-band peak. NIR spectrum of SN~2019ehk shown in blue (WJG20a); these being the only early-time NIR spectra taken of \cas{}. Prominent line transitions are marked in black.  \label{fig:ir_spec}}
\end{figure}

In Figure \ref{fig:ir_spec}, we present the IR spectra of \sng{} at +10~days after second maximum compared to SN~2019ehk at +38~days; these two observations being the only confirmed IR spectra of a \ca{} during the photospheric phase. The IR spectrum of \sng{} shows nearby identical transitions to SN~2019ehk, both objects showing clear P-Cygni profiles of \ion{Ca}{ii}, \ion{He}{i}, \ion{C}{i}, and \ion{Mg}{i}. Furthermore, the expansion velocities of these transitions are consistent with the ejecta velocities derived from optical spectra e.g., $\sim$$1.1\times 10^4~\kms$ for \ion{He}{i} and $\sim$$9000~\kms$ for \ion{Ca}{ii}. 

\begin{figure*}
\centering
\subfigure[]{\includegraphics[width=0.49\textwidth]{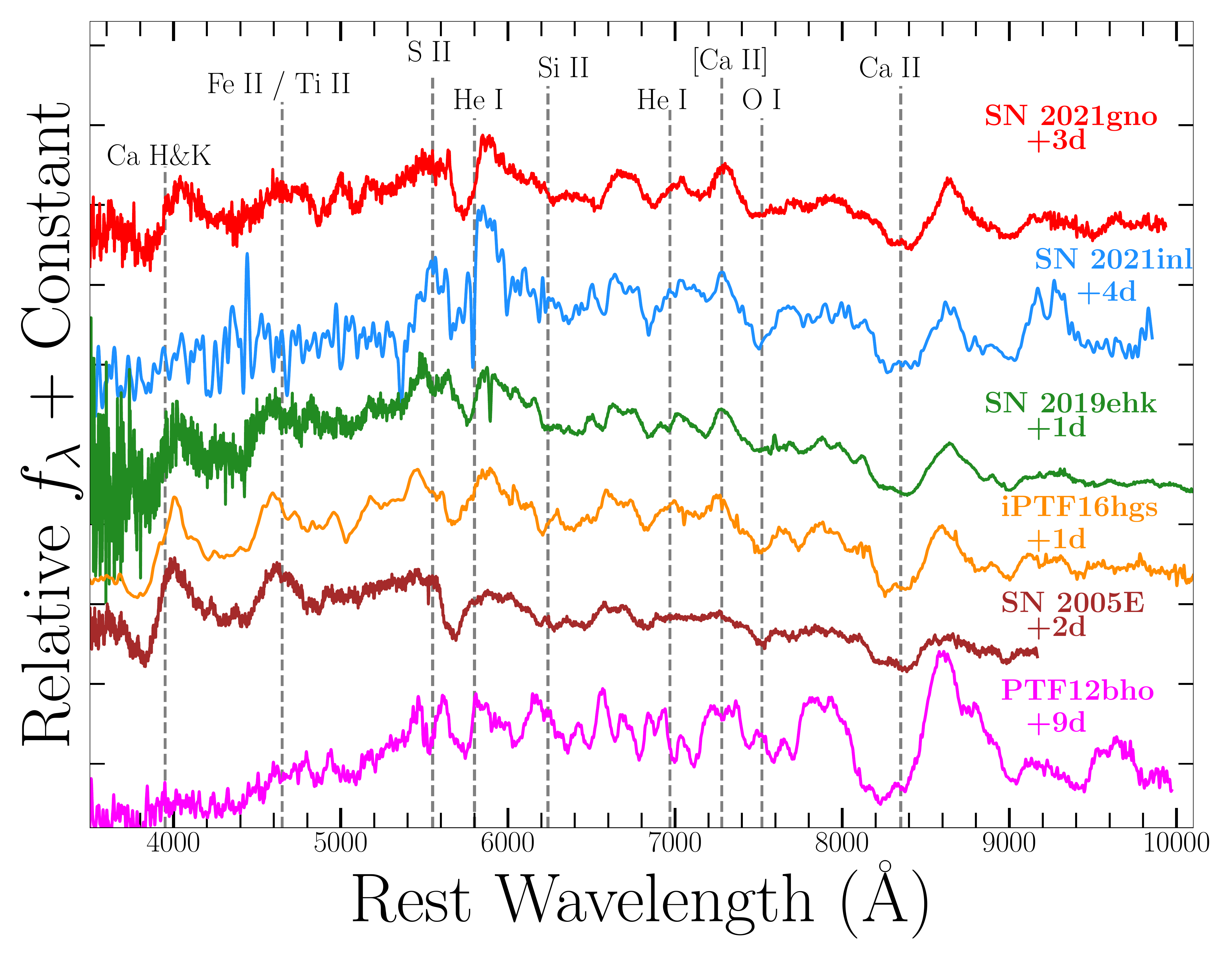}}
\subfigure[]{\includegraphics[width=0.49\textwidth]{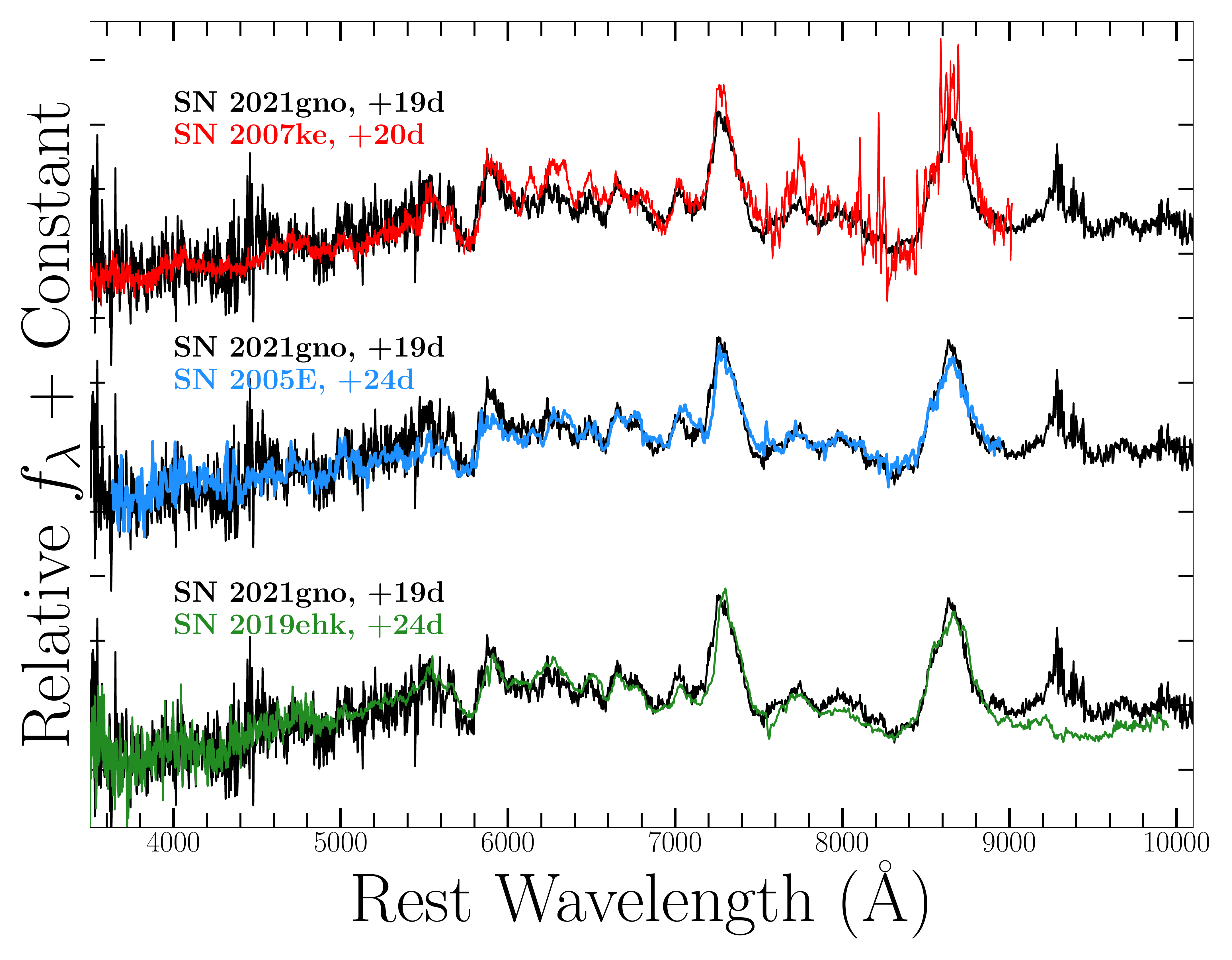}}
\caption{(a) Spectral comparison of \sng\ (red), \sni\ (blue), and other \cas \ near maximum light \citep{perets05, sullivan11, lunnan17, de18}. Common ions are marked by grey lines. (b) Direct spectral comparison of \sng\ (black) and \cas \ SNe~2007ke, 2005E, and 2019ehk at approximately the same phase \citep{perets05, lunnan17}. Almost every line transition is matched between spectra, with \sng\ showing similar \ion{Ca}{ii} emission to all other objects.  \label{fig:carich_comp_spec} }
\end{figure*}

In Figure \ref{fig:carich_comp_spec}(a), we present early-time spectral comparisons of SNe~2021gno and 2021inl to other \cas{} near (second) maximum light. Overall, both SNe show consistent spectral features to all plotted \cas{}, but are most similar to SN~2019ehk and iPTF16hgs at this phase. All four objects show prominent \ion{He}{i} and \ion{Ca}{ii} transitions as well as the fast emergence of a [\ion{Ca}{ii}] emission profile relative to peak. Furthermore, we compare the mid-time spectra of \sng{} at +19~days to SNe~
2005E, 2007ke, and 2019ehk in Figure \ref{fig:carich_comp_spec}(b). At this phase, \sng{} shows nearly identical transitions to these \cas{} such as prominent [\ion{Ca}{ii}] and marginal [\ion{O}{i}]. These spectral comparisons are further indication that both SNe~2021gno and 2021inl are clear members of the \ca{} class. 

\subsection{Inferences from Nebular Phase Spectroscopy}\label{subsec:nebular_gas}

\begin{figure*}[t!]
\centering
\includegraphics[width=\textwidth]{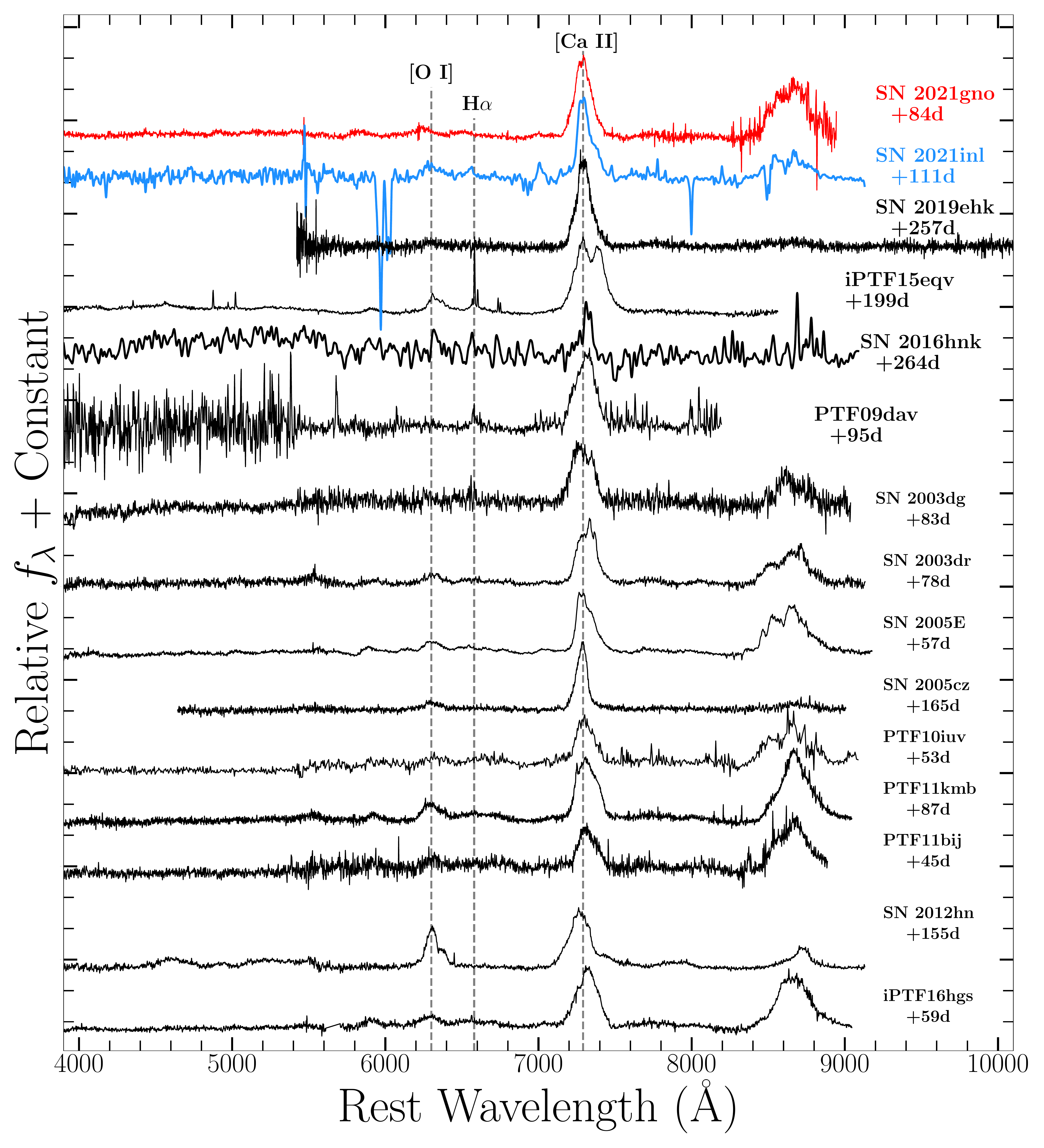}
\caption{Nebular spectra of all confirmed \cas{} \citep{ sullivan11,kasliwal12, foley15, lunnan17, milisavljevic17, wjg20,wjg19}. Nebular spectra of \sng\ at +84~days and \sni\ at +111~days, shown in red and blue, respectively, both spectra further establishing these objects as \cas{}. Prominent [\ion{O}{i}] and [\ion{Ca}{ii}] lines as well as H$\alpha$ marked by dashed grey lines. \label{fig:carich_nebular}  }
\end{figure*}

Similar to other \cas{}, SNe~2021gno and 2021inl show a fast transition from the photospheric to the optically thin regime where their spectra become dominated by forbidden emission lines such as [\ion{Ca}{ii}] and [\ion{O}{i}] (Fig. \ref{fig:carich_nebular}). For \sng{}, the transition to the nebular regime occurs at $\sim$13-18~days after explosion, which is evident from the presence of [\ion{Ca}{ii}] emission in the early-time spectra; this transition then comes to dominate the spectra at later phases (Fig. \ref{fig:spectral_series}). Despite lower cadence spectroscopic observations, a similar behavior is observed in \sni{}, whose first spectrum at +8~days shows marginal evidence for [\ion{Ca}{ii}] emission, which later becomes the dominant transition by +94~days (Fig. \ref{fig:21inl_spec_LC}b). 

Once in the optically thin regime, we calculate [\ion{Ca}{ii}]/[\ion{O}{i}] line flux ratio, which, if greater than 2, is a common classifier of \cas{} and present this quantity in Figure \ref{fig:caii_oi}(a) for both SNe. Based on this metric, we find that both objects are significantly ``rich'' in [\ion{Ca}{ii}] emission as shown by a maximum line flux ratio of [\ion{Ca}{ii}]/[\ion{O}{i}]~$\approx 10$. These flux ratios are consistent with other \cas{} presented in Figure \ref{fig:caii_oi}(a), but neither SN has as large of a [\ion{Ca}{ii}]/[\ion{O}{i}] ratio as SN~2019ehk, which remains the member of \ca{} with largest [\ion{Ca}{ii}] flux relative to [\ion{O}{i}] at all phases. Furthermore, in Figures \ref{fig:caii_oi}(b)/(c), we present the velocity profiles [\ion{Ca}{ii}] and [\ion{O}{i}] of SNe~2021gno and 2021inl, respectively, with the O emission scaled to match the Ca feature. We find that in both objects, these forbidden line transitions are consistent in shape and indicate [\ion{Ca}{ii}] and [\ion{O}{i}] expansion velocities of $\sim 5000-6000~\kms$ based on the FWHM of the emission profiles. 

In order to understand the Ca and O abundance in each explosion, we apply a similar analysis to that outlined in Section 6.3 of WJG20a where the observed luminosities of [\ion{Ca}{ii}] and [\ion{O}{i}] are related to the populations of the excited states, ion number densities ($n_e > 10^7$~cm$^{-3}$), and Einstein A coefficient values of each ion:

\begin{equation}
    L_{\textrm{[\ion{O}{i}]}} = n_{\textrm{\ion{O}{i}}} ~ A_{\textrm{[\ion{O}{i}]}} ~ h\nu_{\textrm{[\ion{O}{i}]}} ~(5/14)~e^{-22000/T}
\end{equation}

\begin{equation}
    L_{\textrm{[\ion{Ca}{ii}]}} = n_{\textrm{\ion{Ca}{ii}}} ~ A_{\textrm{[\ion{Ca}{ii}]}} ~ h\nu_{\textrm{[\ion{Ca}{ii}]}} ~(10/11)~e^{-19700/T}
\end{equation}

\noindent
where $h \nu$ is the photon energy, $n$ is the ion number density, $A_{\textrm{[\ion{Ca}{ii}]}}=2.6$ s$^{-1}$, $A_{\textrm{[\ion{Ca}{ii}]}}\approx390 A_{\textrm{[\ion{O}{i}]}}$, the exponentials are the Boltzmann factors ($T$ is in K), and the numerical factors are statistical weights. To find the ion number densities and subsequent masses in each SN, we first estimate the forbidden line luminosities to be $L_{\textrm{[\ion{O}{i}]}} = 3.9 \times 10^{38}$ erg s$^{-1}$ and $L_{\textrm{[\ion{Ca}{ii}]}} = 3.5 \times 10^{39}$ erg s$^{-1}$ for \sng{} ($\delta t = 84$~days since explosion), and $L_{\textrm{[\ion{O}{i}]}} = 8.2 \times 10^{38}$ erg s$^{-1}$ and $L_{\textrm{[\ion{Ca}{ii}]}} = 3.6 \times 10^{39}$ erg s$^{-1}$ for \sni{} ($\delta t = 111$~days since explosion). In the analytic relations above, we choose to calculate Ca and O masses for a range of temperatures $T = 5000-10^4$~K, for completeness. 

For \sng{}, we calculate O and Ca masses of $M(\rm{O}) = (0.6 - 6) \times 10^{-2}~\Msun$ and $M(\rm{Ca}) = (1 - 9) \times 10^{-4}~\Msun$, for temperatures $T = 10^4 - 5000$~K. Similarly, for \sni{}, we find O and Ca masses of $M(\rm{O}) = 0.01-0.1~\Msun$ and $M(\rm{Ca}) = (1 - 10) \times 10^{-4}~\Msun$, for $T = 10^4 - 5000$~K. These abundances are lower overall, but still somewhat consistent, to those found by WJG20a for SN~2019ehk e.g., $M(\rm{O}) = 0.10~\Msun$ and $M(\rm{Ca}) =4 \times 10^{-3}~\Msun$. However, it should be noted that at these phases both SNe are not fully nebular and therefore the derived masses may be lower than the true elemental masses in the explosion. Nevertheless, these mass estimates continue to prove that the ``richness'' of Ca emission in \cas{} is not due to a larger intrinsic amount of Ca relative to O, but rather it is likely the result of relative abundances and ionization temperatures in the inner, low density ejecta. 

\section{Early-time Flux Excess}\label{sec:flare}

\subsection{Observational Properties}\label{sec:flare_specifics}

\begin{figure*}
\centering
\subfigure[]{\includegraphics[width=0.32\textwidth]{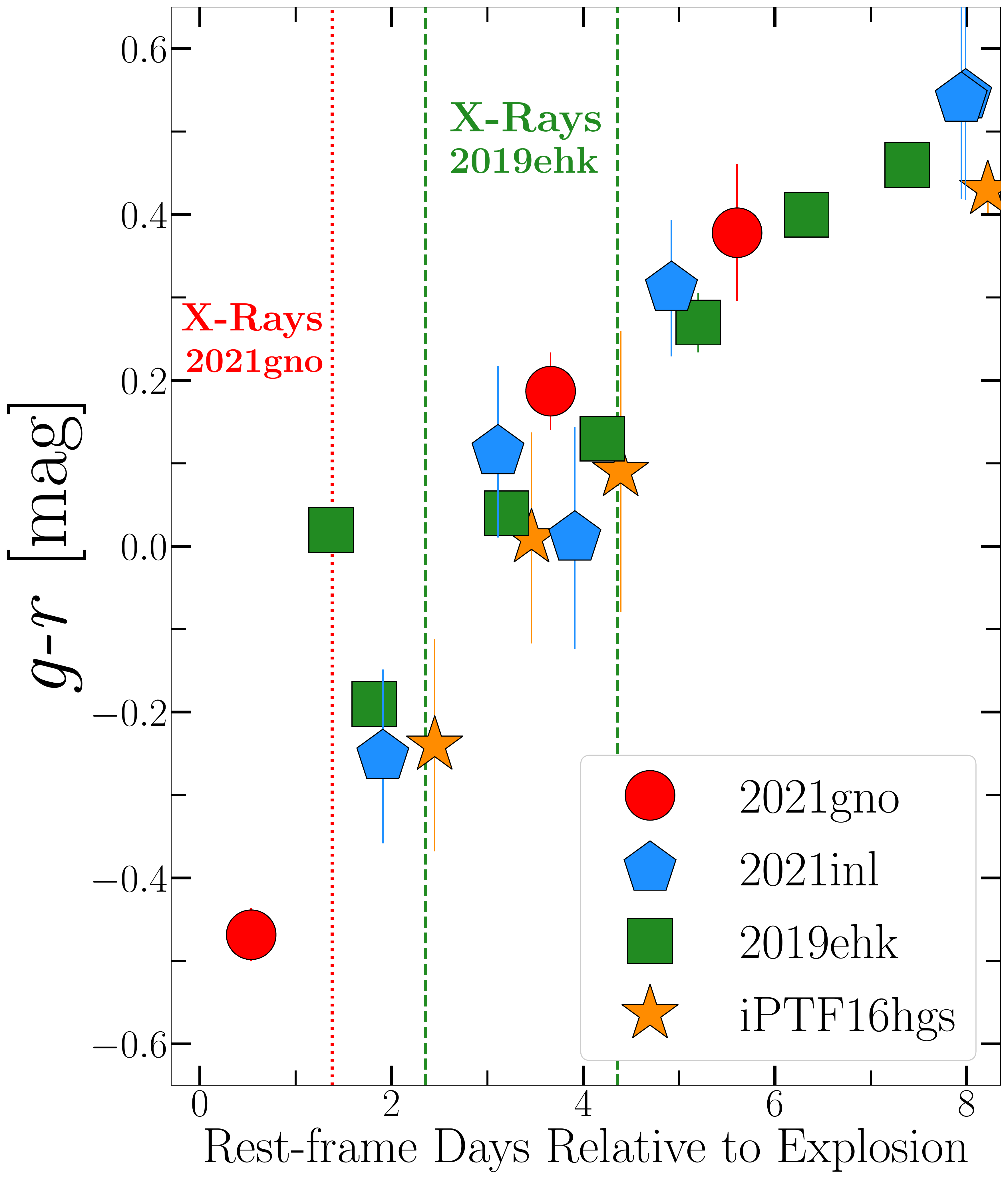}}
\subfigure[]{\includegraphics[width=0.33\textwidth]{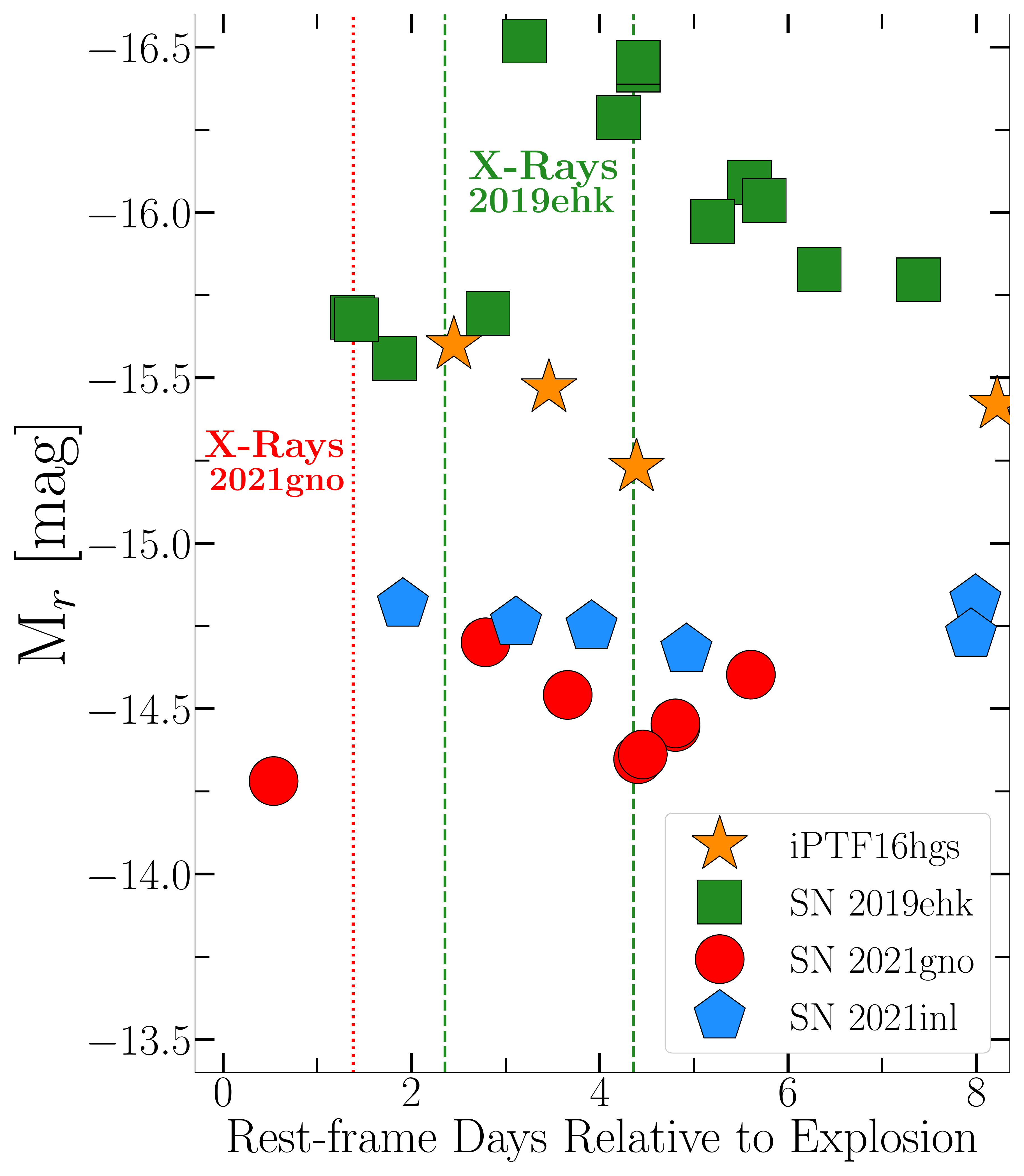}}
\subfigure[]{\includegraphics[width=0.33\textwidth]{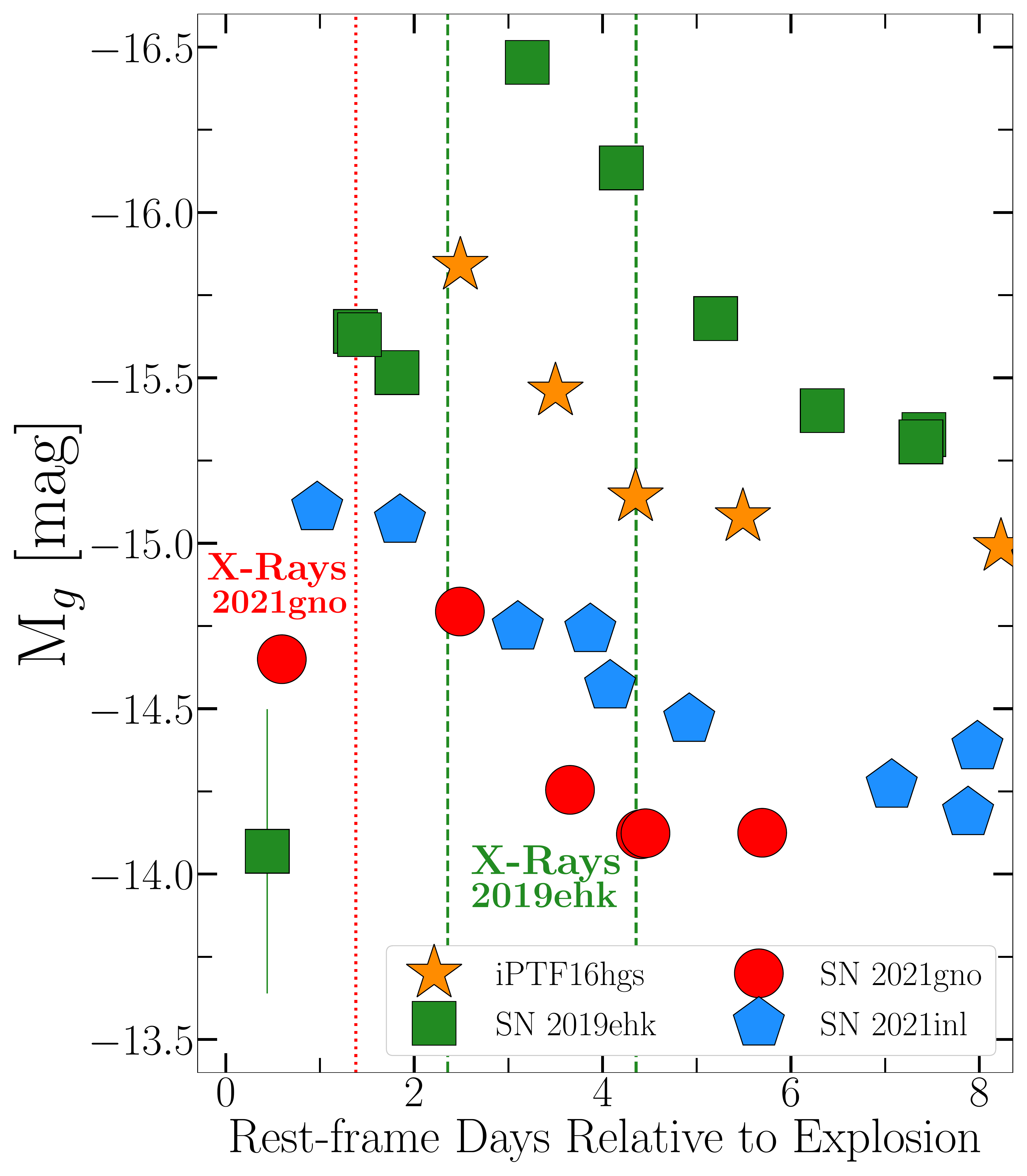}}
\caption{(a) \textit{g-r} color comparison of SNe~2021gno (red circles), 2021inl (blue polygons), 2019ehk (green squares), and iPTF16hgs (orange stars) during the primary light curve peak. Phases of X-ray detections in SNe~2021gno and 2019ehk shown as dotted red and dashed green lines, respectively. (b)/(c) Absolute magnitude $r-$ and $g-$band (left and right) photometry of all four \cas \ during their primary light curve peak. \label{fig:flare_compare} }
\end{figure*}

Similar to other double-peaked \cas{}, the early-time excess in flux above the ${}^{56}\textrm{Ni}$-powered continuum is observed in all available UV/optical/NIR filters used to observe SNe~2021gno and 2021inl. Additionally, these very early-time observations of both SNe represent the only other instances where the initial rise of the primary light curve peak was recorded in a \ca{}, the first being in SN~2019ehk. In Figure \ref{fig:flare_compare}, we present the $g-r$ colors, as well as $r-$ and $g-$band light curves of SNe~2019ehk, 2021gno, 2021inl, and iPTF16hgs during their primary light curve phase. 

For all four double-peaked \cas{} in Figure \ref{fig:flare_compare}(a), the $g-r$ color evolution during the flux excess follows a consistent trend: all objects show a linear increase in color following first detection and all begin with quite blue colors e.g., $g-r < -0.2$~mag. Seemingly, the physical process behind this early-time flux excess is responsible for a retention of high blackbody temperatures and, consequently, blue colors until the SN emission becomes dominated by energy injection from ${}^{56}\textrm{Ni}$ decay. 

As shown in Figures \ref{fig:flare_compare}(b)/(c), SN~2019ehk remains the most luminous double-peaked \ca{}, with its flux excess peaking at $M \approx -16.5$~mag in $g-$ and $r-$bands. SNe~2021gno and 2021inl are lower luminosity events than SN~2019ehk and iPTF16hgs, with their primary $g-$ and $r-$band light curves peaking at $M \approx -14.8$~mag and $M \approx -15.2$~mag, respectively. Furthermore, the light curve slopes during this phase varies between all \cas{}. \sng{} shows a $g-$band decline rate of $\Delta {\rm m(g)}_5 = 0.52$~mag during the $\sim$5~day primary peak duration while \sni{} has a decline rate of $\Delta {\rm m(g)}_7 = 0.64$~mag. Additionally, SN~2019ehk has a very  fast decline rate of $\Delta {\rm m(g)}_5 = 1.1$~mag during its largest flare in early-time flux, while iPTF16hgs has a similarly rapid decline of $\Delta {\rm m(g)}_3 = 0.75$~mag. 

\subsection{Shock Breakout and Envelope Cooling Model}\label{subsec:flare_shockcool}

For stellar progenitors with an extended envelope, the energy deposited by the passage of a shock through their envelopes manifests in detectable shock cooling emission (SCE) on a timescale of $t \lesssim$ days after shock breakout. This process has been modeled both analytically (e.g., \citealt{nakar14, piro15}) and numerically (e.g., \citealt{sapir17, piro17, piro21}), these models being highly effective at reproducing the early time double-peaked light curves of SNe~IIb (e.g., SNe~1993J, 2011dh, 2016gkg, 2017jgh; \citealt{wheeler93, arcavi11, arcavi17, piro17, Armstrong21}), super-luminous SNe (e.g., DES14X3taz; \citealt{smith16}), SNe Ic (e.g., SNe~2014ft, 2020bvc, 2020oi; \citealt{de18sci, ho20, gagliano22}), fast-risers (e.g., 2019dge; \citealt{yao20} and \cas{} (e.g., iPTF16hgs, SN~2019ehk; \citealt{de18, nakaoka21, wjg20}). Furthermore, by fitting the primary light curve peaks of these double-peaked SNe, information about the extended material around the progenitor star at the time of explosion can be derived, such as the envelope mass and radius, as well as the shock velocity. 

In order to understand the physical origin of their early-time flux excess, we fit the primary light curve peaks of SNe~2021gno and 2021inl with models for SCE of extended material. We apply four models to fit the SN light curves: the original SCE model by \cite{piro15} as well as the revised two-component formalism presented in \cite{piro21}, in addition to the models of \cite{sapir17} who numerically model SCE from both red and blue supergiant, H-rich envelopes (polytropic index of n = 3/2 and n = 3, respectively). Presentation of the analytic expressions behind these models can be found in \cite{arcavi17} or Section 7.3 of WJG20a. Following shock breakout, each model produces constraints on the envelope mass, $M_e$, envelope radius, $R_e$, the velocity of the envelope, $v_e$, and the time offset from explosion $t_o$ (consistent with our explosion time estimate). In this analysis, we use \texttt{emcee}, a Python-based application of an affine invariant MCMC with an ensemble sampler \citep{foreman-mackey13}. We compile the best fit parameter estimates from each model in Table \ref{tbl:shocktable}.

\begin{figure*}
\centering
\includegraphics[width=\textwidth]{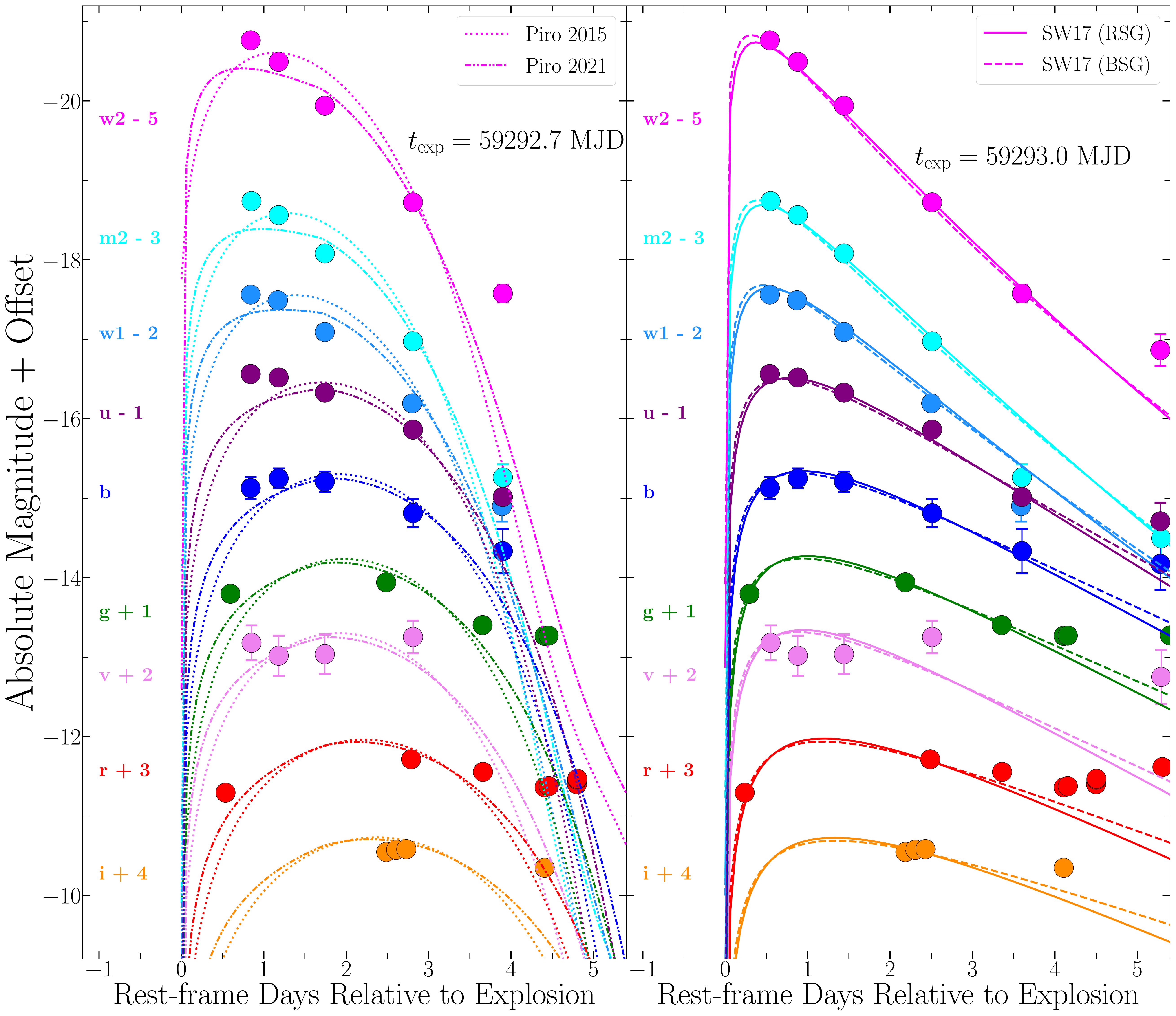}
\caption{Multi-color shock cooling model fits to the first light curve peak in \sng\ assuming a blackbody SED. \textit{Left:} \cite{piro15} and \cite{piro21} models are presented as dotted and dot-dash lines, respectively. \textit{Right:} \cite{sapir17} models shown as dashed (n=3) and solid (n=3/2) lines. Model specifics are discussed in \S\ref{subsec:flare_shockcool} and physical parameters are presented in Table \ref{tbl:shocktable}. \label{fig:SC_21gno}  }
\end{figure*}

\begin{figure}[t!]
\centering
\includegraphics[width=0.45\textwidth]{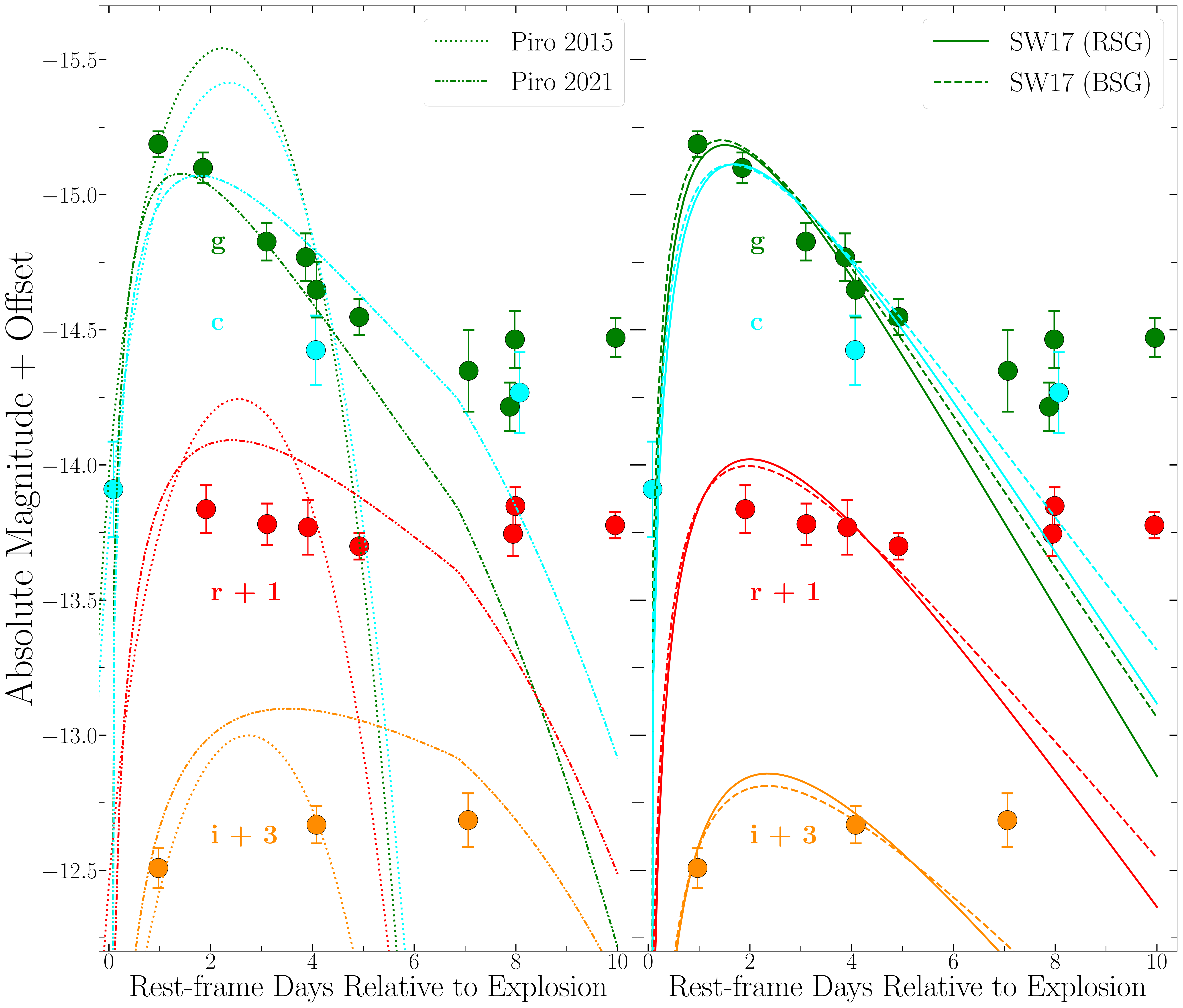}
\caption{Multi-color shock cooling model fits to the first light curve peak in \sni\ assuming a blackbody SED. \textit{Left:} \cite{piro15} and \cite{piro21} models are presented as dotted and dot-dash lines, respectively. \textit{Right:} \cite{sapir17} models shown as dashed (n=3) and solid (n=3/2) lines. Model specifics are discussed in \S\ref{subsec:flare_shockcool} and physical parameters are presented in Table \ref{tbl:shocktable}. \label{fig:SC_21inl}}
\end{figure}

In Figures \ref{fig:SC_21gno} \& \ref{fig:SC_21inl}, we present the best-fitting multi-color light curves of the aforementioned models for SNe~2021gno and 2021inl, respectively. We also present model bolometric light curves, as well as their blackbody temperatures and radii, in Figure \ref{fig:BB_models} with respect to the SNe~2021gno and 2021inl data. In general, we find that SCE can accurately reproduce the early-time flux excess in both objects, with the models of \cite{sapir17} providing the best fit and lowest $\chi^2$ value overall. From all four model fits, the SN~2021gno light curve is best reproduced by an extended mass $M_e \approx 0.013 - 0.47~\Msun$ with radius $R_e \approx 27.5 - 385~\Rsun$ and shock velocity $v_e \approx (4.5 - 7.8) \times 10^4~\kms$. Furthermore, we find two best fitting times of explosions: 59292.3 MJD for \cite{piro15,piro21} models and 59293.01 MJD for \cite{sapir17} models, both values being consistent with the model-independent estimate of $t_{\rm exp} = 59292.7 \pm 0.55$~MJD. Additionally, for SN~2021inl, we find best fitting SCE model parameters of $M_e \approx 0.02 - 1.61~\Msun$, $R_e \approx 20.5 - 207~\Rsun$ and $v_e \approx (4.5 - 7.8) \times 10^4~\kms$; there is no change to the original explosion date estimate. Lastly, we caution against using the $M_e$ derived from the blue supergiant SCE model by \cite{sapir17} to best understand the progenitor environments of these \cas{} given that the estimated envelope mass is larger than the ejecta mass in SN~2021inl and a significant fraction of the mass of SN~2021gno, both scenarios being unphysical in nature. We therefore conclude that the most physical range of best fitting extended masses for both objects are $M_e = (1.5 - 4.5) \times 10^{-2}~\Msun$ for \sng{} and $M_e \approx 2.3 \times 10^{-2}~\Msun$ for \sni{}; only the \cite{piro15} model returned a mass that was not comparable in size, and consequently unphysical, to SN~2021inl's total ejecta mass.

\begin{figure}[t!]
\centering
\includegraphics[width=0.45\textwidth]{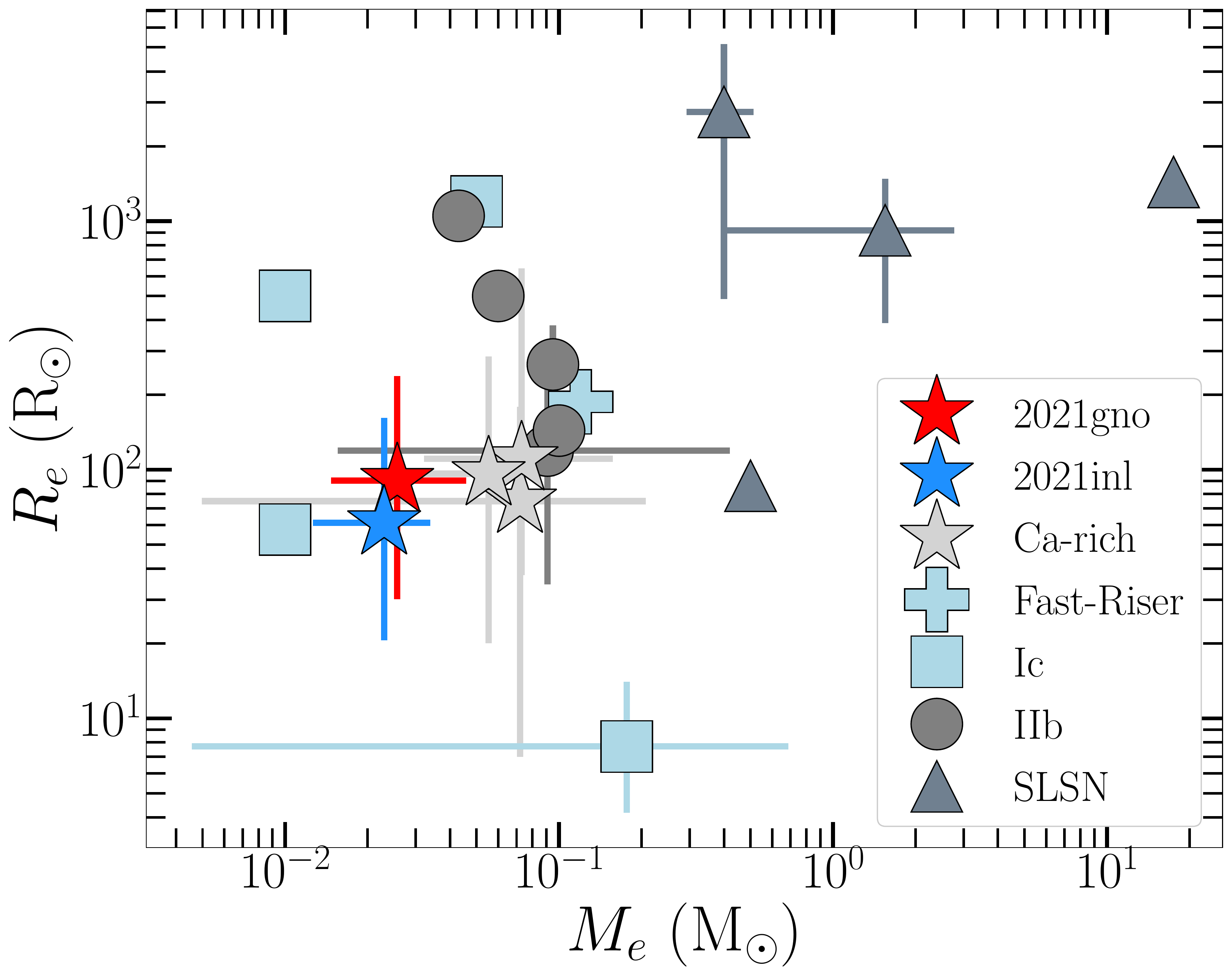}
\caption{Radius versus mass of extended material derived from light curve modeling with shock cooling emission models (e.g., see \S\ref{subsec:flare_shockcool} for discussion of representative sample). Double-peaked \cas{} shown as stars (2021gno in red, 2021inl in blue), fast-rising transients as blue plus signs, SNe~Ic as blue squares, SNe~IIb as grey circles, and super-luminous SNe as grey triangles. Note that not all of these double-peaked SN light curves were fit with exactly the same SCE models. \label{fig:sc_compare}}
\end{figure}

In Figure \ref{fig:sc_compare}, we attempt to compare the radius and mass of the extended material estimated from the SCE modeling of SNe~2021gno and 2021inl to other double-peaked events whose primary light curve peak was modeled in a similar fashion. As shown in the plot, the SCE parameter space of all five double-peaked \cas{} is highly consistent: on average, these objects can be modeled with SCE from extended material that has a compact radius of $\sim$50-120~$\Rsun$ and mass of $\sim$0.05-0.1~$\Msun$. Compared to SCE model parameters presented in the literature, \cas{} show a similar extended mass to fast-rising events such as SN~2019dge \citep{yao20} and SNe~IIb \citep{wheeler93, arcavi11, arcavi17, piro17, Armstrong21}, the latter typically exhibiting larger extended radii, likely indicating a more massive progenitor star than what produces \cas{}. Furthermore, the SCE parameter space of \cas{} is unlike that of SNe~Ic \citep{de18sci, ho20, gagliano22} and super-luminous SNe \citep{smith16}, the former showing a much larger range of radii and smaller masses, while the latter is best fit by a much larger extended material mass and radius. However, we note that the parameters derived for all \cas{} presented were done using four separate SCE models (e.g., \citealt{piro15, piro21, sapir17}, while other objects shown were only modeled with one of these formalisms. Therefore, direct comparison of the SCE parameters may not be completely accurate. 

\subsection{CSM Interaction Model}\label{subsec:flare_csm}
In addition to the SCE model, we explore interaction of the explosion's shock with a circumstellar medium as a mechanism to explain the primary light curve peaks of SNe 2021gno and 2021inl. We model the interaction as homologously expanding ejecta interacting with a detached CSM shell. In this picture, the CSM is sufficiently optically thick that the radiation becomes visible only after shock breakout from the outer edge of the CSM. The light curve is then powered by the resulting shock cooling emission of the swept up CSM and ejecta.

In this model, we assume a broken power-law ejecta with density profile $\rho_{\rm ej}\propto r^{-1}$ and $\rho_{\rm ej}\propto r^{-10}$ in the inner and outer ejecta, respectively; and assume the ejecta is expanding homologously with a kinetic energy of $E_{\rm sn}$. The CSM of mass $M_{\rm csm}$ extends from an inner radius $R_{\rm csm}$ with a width of $\Delta R_{\rm csm}$. The density profile follows a $\rho_{\rm csm}\propto r^{-2}$ profile out to $R_{\rm csm}$+$\Delta R_{\rm csm}$.

We run numerical simulations using the radiation hydrodynamics code Sedona \citep{kasen06}. The equations of radiation hydrodynamics are solved in one-dimensional spherical symmetry using implicit Monte Carlo radiative transfer \citep{roth15} coupled to a moving mesh hydrodynamics code based off of \cite{duffell16}. We assume a grey electron scattering opacity of $\kappa_{\rm es}=0.1$ cm$^2$ g$^{-1}$ and an absorptive opacity of $\kappa_{\rm abs}=\epsilon\kappa_{\rm es}$, with $\epsilon=10^{-3}$ to account for Compton thermalization. We assume $M_{\rm ej}=0.3-0.6~\Msun$ (i.e., $M_{\rm ej}$ for \sng{} and \sni{}, respectively) , $E_{\rm sn}=(1-2)\times 10^{50}$ erg s$^{-1}$ (i.e., $E_{k}$ for \sng{} and \sni{}, respectively), $M_{\rm csm}=0.02~\Msun$, $R_{\rm csm}=10^{13}$ cm, $\Delta R_{\rm csm} = 10^{13}$ cm. These CSM properties are based on SCE model parameters (\S\ref{subsec:flare_shockcool}) and X-ray/radio observations (\S\ref{Sec:Radio_Xray_Modeling}), and allow us to create a fiducial model for both comparison to observations as well as rough estimation of CSM properties in both objects.

These models are presented with respect to both object's bolometric luminosity during the early-time flux excess in Figure \ref{fig:BB_models}. As shown in the plot, both model light curves over-estimate the total luminosity in the primary light curve peak for both SNe; this indicates that the CSM mass is likely lower than that used in the simulations i.e., $M_{\rm csm} \lesssim 0.02~\Msun$. For reference, we also plot the CSM interaction model designed for SN~2019ehk's early-time excess ($M_{\rm csm}=1.5\times 10^{-3}~\Msun$, $R_{\rm csm}=10^{14}$ cm), which yields a better match to the light curve peak in SNe~2021gno and 2021inl. Furthermore, based on these model comparisons, the inferred CSM properties in \sng{} are consistent with the CSM mass independently inferred from X-ray modeling in \S\ref{Sec:Radio_Xray_Modeling}.

\begin{figure}[h!]
\centering
\subfigure[]{\includegraphics[width=0.45\textwidth]{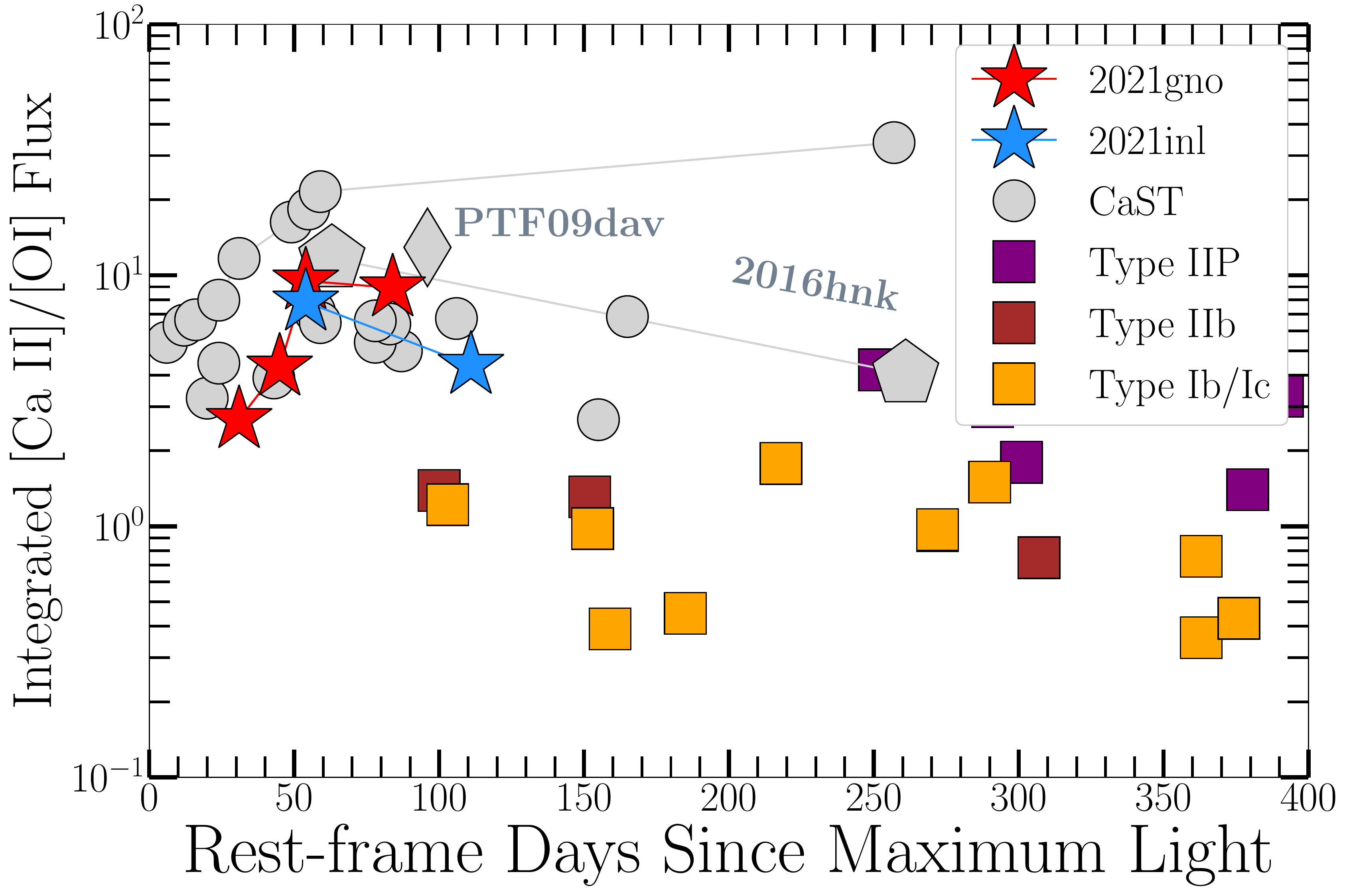}}
\subfigure[]{\includegraphics[width=0.45\textwidth]{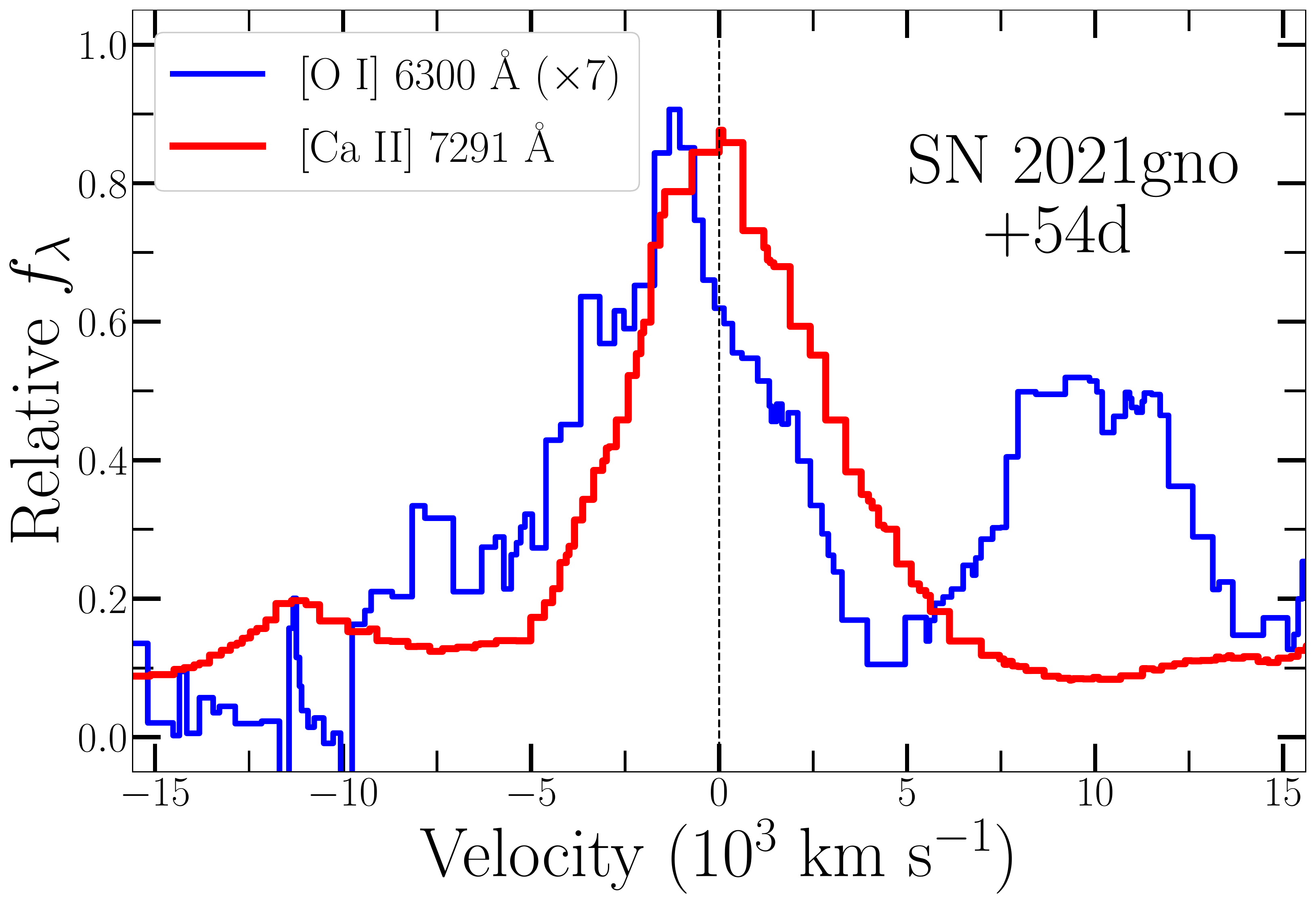}}
\subfigure[]{\includegraphics[width=0.45\textwidth]{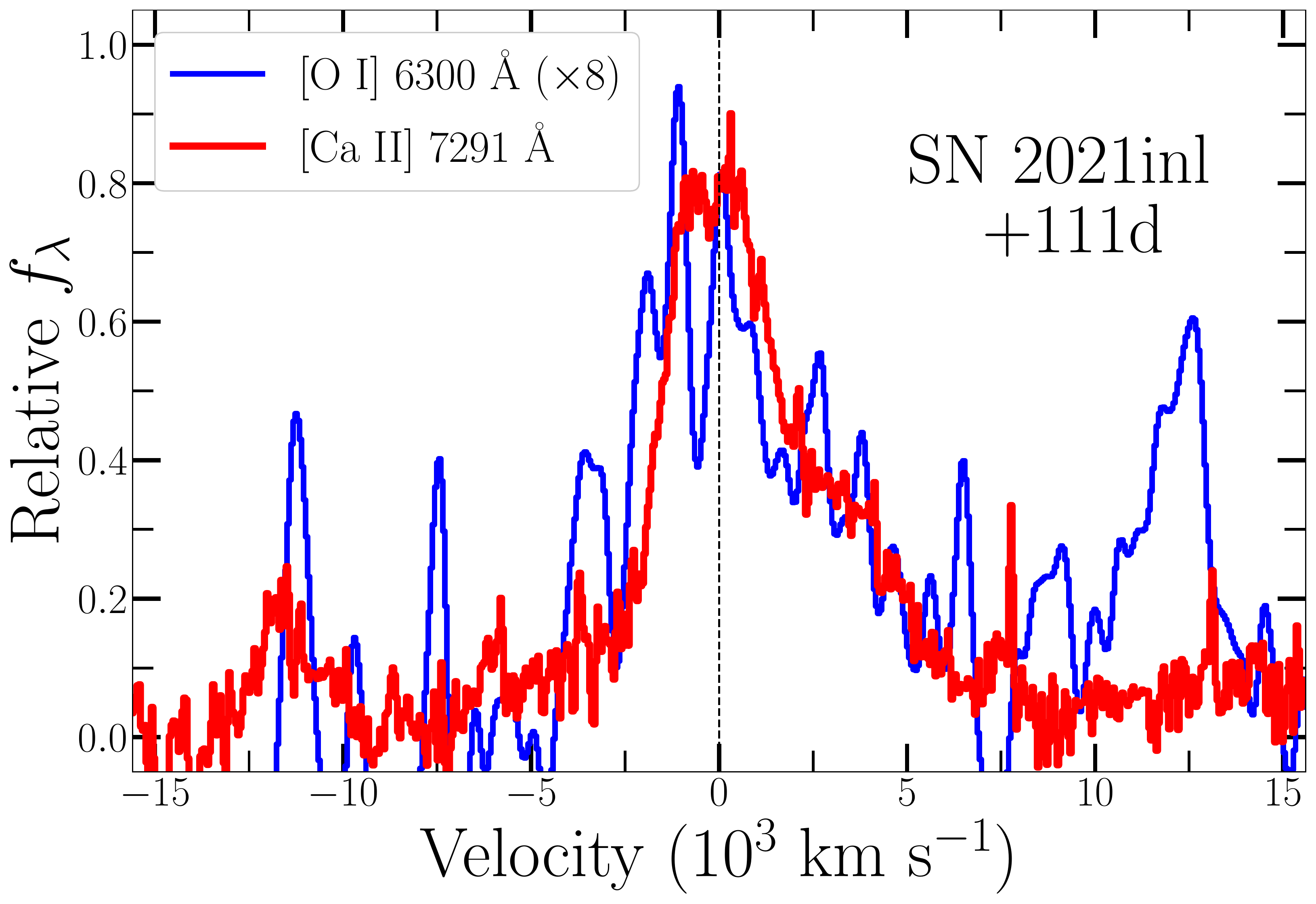}}
\caption{(a) Ratio of integrated [\ion{Ca}{ii}] and [\ion{O}{i}] flux with respect to phase for \sng\ (red stars), \sni\ (blue stars), and current sample of \cas\ (gray circles, diamonds and polygons), and assorted types of core-collapse SNe. All \cas{}, including SNe~2021gno and 2021inl, show [\ion{Ca}{ii}]/[\ion{O}{i}] > 2. [\ion{Ca}{ii}]/[\ion{O}{i}] values for all Type II/Ibc objects from \cite{milisavljevic17}. (b)/(c) Velocity profiles of [\ion{Ca}{ii}] $\lambda\lambda$ 7291,7324 (red) and scaled [\ion{O}{i}] $\lambda\lambda$ 6300, 6364 (blue) in SN~2021gno at +54 days and SN~2021inl at +111 days post-explosion. \label{fig:caii_oi}}
\end{figure}

\section{CSM Constraints from X-ray/Radio Emission in SN 2021gno} \label{Sec:Radio_Xray_Modeling}

\sng{} is the second \ca{}, after SN~2019ehk (WJG20a), to show luminous X-ray emission ($L_x \approx 5\times 10^{41}$~erg s$^{-1}$) at very early-time phases ($\delta t \approx 1$~days), as shown in Figure \ref{fig:xray_radio_LC}(a). Given the consistency with a rapidly-decaying X-ray emission ($L_x\propto t^{-3}$) and the hard spectrum (\S\ref{SubSec:XRT}), we suggest that, like SN~2019ehk, the X-ray luminosity observed in \sng{} is most likely derived from thermal bremsstrahlung emission from shocked CSM gas in adiabatic expansion. Emission measure goes as $EM = n^2 V$ and we can derive properties of the local CSM density in SN~2021gno by the following relation: 


\begin{equation}
    n = \Big [ (EM)(\mu_e \mu_I)(4\pi R^2 \Delta R f)^{-1}  \Big ]^{1/2}
\end{equation}
\noindent
where $\mu_e, \mu_I$ are the electron and ion molecular weights, respectively, $R$ is the radius of the CSM, $\Delta R$ is the CSM thickness, and $f$ is the filling factor (i.e., how homogeneous the shell of CSM is around the progenitor). For the expression above, we use an emission measure of $EM = (1.8 \pm 0.7) \times 10^{64}$~cm$^{-3}$, filling factor $f = 1$, $\mu_e = \mu_I$ assuming H-rich CSM, and that the CSM thickness goes as $\Delta R \approx R$. Because the exact CSM extent is unknown, we calculate multiple possible particle densities based on different CSM geometries. For a shock travelling with speed $v_s = 0.1c$, the location of the blastwave at the time of X-ray emission ($\delta t \approx 1$~day) gives $R_{\rm CSM} = 3 \times 10^{14}$~cm and thus a particle density of $n = (9.4 \pm 1.9 ) \times 10^{9}$~cm$^{-3}$. Assuming a H-rich CSM composition, this yields a CSM density of $\rho_{\rm CSM} = (1.6 \pm 0.3) \times 10^{-14} $~g cm$^{-3}$ and mass of $M_{\rm CSM} = (1.6 \pm 0.3) \times 10^{-3}~\Msun$, assuming a spherical geometry. If the CSM radius is in fact comparable to the blackbody radius at $\delta t \approx 1$~day (e.g., $R_{\rm CSM} = 9 \times 10^{13}$~cm), as was done for the X-ray analysis of SN~2019ehk (WJG20a), we find a CSM density and mass of $\rho_{\rm CSM} = (7.4 \pm 1.5) \times 10^{-14} $~g cm$^{-3}$ and $M_{\rm CSM} = (3.4 \pm 0.7) \times 10^{-4}~\Msun$, respectively. 

At larger distances from the progenitor, we interpret the radio upper limits discussed in \S\ref{SubSec:radio} ($\delta t = 35 - 245$~days) in the context of synchrotron emission from electrons accelerated to relativistic speeds at the explosion's forward shock, as the SN shock expands into the medium. To derive parameters of the medium, we adopt the synchrotron self-absorption (SSA) formalism by \cite{Chevalier98} and we self-consistently account for free-free absorption (FFA) following \cite{Weiler02} (however, see \cite{terreran21} and \cite{wjg22} for application and additional details on these derivations). For the calculation of the free-free optical depth $\tau_{\rm ff}(\nu)$, we adopt a wind-like density profile $\rho_{\rm{csm}}\propto r^{-2}$ in front of the shock, and we conservatively assume a gas temperature $T=10^4\,\rm{K}$ (higher gas temperatures  would lead to tighter density constraints). The resulting SSA+FFA synchrotron spectral energy distribution depends on the radius of the emitting region, the magnetic field, the environment density and on the shock microphysical parameters $\epsilon_B$ and $\epsilon_e$ (i.e.~the fraction of post-shock energy density in magnetic fields and relativistic electrons, respectively).

At the time of the latest radio non-detection in \sng{}, the shock will have probed distances of $r \approx 2 \times 10^{16}$~cm for $v_s = 10^4~\kms$; however this distance could vary based on the chosen shock speed. We find that, for typical microphysical parameters $\epsilon_B = 0.01$ and $\epsilon_e = 0.1$ (same as for \ca{} SN~2019ehk; WJG20a), the lack of radio emission indicates a low density medium that corresponds to a progenitor mass loss rate of $\dot M<10^{-4}\,\rm{M_{\sun}\,yr^{-1}}$, for an adopted wind speed of $v_w = 500~\kms$. This $v_w$ value is the same as in SN~2019ehk, which had direct detections of CSM velocity based on shock-ionized emission lines in the early-time spectra (WJG20a). Overall, the mass loss limits of both SNe~2021gno and 2019ehk are consistent with one another, the latter being more constraining given the depth of the radio observations.  


\section{Discussion} \label{Sec:discussion}

\subsection{A Physical Progenitor Model} \label{subsec:model}

\begin{figure}[t!]
\centering
\includegraphics[width=0.45\textwidth]{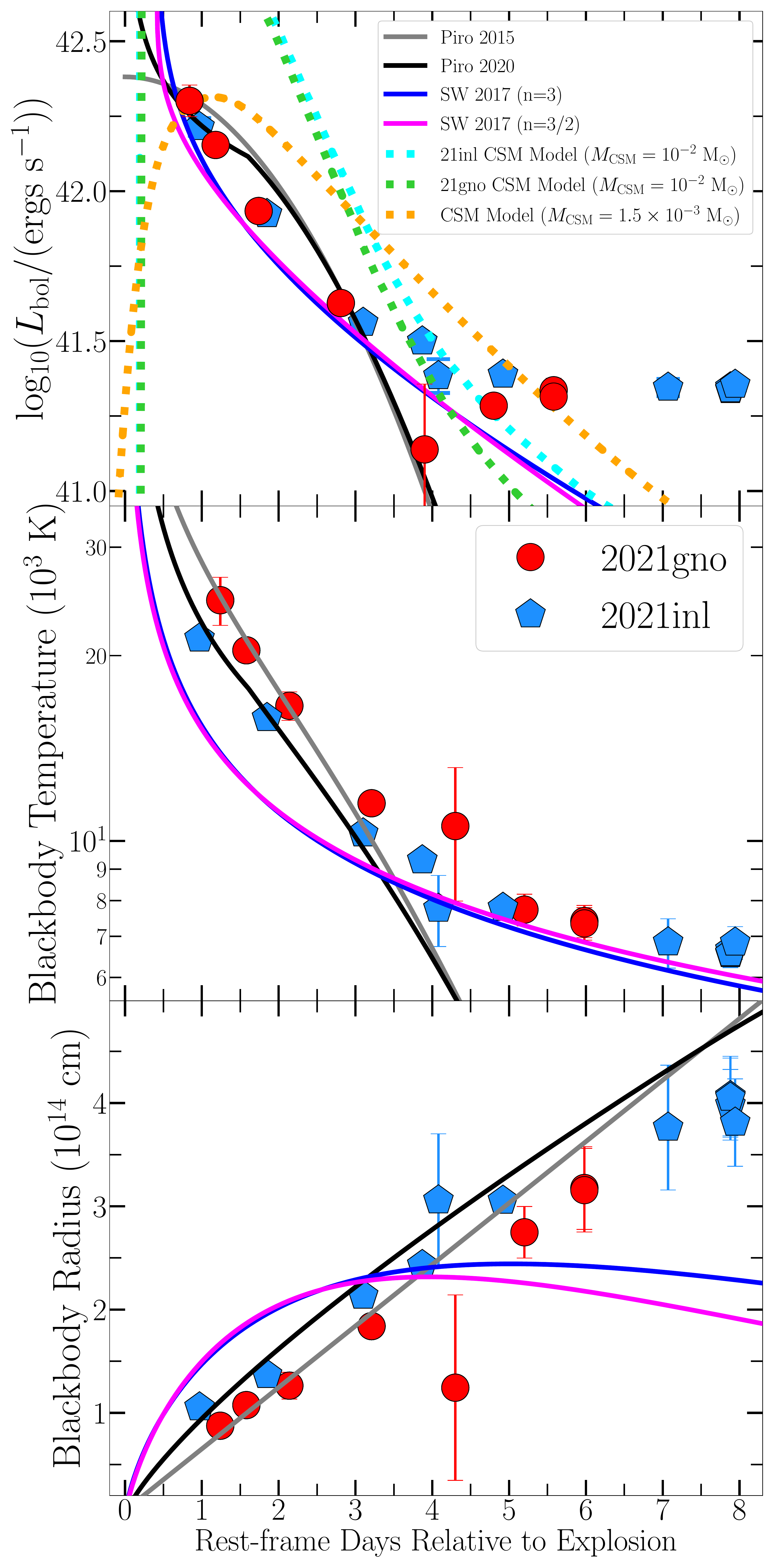}
\caption{\textit{Top:} Bolometric luminosity during the primary light curve peaks in SNe~2021gno and 2021inl. Shock interaction models plotted as orange/cyan/green dashed lines (see \S\ref{subsec:flare_csm}). Shock cooling models are plotted as solid lines: \cite{piro15} in grey, \cite{piro21} in black, \cite{sapir17} n = 3/2[3] in pink[blue]. SCE model parameters: $M_e = 0.02 - 0.5~\Msun$, $R_e = 40 - 230~\Rsun$, and $v_e = (7-9)\times 10^{4}~\kms$. \textit{Middle:} Blackbody temperatures during the primary light curve peak. For the interaction model we show the effective blackbody temperature.  \textit{Bottom:} Blackbody radii during the primary light curve peak. The shock interaction model presents the radius of the emitting region. \label{fig:BB_models}}
\end{figure}

The high-cadence, multi-wavelength follow-up of SNe~2021gno and 2021inl allows for some of the best constraints to be made on \ca{} progenitor systems to date. For \sng{}, modeling of the bolometric light curve has revealed that the explosion was low energy ($E_k \approx 1.3\times 10^{50}$~erg), which produced $\sim$0.6~$\Msun$ of ejecta and synthesized $\sim$0.012~$\Msun$ of ${}^{56}\textrm{Ni}$. Furthermore, the multi-band light curve revealed a flux excess above the radioactive decay powered continuum emission that lasted $\sim 5$~days post-explosion. Modeling of this primary light curve peak (e.g., \S \ref{subsec:flare_shockcool}) suggests that the progenitor star could have had an extended envelope of material with radius $R_e = 30-230~\Rsun$ and mass $M_e = (1.5 - 4.5) \times 10^{-2}~\Msun$. Additionally, modeling of the luminous X-ray emission detected in \sng{} at $\sim$1~day after explosion indicates that the progenitor system also contained a shell of CSM that extended to $R \approx (0.9-3)\times 10^{14}$~cm and was comprised of $\sim$$(0.3 - 1.6)\times 10^{-3}~\Msun$ of H- and/or He-rich gas, if the CSM composition is similar to SN~2019ehk. In Figure \ref{fig:BB_models}, we show that this amount of CSM can also be the power-source behind the multi-band primary light curve peak; this material being ejected by the progenitor star in the final months before explosion for a possible wind velocity of $\sim 500~\kms$. Lastly, radio non-detections at late-times suggest a relatively clean progenitor environment at distances of $10^{16-17}$~cm and a progenitor mass loss rate in the final year(s) before explosion of $\dot M<10^{-4}\,\rm{M_{\sun}\,yr^{-1}}$. 

The above information allows for decent constraints to be made on the potential progenitor star(s) responsible for \sng{}. One progenitor scenario is that \sng{} resulted from a low-mass, massive star ($\sim$8-11~$\Msun$) that experienced enough enhanced mass loss prior to explosion to remove all stellar H-rich material as well as to place $\sim$$(1 - 4) \times 10^{-2}~\Msun$ of extended material/envelope at distances $\lesssim$230~$\Rsun$ and/or $\sim$$(0.3 - 1.6)\times 10^{-3}~\Msun$ CSM at $r \lesssim 3\times 10^{14}$~cm. A massive star progenitor is also consistent with the location of \sng{} in the spiral arm of a star-forming host galaxy. However, increased mass loss in such a progenitor could only have taken place in the final months before explosion given the low mass loss rate of $\dot M<10^{-4}\,\rm{M_{\sun}\,yr^{-1}}$ in the progenitor's final year(s). Compared to  simulations, SN~2021gno's ejecta mass is consistent with the collapse of a  $9-10~\Msun$ progenitor \citep{wanajo18}, but the total synthesized Ni mass is, on average, an order of magnitude lower in these models. Similarly, from \texttt{BPASS} library \citep{eldridge17}, all massive star explosions occurring in binary systems in the lowest mass bins (e.g., 8-11~$\Msun$) produce >1.5~$\Msun$ of ejecta and synthesize a total ${}^{56}\textrm{Ni}$ mass that is inconsistent with \sng{}. Furthermore, while ultra-stripped SN (USSN) progenitor models (e.g., \citealt{yoshida17, moriya17}) produce $M_{\rm ej} \lesssim 0.2~\Msun$, they can reproduce the Ni yield observed in \sng{}, but it is unclear whether these progenitors can retain enough of a He-rich envelope to produce SCE as well as CSM capable of luminous X-ray emission via shock interaction. Lastly, a promising progenitor candidate is a He-star binary system capable of producing a type Ib-like explosion (e.g., see \citealt{yoon17, Jung21}). Based on the models presented in Table A6 of WJG20a, the ejecta mass of \sng{} is consistent with a He-star with artificial envelope removal (e.g., models \#2, 4) and
a He-star + NS binary (models \#7,8), both ending in O core burning. However, all of these models would be ruled out if the X-ray emission is derived from H-rich CSM. Furthermore, it is difficult to reconcile a massive star progenitor with the non-detection of a star forming region at the \sng{} explosion site (\S\ref{sec:host}) and a SFR of $< 3.4 \times 10^{-5}~\Msun$ yr$^{-1}$. Given these constraints on a massive star progenitor for \sng{}, it may be the case that a WD system is better suited to reproduce the SN observables, as discussed below for \sni{}. 

In terms of constraining the progenitor of \sni{}, the large offset of the SN from an early-type elliptical galaxy makes a massive star progenitor system very unlikely. Consequently, a more plausible scenario is one that involves the explosion of a WD in a binary system. However, such an explosion needs to produce $\sim$0.3~$\Msun$ of ejecta and $\sim$0.012~$\Msun$ of ${}^{56}\textrm{Ni}$, as well as allow for either SCE from a confined ($\sim$20-150~$\Rsun$) and low mass ($\sim$0.02~$\Msun$) extended envelope and/or $\lesssim$$10^{-3}~\Msun$ of CSM. While such a SN is unable to be formed by typical SN~Ia explosion channels, the formation of a confined, extended envelope can occur during the ejection of material in ``tidal tails'' that then ``settles'' around the primary WD prior to merger \citep{raskin13, shen12, schwab16}. 

\begin{figure}[t!]
\centering
\includegraphics[width=0.45\textwidth]{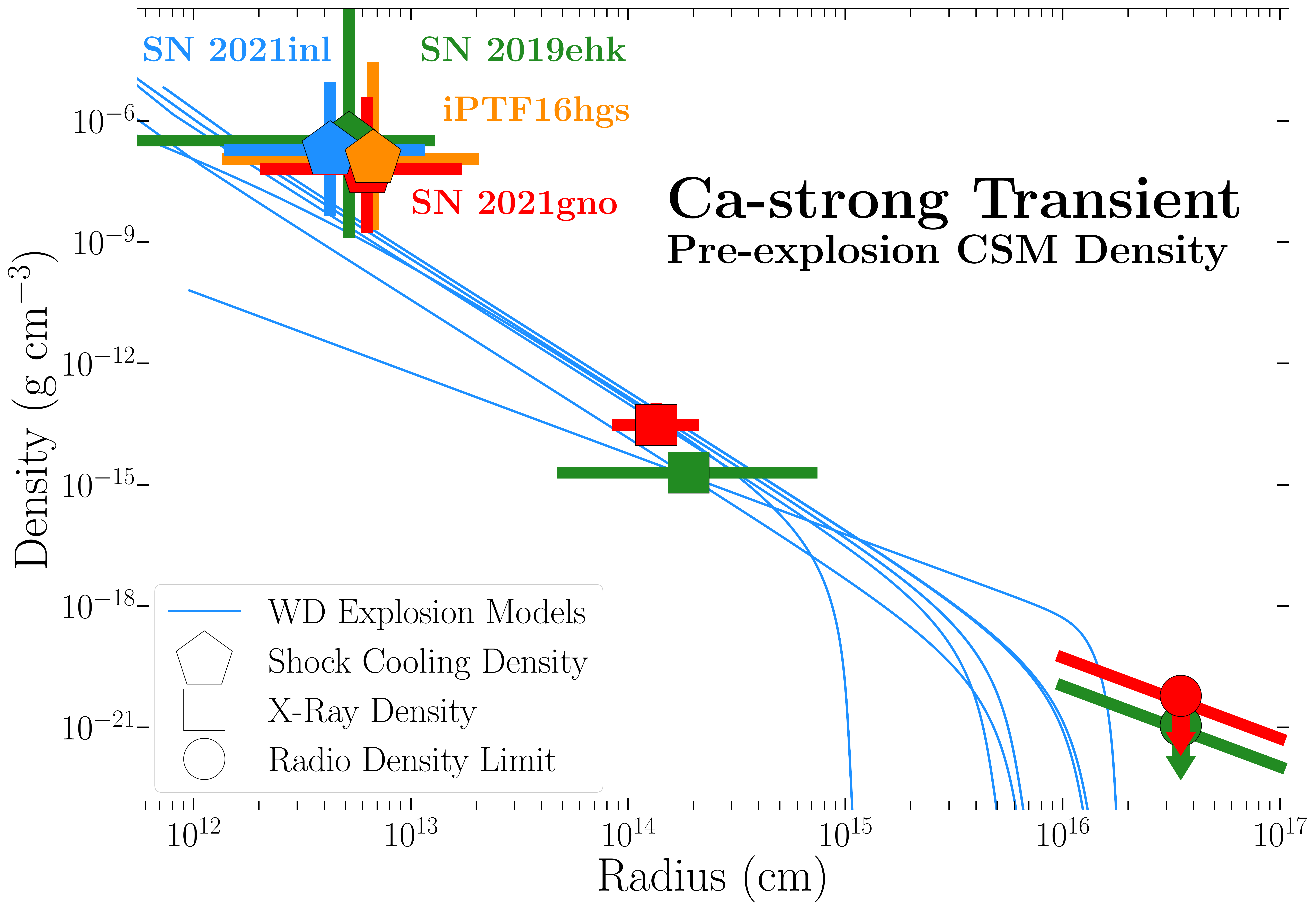}
\caption{Density profile of the explosion environment of \cas\ 2021gno (red), 2021inl (blue), 2019ehk (green), and iPTF16hgs (orange). Densities plotted as polygons are derived from shock cooling modeling. Shown as red and green squares are density limits derived from X-ray detections in SNe~2021gno and 2019ehk, respectively. The red and green circles are the density limit derived from modeling of the radio non-detections in SNe~2021gno and 2019ehk (WJG20a), respectively. Blue lines are CSM models for WD mergers at the time of explosion (see \S\ref{subsec:model}) \label{fig:density_plot}}
\end{figure}

We explore the possibility that \sni{} resulted from a double degenerate system containing a hybrid and a CO WD, which was initially presented by WJG20a to explain SN~2019ehk. In this scenario, the tidal disruption of the hybrid HeCO WD by lower-mass CO WD (or another hybrid WD) can induce a He-detonation that can lead to \ca-like transient \citep{Bobrick2017, perets19, Zenati2019, zenati2019b}. Furthermore, prior to the disruption, significant mass transfer ($\dot M\approx10^{-2}\,\rm{M_{\sun}\,yr^{-1}}$) will place CSM in the local environment, capable of powering the initial light curve peak observed in \sni{}. We note that such a system could also reproduce the observables in \sng{} such as $M_{\rm ej}$ and $M({}^{56}\textrm{Ni})$ (e.g., see Table A4 of \citealt{wjg21}), as well as X-ray emission from CSM interaction and the lack of detectable star formation at the explosion site. In Figure \ref{fig:density_plot}, we show the density profiles in the pre-explosion environments of SNe~2019ehk, 2021gno, 2021inl, and iPTF16hgs derived from SCE models as well as X-ray and radio modeling for SNe~2019ehk and 2021gno specifically. We show that the pre-explosion environments are consistent with the CSM density profiles of WD disruption models discussed above, further indicating that this scenario may be a plausible model to explain these \cas{}. 

\subsection{SNe~2021gno and 2021inl in the ``Calcium-strong'' Class} \label{subsec:ca-rich_sne}

Based on the observational properties of both SNe~2021gno and 2021inl, it is evident that these objects fit within the confines of the \ca{} observational class. As discussed in \S\ref{subsec:phot_properties}, both SNe display low luminosity ($M_{\rm r,peak} \approx -15$~mag) and rapidly evolving ($t_r \approx 8 - 15$~days) light curves whose color evolution is consistent with other confirmed \cas{} (e.g., Fig. \ref{fig:colors}). Furthermore, the spectroscopic evolution of SNe~2021gno and 2021inl also solidifies their place in this observational class: both objects showing type I spectra near peak, which quickly transitions to an optically thin regime where all nebular emission is dominated by [\ion{Ca}{ii}] emission and weak [\ion{O}{i}] (e.g., [\ion{Ca}{ii}]/[\ion{O}{i}] > 2; Fig. \ref{fig:caii_oi}). 

Nonetheless, these SNe do appear to deviate from normative \ca{} characteristics based on their double-peaked light curves and pre-explosion environments. With regards to the former, SNe~2021gno and 2021inl now represent the 4th and 5th confirmed \cas{} with an early-time flux excess, the other events being iPTF16hgs, SN~2018lqo, and SN~2019ehk. The observation of this primary light curve peak confirms the presence of an extended stellar envelope capable of producing SCE and/or a dense CSM that powers the initial flux excess through SN shock interaction in at least some \ca{} progenitor systems. Of the 9 \cas{} discovered $<$3~days after explosion and with $<$2~day photometric cadence, 5 ($55\%$) show a clear early-time flux excess: iPTF16hgs, SNe~2018lqo, 2019ehk, 2021gno and 2021inl. However, it is possible that $100\%$ of these objects show this signature given the marginal detections of a very early-time flux excess in the remaining sub-sample objects PTF11kmb, PTF12bho, SN~2018kyj and SN~2019hty, indicating that this feature is potentially ubiquitous to \cas{}. 


In terms of their progenitor environments, both SNe exist in visibly different host galaxies, the large projected offset of \sni{} from its early-type host galaxy being most similar to the environments of many other \cas{} \citep{perets10,perets11, kasliwal12, lym+13,lyman14, per+14, foley15, lunnan17, de20_carich,per+21}. Additionally, despite the fact that \sng{} exploded in a star-forming, late-type galaxy, there is no evidence for star formation at the explosion site (similar to SN~2019ehk), which makes this event quite similar to other \cas{} in spite of the visibly different host galaxy type. However, a number of \cas{} have been discovered in spiral host galaxies with explosion site star formation (e.g., iPTF15eqv, iPTF16hgs, SN~2016hnk; \citealt{milisavljevic17, de18, galbany19}), as well as in, or offset from, disk galaxies (e.g.,  PTF09dav, SN~2001co, SN~2003H, SN~2003dr, SN~2003dg; \citealt{sullivan11, kasliwal12, Perets2014, foley15}). This spread in \ca{} host environments continues to indicate that the progenitor systems responsible for these transients are likely heterogeneous, some arising from certain types of massive stars and others coming from the explosion of compact stars such as WDs. 

Lastly, \sng{} is now the second confirmed \ca{} to be detected in X-rays, a novel observational probe for this explosion class. The X-ray emission in \sng{} was detected earlier than SN~2019ehk (WJG20a), in addition to being more luminous, but was nonetheless consistent with a rapid decline rate of $L_x \approx t^{-3}$. Now that this behavior has been confirmed in more than one \ca{}, it is more likely that X-ray emission from shock interaction in a dense, confined shell of CSM is a trait that could be more common to the \ca{} class as a whole. However, detecting future \cas{} at X-ray wavelengths requires the discovery of future objects at $D \lesssim 40$~Mpc and the follow-up of these transients with X-ray telescopes in the first $\sim$day after explosion. Given that X-rays were detected in both SNe~2019ehk and 2021gno, the only two \cas{} where \textit{Swift}-XRT was re-pointed at very early phases ($\delta t < 4$~days), it is highly likely that X-ray emission is present in all double-peaked, and possibly all single-peaked, \cas{} directly after explosion.


\section{Conclusions} \label{sec:conclusion}

In this paper, we have presented multi-wavelength observations of two new \cas{}, SNe~2021gno and 2021inl. Despite their unique double-peaked light curves, both objects are photometrically and spectroscopically consistent with prototypical \cas{} throughout their evolution, which solidifies their place in this explosion class. Below we list the primary findings that make SNe~2021gno and 2021inl significant and novel additions to our understanding of these peculiar explosions:

\begin{itemize}
    \item \sng{} was first detected within 0.5~days of explosion and is located on the outer edge of the star-forming spiral host galaxy NGC~4165. \sni{} was first detected within 0.1~days of explosion and is located at a large offset from early-type host galaxy NGC~4923. 
    \item Based on their fast light curve evolution ($t_r \lesssim 15$~days), low overall luminosities ($M_{\rm g, peak} > -15$~mag), and dominant [\ion{Ca}{ii}] emission lines (e.g., [\ion{Ca}{ii}]/[\ion{O}{i}] $\approx 10$), both SNe can be confidently classified as \cas{}. Furthermore, the ratio of ion masses derived for both SNe in \S\ref{subsec:nebular_gas} (e.g., $M({\rm O}) > M({\rm Ca})$) continues to indicate that these explosions are not ``rich'' in Ca abundance but rather are ``rich'' in Ca emission i.e., ``Calcium-strong.''
    \item Despite visibly different host galaxies, modeling of the host spectra reveals that the explosion sites of both SNe had very little ($\lesssim$$10^{-5} \ \Msun$~yr$^{-1}$) or no recent star formation, which strongly suggests that neither SN came from a massive star progenitor.  
    \item \sng{} is the second \ca{} with confirmed, luminous X-ray emission ($L_x \approx 5 \times 10^{41}$~erg s$^{-1}$) as detected by \textit{Swift}-XRT at $\delta t \approx 0.8$~days post-explosion. Based on the rapid fading and modeling of the X-ray spectrum, we conclude that this emission was derived from shocked CSM gas comprised of $M_{\rm CSM} = (0.3 - 1.6)\times 10^{-3}~\Msun$ of shocked gas that extended to distances $R = (0.9-3)\times 10^{14}$~cm, possibly comprised of H- and/or He-rich material. At larger distances from the progenitor star (e.g., $\sim 10^{16-17}$~cm), modeling of \sng{} radio observations indicates a progenitor mass loss rate of $\dot M<10^{-4}\,\rm{M_{\sun}\,yr^{-1}}$ ($v_w = 500~\kms$) in the final year(s) before explosion. 
    \item SNe~2021gno and 2021inl are the fourth and fifth \cas{} with multi-color, double-peaked light curves. We model the initial flux excess using four analytic formalisms for shock cooling emission from extended material to derive best fit parameters of this material (\S\ref{subsec:flare_shockcool}). For \sng{}, we find that a radius and mass of extended material ranging from $R_e \approx 30 - 230 \ \Rsun$ and $M_e \approx 0.02 - 0.05 \ \Msun$, respectively, can reproduce the early-time emission. Similarly, for \sni{}, we derive radius and mass of extended material of $R_e \approx 20 - 150 \ \Rsun$ and $M_e \approx 0.02 \ \Msun$, respectively. 
    \item Given the direct evidence for CSM interaction in \sng{}, we also model the primary light curve peak in both SNe with numerical models for shock interaction with confined CSM (\S\ref{subsec:flare_csm}). We find that the observed flux excess in \sng{} can be fit with $R_{\rm CSM} = 10^{13-14}$~cm and $M_{\rm CSM} \lesssim 10^{-2}~\Msun$, both properties being consistent with X-ray modeling. For \sni{}, we find a similar best fit CSM radius and mass.
    \item Using a combination of shock cooling, shock interaction, X-ray, and radio modeling, as well as host galaxy SFR, we are able to place some of the tightest constraints to date on the density profile of the local \ca{} progenitor environment (Fig. \ref{fig:density_plot}). For both SNe~2021gno and 2021inl, as well as other double-peaked \cas{} SN~2019ehk and iPTF16hgs, the progenitor CSM density is consistent with models for the merger of low-mass, hybrid WDs. For SNe~2019ehk, 2021gno and 2021inl specifically, this is supported by the lack of host galaxy star formation at the explosion sites of these  events.
\end{itemize}


Future multi-wavelength (X-ray to radio) observations of double-peaked \cas{} at very early-time phases will be instrumental in filling out the progenitor environment phase space and constraining the progenitor channel of these peculiar explosions. Multi-color transient surveys with higher limiting magnitudes ($> 21$~mag) such as YSE currently and LSST in the future will greatly increase the number of \cas{} discovered within a day of explosion. 

\section{Acknowledgements} \label{Sec:ack}

Research at UC Berkeley is conducted on the territory of Huichin, the ancestral and unceded land of the Chochenyo speaking Ohlone people, the successors of the sovereign Verona Band of Alameda County. Keck I/II, ATLAS, and PS1 observations were conducted on the stolen land of the k\={a}naka `\={o}iwi people. We stand in solidarity with the Pu'uhonua o Pu'uhuluhulu Maunakea in their effort to preserve these sacred spaces for native Hawai`ians. MMT observations were conducted on the stolen land of the Tohono O'odham and Hia-Ced O'odham nations; the Ak-Chin Indian Community, and Hohokam people. ZTF observations were conducted on the stolen land of the Pauma and Cupe\~{n}o tribes; the Kumeyaay Nation and the Pay\'{o}mkawichum (Luise\~{n}o) people. Shane 3-m observations were conducted on the stolen land of the Ohlone (Costanoans), Tamyen and Muwekma Ohlone tribes.

The Young Supernova Experiment and its research infrastructure is supported by the European Research Council under the European Union's Horizon 2020 research and innovation programme (ERC Grant Agreement No.\ 101002652, PI K.\ Mandel), the Heising-Simons Foundation (2018-0913, PI R.\ Foley; 2018-0911, PI R.\ Margutti), NASA (NNG17PX03C, PI R.\ Foley), NSF (AST-1720756, AST-1815935, PI R.\ Foley; AST-1909796, AST-1944985, PI R.\ Margutti), the David \& Lucille Packard Foundation (PI R.\ Foley), VILLUM FONDEN (project number 16599, PI J.\ Hjorth), and the Center for AstroPhysical Surveys (CAPS) at the National Center for Supercomputing Applications (NCSA) and the University of Illinois Urbana-Champaign.

W.J-G is supported by the National Science Foundation Graduate Research Fellowship Program under Grant No.~DGE-1842165. W.J-G acknowledges support through NASA grants in support of {\it Hubble Space Telescope} program GO-16075 and 16500. This research was supported in part by the National Science Foundation under Grant No. NSF PHY-1748958.  R.M. acknowledges support by the National Science Foundation under Award No. AST-1909796 and AST-1944985. R.M. is a CIFAR Azrieli Global Scholar in the Gravity \& the Extreme Universe Program, 2019. The Margutti's team at Northwestern and UC Berkeley is partially funded by the Heising-Simons Foundation under grant \# 2018-0911 and \#2021-3248 (PI: Margutti).

V.A.V acknowledges support by the National Science Foundation under Award No.AST-2108676. C. R. A. was supported by grants from VILLUM FONDEN (project numbers 16599 and 25501). Parts of this research were supported by the Australian Research Council Centre of Excellence for All Sky Astrophysics in 3 Dimensions (ASTRO 3D), through project number CE170100013.

The UCSC team is supported in part by NASA grant 80NSSC20K0953, NSF grant AST--1815935, the Gordon \& Betty Moore Foundation, the Heising-Simons Foundation, and by a fellowship from the David and Lucile Packard Foundation to R.J.F.

Some of the data presented herein were obtained at the W. M. Keck Observatory, which is operated as a scientific partnership among the California Institute of Technology, the University of California, and NASA. The Observatory was made possible by the generous financial support of the W. M. Keck Foundation. The authors wish to recognize and acknowledge the very significant cultural role and reverence that the summit of Maunakea has always had within the indigenous Hawaiian community. We are most fortunate to have the opportunity to conduct observations from this mountain.

A major upgrade of the Kast spectrograph on the Shane 3~m telescope at Lick Observatory was made possible through generous gifts from the Heising-Simons Foundation as well as William and Marina Kast. Research at Lick Observatory is partially supported by a generous gift from Google.

Based in part on observations obtained with the Samuel Oschin 48-inch Telescope at the Palomar Observatory as part of the Zwicky Transient Facility project. ZTF is supported by the NSF under grant AST-1440341 and a collaboration including Caltech, IPAC, the Weizmann Institute for Science, the Oskar Klein Center at Stockholm University, the University of Maryland, the University of Washington, Deutsches Elektronen-Synchrotron and Humboldt University, Los Alamos National Laboratories, the TANGO Consortium of Taiwan, the University of Wisconsin at Milwaukee, and the Lawrence Berkeley National Laboratory. Operations are conducted by the Caltech Optical Observatories (COO), the Infrared Processing and Analysis Center (IPAC), and the University of Washington (UW).

The Pan-STARRS1 Surveys (PS1) and the PS1 public science archive have been made possible through contributions by the Institute for Astronomy, the University of Hawaii, the Pan-STARRS Project Office, the Max-Planck Society and its participating institutes, the Max Planck Institute for Astronomy, Heidelberg and the Max Planck Institute for Extraterrestrial Physics, Garching, The Johns Hopkins University, Durham University, the University of Edinburgh, the Queen's University Belfast, the Harvard-Smithsonian Center for Astrophysics, the Las Cumbres Observatory Global Telescope Network Incorporated, the National Central University of Taiwan, STScI, NASA under grant NNX08AR22G issued through the Planetary Science Division of the NASA Science Mission Directorate, NSF grant AST-1238877, the University of Maryland, Eotvos Lorand University (ELTE), the Los Alamos National Laboratory, and the Gordon and Betty Moore Foundation.

IRAF is distributed by NOAO, which is operated by AURA, Inc., under cooperative agreement with the National Science Foundation (NSF).

\facilities{Neil Gehrels \emph{Swift} Observatory, AMI, Zwicky Transient Facility, ATLAS, YSE/PS1, Shane (Kast), MMT (Binospec), Keck I/II (LRIS)}

\software{IRAF (Tody 1986, Tody 1993),  photpipe \citep{Rest+05}, DoPhot \citep{Schechter+93}, HOTPANTS \citep{becker15}, HEAsoft (v6.22; HEASARC 2014), CASA (v6.1.2; \citealt{McMullin07}), Sedona \citep{kasen06}, Castro \citep{almgren10}}

\bibliographystyle{aasjournal} 
\bibliography{references} 


\clearpage
\appendix

\renewcommand\thetable{A\arabic{table}} 
\setcounter{table}{0}

\begin{deluxetable*}{ccccc}[h!]
\tablecaption{X-ray Observations of SN~2021gno \label{tab:xray_obs}}
\tablecolumns{6}
\tablewidth{0pt}
\tablehead{\colhead{MJDs} & \colhead{Phase\tablenotemark{a}} & \colhead{Photon Index } & \colhead{0.3-10 keV Unabsorbed Flux} &  \colhead{Instrument} \\
\colhead{} & \colhead{(days)} & \colhead{($\Gamma$)} & \colhead{($10^{-12}$~erg~s$^{-1}$~cm$^{-2}$)} & \colhead{}}
\startdata
59293.51, 59293.84 & 0.81 -- 1.14 & $0.74^{+0.50}_{-0.52}$ & $4.1^{+2.2}_{-2.1}$ &\textit{Swift-}XRT \\
59294.36, 59295.44, 59296.23 & 1.66 -- 3.53 & -- & $<1.6\tablenotemark{b}$&\textit{Swift-}XRT \\
59298.28, 59302.66, 59303.14, 59308.71 & 5.58 -- 16.01 & -- & $<4.2$ &\textit{Swift-}XRT \\
59313.30, 59317.86, 59325.64, 59414.17, 59524.21 & 20.6 -- 231.51 & -- & $<7.3$ &\textit{Swift-}XRT \\
\enddata
\tablenotetext{a}{Relative to explosion (MJD 59292.7).}
\tablenotetext{b}{Flux calibration performed assuming same spectral parameters inferred at $t=+1.21 - 1.54$ d.}

\end{deluxetable*}

\begin{deluxetable*}{cccccc}[h!]
\tablecaption{Optical Spectroscopy of SN~2021gno \label{tab:spec_table1}}
\tablecolumns{5}
\tablewidth{0.45\textwidth}
\tablehead{
\colhead{UT Date} & \colhead{MJD} &
\colhead{Phase\tablenotemark{a}} &
\colhead{Telescope} & \colhead{Instrument} & \colhead{Wavelength Range}\\
\colhead{} & \colhead{} & \colhead{(days)} & \colhead{} & \colhead{} & \colhead{(\AA)}
}
\startdata
2021-03-22 & 59295 & $+2.3$ & NOT & ALFOSC & 3600--9000 \\
2021-03-22 & 59295 & $+2.3$ & Shane & Kast & 3300--10200 \\
2021-03-30 & 59303 & $+10.3$ &  ANU 2.3m & WiFES &  3800--7000 \\
2021-04-01 & 59305 & $+12.4$ & Livermore Telescope & SPRAT & 3800--8000 \\
2021-04-02 & 59306 & $+13.3$ & NOT & ALFOSC & 3600--9000 \\
2021-04-06 & 59310 & $+17.3$ & Shane & Kast & 3300--10200\\
2021-04-10 & 59314 & $+21.3$ & IRTF & SpeX & 7000--25000 \\
2021-04-13 & 59317 & $+24.3$ & Shane & Kast & 3300--10200 \\
2021-04-19 & 59323 & $+30.3$ & Shane & Kast & 3300--10200 \\
2021-05-03 & 59337 & $+44.3$ & Shane & Kast & 3300--10200 \\
2021-05-12 & 59346 & $+53.3$ & Keck & LRIS & 3000--10000 \\
2021-04-09 & 59376 & $+83.3$ & MMT & Binospec & 3800--9200 \\
\enddata
\tablenotetext{a}{Relative to explosion (MJD 59292.7)}
\end{deluxetable*}

\begin{deluxetable*}{cccccc}[h!]
\tablecaption{Optical Spectroscopy of SN~2021inl \label{tab:spec_table2}}
\tablecolumns{5}
\tablewidth{0.45\textwidth}
\tablehead{
\colhead{UT Date} & \colhead{MJD} &
\colhead{Phase\tablenotemark{a}} &
\colhead{Telescope} & \colhead{Instrument} & \colhead{Wavelength Range}\\
\colhead{} & \colhead{} & \colhead{(days)} & \colhead{} & \colhead{} & \colhead{(\AA)}
}
\startdata
2021-04-13 & 59317 & $+7.7$ & Shane & Kast & 3300--10200\\
2021-04-19 & 59323 & $+13.7$ & Shane & Kast & 3300--10200 \\
2021-05-12 & 59346 & $+36.7$ & Keck & LRIS & 3000--10000 \\
2021-07-08 & 59403 & $+93.7$ & Keck & LRIS & 3000--10000 \\
\enddata
\tablenotetext{a}{Relative to explosion (MJD 59309.4)}
\end{deluxetable*}

\begin{table*}
\centering
    \caption{VLA radio observations of SN~2021gno.}
    \label{tab:radio}
    \begin{tabular}{ccccc}
    \hline
    \hline
  Start Date & Time\footnote{Relative to explosion (MJD 59292.7)} & Frequency & Bandwidth & Flux Density\footnote{\label{Tab_radio_1}Upper-limits are quoted at $3\sigma$.} \\
(UT) & (days) & (GHz) & (GHz) & ($\mu$Jy/beam)\\
    \hline
2021-04-21 & 35 & 15.5 & 5 & $\leq258$\\
2021-04-25 & 39 & 15.5 & 5 & $\leq276$\\
2021-11-19 & 245 & 15.5 & 5 & $\leq285$\\
\hline
\end{tabular}
\end{table*}

\begin{deluxetable*}{cccccccc}\label{tbl:shocktable}
\tablecaption{Shock Cooling Models}
\tablecolumns{8}
\tablewidth{\textwidth}
\tablehead{
\colhead{Model} & \colhead{SN} & \colhead{Phase Range} & \colhead{$R_e$} & \colhead{$M_e$} & \colhead{$v_{e}$} & \colhead{$t_{\rm off}$} & \colhead{$\chi^2_{\nu}$}\\
\colhead{} & \colhead{} & \colhead{days} & \colhead{$\Rsun$} & \colhead{$[\times 10^{-2}] \ \Msun$} & \colhead{$[\times 10^3] \ \kms$} & \colhead{days} & \colhead{}}
\startdata
\cite{piro15} & 2021gno & $t<5$ & $62.0^{+0.70}_{-0.69}$ & $1.72^{+0.015}_{-0.013}$ & $6.6 \pm 0.40$ & $0.001^{+0.002}_{-0.001}$ & 105 \\
\cite{piro21} & 2021gno & $t<5$ & $231.5^{+7.8}_{-8.2}$ & $1.51^{+0.03}_{-0.03}$ & $6.19^{+0.06}_{-+0.05}$ & $0.01^{+0.001}_{-0.001}$ & 51.3 \\
\cite{sapir17} [n=3/2] & 2021gno & $t<5$ & $31.1^{+1.12}_{-1.10}$ & $4.47^{0.10}_{-0.010}$ & $7.07^{+0.14}_{-0.14}$ & $0.28^{+0.010}_{-0.010}$ & 10.8\\
\cite{sapir17} [n=3] & 2021gno & $t<5$ & $37.4^{+1.39}_{-1.25}$ & $50.4^{+1.30}_{-1.26}$ & $8.57^{+0.19}_{-0.19}$ & $0.31^{+0.0091}_{-0.01}$ & 10.2\\
\cite{piro15} & 2021inl & $t<6$ & $156.8^{+6.93}_{-6.1}$ & $2.29^{+0.10}_{-0.10}$ & $5.2 \pm 0.20$ & $0.01^{+0.002}_{-0.001}$ & 40.1 \\
\cite{piro21} & 2021inl & $t<6$ & $21.2^{+1.71}_{-1.50}$ & $37.1^{+24.7}_{-12.6}$ & $9.46^{+0.63}_{-0.67}$ & $0.003^{+0.004}_{-0.002}$ & 7.91 \\
\cite{sapir17} [n=3/2] & 2021inl & $t<6$ & $24.7^{+5.34}_{-4.03}$ & $18.7^{+4.20}_{-3.50}$ & $5.26^{+0.23}_{-0.23}$ & $0.003^{+0.003}_{-0.002}$ & 5.57\\
\cite{sapir17} [n=3] & 2021inl & $t<6$ & $42.7^{+7.13}_{-4.83}$ & $170^{+19.5}_{-27.3}$ & $5.15^{+0.23}_{-0.25}$ & $0.004^{+0.004}_{-0.002}$ & 3.71\\
\enddata
\end{deluxetable*}

\begin{deluxetable}{cccccc}[h!]
\tablecaption{Optical Photometry of SN~2021inl \label{tbl:phot_table_s}}
\tablecolumns{6}
\tablewidth{0.45\textwidth}
\tablehead{
\colhead{MJD} &
\colhead{Phase\tablenotemark{a}} &
\colhead{Filter} & \colhead{Magnitude} & \colhead{Uncertainty} & \colhead{Instrument}
}
\startdata
59310.36 & +0.97 & $g$ & 19.51 & 0.05 & PS1 \\
59312.49 & +3.10 & $g$ & 19.87 & 0.07 & PS1 \\
59313.47 & +4.08 & $g$ & 20.05 & 0.10 & PS1 \\
59314.31 & +4.92 & $g$ & 20.15 & 0.07 & PS1 \\
59316.46 & +7.07 & $g$ & 20.34 & 0.15 & PS1 \\
59317.37 & +7.98 & $g$ & 20.23 & 0.10 & PS1 \\
59319.35 & +9.96 & $g$ & 20.22 & 0.07 & PS1 \\
59321.45 & +12.06 & $g$ & 20.42 & 0.08 & PS1 \\
59322.49 & +13.10 & $g$ & 20.56 & 0.11 & PS1 \\
59323.36 & +13.97 & $g$ & 20.71 & 0.17 & PS1 \\
59350.26 & +40.87 & $g$ & 21.47 & 0.43 & PS1 \\
59353.37 & +43.98 & $g$ & 22.12 & 0.63 & PS1 \\
59312.50 & +3.11 & $r$ & 19.75 & 0.08 & PS1 \\
59314.31 & +4.92 & $r$ & 19.84 & 0.05 & PS1 \\
59317.38 & +7.99 & $r$ & 19.69 & 0.07 & PS1 \\
59319.34 & +9.95 & $r$ & 19.76 & 0.05 & PS1 \\
59321.46 & +12.07 & $r$ & 19.81 & 0.04 & PS1 \\
59326.30 & +16.91 & $r$ & 20.21 & 0.14 & PS1 \\
59332.44 & +23.05 & $r$ & 20.64 & 0.22 & PS1 \\
59334.43 & +25.04 & $r$ & 21.09 & 0.55 & PS1 \\
59334.44 & +25.05 & $r$ & 20.92 & 0.20 & PS1 \\
59336.44 & +27.05 & $r$ & 21.08 & 0.24 & PS1 \\
59344.27 & +34.88 & $r$ & 21.47 & 0.17 & PS1 \\
59352.43 & +43.04 & $r$ & 21.50 & 0.33 & PS1 \\
59360.27 & +50.88 & $r$ & 22.04 & 0.69 & PS1 \\
59364.40 & +55.01 & $r$ & 22.28 & 0.39 & PS1 \\
59375.37 & +65.98 & $r$ & 22.52 & 0.88 & PS1 \\
59414.28 & +104.89 & $r$ & 22.37 & 0.87 & PS1 \\
59310.36 & +0.97 & $i$ & 20.02 & 0.07 & PS1 \\
59313.47 & +4.08 & $i$ & 19.86 & 0.07 & PS1 \\
59316.45 & +7.06 & $i$ & 19.84 & 0.10 & PS1 \\
59322.49 & +13.10 & $i$ & 19.77 & 0.05 & PS1 \\
59323.35 & +13.96 & $i$ & 19.72 & 0.06 & PS1 \\
59326.30 & +16.91 & $i$ & 19.66 & 0.10 & PS1 \\
59334.42 & +25.03 & $i$ & 20.71 & 0.51 & PS1 \\
59342.27 & +32.88 & $i$ & 20.91 & 0.18 & PS1 \\
59346.40 & +37.01 & $i$ & 21.25 & 0.21 & PS1 \\
59349.45 & +40.06 & $i$ & 21.56 & 0.51 & PS1 \\
59351.38 & +41.99 & $i$ & 20.82 & 0.26 & PS1 \\
59362.39 & +53.00 & $i$ & 21.82 & 0.50 & PS1 \\
59368.34 & +58.95 & $i$ & 21.98 & 0.54 & PS1 \\
59373.33 & +63.94 & $i$ & 22.25 & 0.33 & PS1 \\
59332.44 & +23.05 & $z$ & 20.85 & 0.34 & PS1 \\
59334.44 & +25.05 & $z$ & 21.10 & 0.33 & PS1 \\
59336.44 & +27.05 & $z$ & 20.56 & 0.28 & PS1 \\
59353.37 & +43.98 & $z$ & 21.01 & 0.31 & PS1 \\
59360.26 & +50.87 & $z$ & 21.40 & 0.47 & PS1 \\
59364.41 & +55.02 & $z$ & 21.40 & 0.31 & PS1 \\
59366.41 & +57.02 & $z$ & 21.55 & 0.38 & PS1 \\
59414.28 & +104.89 & $z$ & 21.79 & 0.66 & PS1 \\
\enddata
\tablenotetext{a}{Relative to explosion (MJD 59309.4)}
\end{deluxetable}

\begin{deluxetable}{cccccc}[h!]
\tablecaption{Optical Photometry of SN~2021inl \label{tbl:phot_table_s}}
\tablecolumns{6}
\tablewidth{0.45\textwidth}
\tablehead{
\colhead{MJD} &
\colhead{Phase\tablenotemark{a}} &
\colhead{Filter} & \colhead{Magnitude} & \colhead{Uncertainty} & \colhead{Instrument}
}
\startdata
59311.24 & +1.85 & $g$ & 19.44 & 0.06 & ZTF \\
59313.26 & +3.87 & $g$ & 19.77 & 0.09 & ZTF \\
59317.27 & +7.88 & $g$ & 20.33 & 0.09 & ZTF \\
59321.37 & +11.98 & $g$ & 20.50 & 0.11 & ZTF \\
59323.27 & +13.88 & $g$ & 20.92 & 0.24 & ZTF \\
59311.30 & +1.91 & $r$ & 19.70 & 0.09 & ZTF \\
59313.30 & +3.91 & $r$ & 19.76 & 0.10 & ZTF \\
59317.33 & +7.94 & $r$ & 19.79 & 0.08 & ZTF \\
59321.30 & +11.91 & $r$ & 19.74 & 0.06 & ZTF \\
59323.34 & +13.95 & $r$ & 19.92 & 0.09 & ZTF \\
59325.27 & +15.88 & $r$ & 20.04 & 0.12 & ZTF \\
59338.25 & +28.86 & $r$ & 21.59 & 0.34 & ZTF \\
59340.30 & +30.91 & $r$ & 21.26 & 0.25 & ZTF \\
59342.30 & +32.91 & $r$ & 21.55 & 0.30 & ZTF \\
59309.48 & +0.09 & $c$ & 20.63 & 0.18 & ATLAS \\
59313.46 & +4.07 & $c$ & 20.11 & 0.13 & ATLAS \\
59317.46 & +8.07 & $c$ & 20.27 & 0.15 & ATLAS \\
59321.45 & +12.06 & $c$ & 20.25 & 0.14 & ATLAS \\
59343.35 & +33.96 & $c$ & 22.36 & 1.02 & ATLAS \\
59347.40 & +38.01 & $c$ & 21.17 & 0.37 & ATLAS \\
59373.36 & +63.97 & $c$ & 22.41 & 0.95 & ATLAS \\
59319.39 & +10.00 & $o$ & 19.71 & 0.11 & ATLAS \\
59323.37 & +13.98 & $o$ & 19.81 & 0.14 & ATLAS \\
59327.32 & +17.93 & $o$ & 20.29 & 0.49 & ATLAS \\
59333.46 & +24.07 & $o$ & 20.07 & 0.20 & ATLAS \\
59335.40 & +26.01 & $o$ & 21.17 & 0.60 & ATLAS \\
59337.53 & +28.14 & $o$ & 20.42 & 0.26 & ATLAS \\
59339.53 & +30.14 & $o$ & 20.92 & 0.45 & ATLAS \\
59341.53 & +32.14 & $o$ & 21.21 & 0.54 & ATLAS \\
59345.38 & +35.99 & $o$ & 20.98 & 0.38 & ATLAS \\
59349.34 & +39.95 & $o$ & 20.71 & 0.29 & ATLAS \\
59351.34 & +41.95 & $o$ & 21.09 & 0.51 & ATLAS \\
59353.36 & +43.97 & $o$ & 21.30 & 0.47 & ATLAS \\
59357.34 & +47.95 & $o$ & 20.30 & 0.38 & ATLAS \\
59359.32 & +49.93 & $o$ & 21.11 & 0.90 & ATLAS \\
59371.35 & +61.96 & $o$ & 21.17 & 0.45 & ATLAS \\
59398.34 & +88.95 & $B$ & $>$25.92 & -- & Keck \\
59398.34 & +88.95 & $B$ & $>$25.41 & -- & Keck \\
59398.34 & +88.95 & $B$ & 24.08 & 0.21 & Keck \\
59398.34 & +88.95 & $B$ & 22.17 & 0.09 & Keck \\
\enddata
\tablenotetext{a}{Relative to explosion (MJD 59309.4)}
\end{deluxetable}

\begin{deluxetable}{cccccc}[h!]
\tablecaption{Optical Photometry of SN~2021gno \label{tbl:phot_table_s}}
\tablecolumns{6}
\tablewidth{0.45\textwidth}
\tablehead{
\colhead{MJD} &
\colhead{Phase\tablenotemark{a}} &
\colhead{Filter} & \colhead{Magnitude} & \colhead{Uncertainty} & \colhead{Instrument}
}
\startdata
59293.29 & +0.59 & $g$ & 17.74 & 0.02 & ZTF \\
59296.35 & +3.65 & $g$ & 18.13 & 0.03 & ZTF \\
59298.39 & +5.69 & $g$ & 18.26 & 0.06 & ZTF \\
59302.35 & +9.65 & $g$ & 17.64 & 0.04 & ZTF \\
59304.26 & +11.56 & $g$ & 17.57 & 0.03 & ZTF \\
59307.31 & +14.61 & $g$ & 17.55 & 0.01 & ZTF \\
59309.31 & +16.61 & $g$ & 17.80 & 0.02 & ZTF \\
59311.23 & +18.53 & $g$ & 18.10 & 0.02 & ZTF \\
59313.26 & +20.56 & $g$ & 18.48 & 0.03 & ZTF \\
59315.28 & +22.58 & $g$ & 18.64 & 0.10 & ZTF \\
59321.25 & +28.55 & $g$ & 19.43 & 0.05 & ZTF \\
59323.28 & +30.58 & $g$ & 19.47 & 0.08 & ZTF \\
59325.32 & +32.62 & $g$ & 19.48 & 0.11 & ZTF \\
59335.25 & +42.55 & $g$ & 19.91 & 0.12 & ZTF \\
59338.31 & +45.61 & $g$ & 20.10 & 0.11 & ZTF \\
59340.21 & +47.51 & $g$ & 19.98 & 0.08 & ZTF \\
59342.23 & +49.53 & $g$ & 20.18 & 0.09 & ZTF \\
59344.26 & +51.56 & $g$ & 20.32 & 0.11 & ZTF \\
59350.21 & +57.51 & $g$ & 20.46 & 0.16 & ZTF \\
59353.23 & +60.53 & $g$ & 20.62 & 0.31 & ZTF \\
59362.18 & +69.48 & $g$ & 20.95 & 0.19 & ZTF \\
59365.19 & +72.49 & $g$ & 21.22 & 0.30 & ZTF \\
59367.25 & +74.55 & $g$ & 20.88 & 0.23 & ZTF \\
59369.23 & +76.53 & $g$ & 20.91 & 0.20 & ZTF \\
59371.19 & +78.49 & $g$ & 21.28 & 0.29 & ZTF \\
59373.18 & +80.48 & $g$ & 21.07 & 0.22 & ZTF \\
59376.22 & +83.52 & $g$ & 20.96 & 0.31 & ZTF \\
59378.19 & +85.49 & $g$ & 21.18 & 0.31 & ZTF \\
59293.24 & +0.54 & $r$ & 18.20 & 0.03 & ZTF \\
59296.36 & +3.66 & $r$ & 17.94 & 0.03 & ZTF \\
59298.30 & +5.60 & $r$ & 17.88 & 0.06 & ZTF \\
59302.26 & +9.56 & $r$ & 17.33 & 0.03 & ZTF \\
59304.30 & +11.60 & $r$ & 17.16 & 0.02 & ZTF \\
59307.25 & +14.55 & $r$ & 17.05 & 0.01 & ZTF \\
59309.27 & +16.57 & $r$ & 17.08 & 0.01 & ZTF \\
59311.31 & +18.61 & $r$ & 17.22 & 0.01 & ZTF \\
59313.30 & +20.60 & $r$ & 17.39 & 0.02 & ZTF \\
59315.25 & +22.55 & $r$ & 17.53 & 0.22 & ZTF \\
59317.21 & +24.51 & $r$ & 17.83 & 0.02 & ZTF \\
59321.29 & +28.59 & $r$ & 18.08 & 0.02 & ZTF \\
59323.32 & +30.62 & $r$ & 18.23 & 0.03 & ZTF \\
59325.26 & +32.56 & $r$ & 18.32 & 0.03 & ZTF \\
59335.21 & +42.51 & $r$ & 18.72 & 0.05 & ZTF \\
59338.23 & +45.53 & $r$ & 18.90 & 0.04 & ZTF \\
59340.23 & +47.53 & $r$ & 19.08 & 0.05 & ZTF \\
59342.28 & +49.58 & $r$ & 19.15 & 0.05 & ZTF \\
59344.23 & +51.53 & $r$ & 19.18 & 0.04 & ZTF \\
59346.31 & +53.61 & $r$ & 19.34 & 0.07 & ZTF \\
59348.19 & +55.49 & $r$ & 19.33 & 0.05 & ZTF \\
\enddata
\tablenotetext{a}{Relative to explosion (MJD 59292.7 )}
\end{deluxetable}

\begin{deluxetable}{cccccc}[h!]
\tablecaption{Optical Photometry of SN~2021gno \label{tbl:phot_table_s}}
\tablecolumns{6}
\tablewidth{0.45\textwidth}
\tablehead{
\colhead{MJD} &
\colhead{Phase\tablenotemark{a}} &
\colhead{Filter} & \colhead{Magnitude} & \colhead{Uncertainty} & \colhead{Instrument}
}
\startdata
59353.19 & +60.49 & $r$ & 19.52 & 0.09 & ZTF \\
59359.19 & +66.49 & $r$ & 19.88 & 0.23 & ZTF \\
59362.23 & +69.53 & $r$ & 19.87 & 0.13 & ZTF \\
59367.22 & +74.52 & $r$ & 20.09 & 0.11 & ZTF \\
59369.21 & +76.51 & $r$ & 20.34 & 0.14 & ZTF \\
59371.23 & +78.53 & $r$ & 20.26 & 0.13 & ZTF \\
59373.21 & +80.51 & $r$ & 20.73 & 0.25 & ZTF \\
59376.26 & +83.56 & $r$ & 20.31 & 0.18 & ZTF \\
59378.25 & +85.55 & $r$ & 20.59 & 0.21 & ZTF \\
59380.21 & +87.51 & $r$ & 20.50 & 0.35 & ZTF \\
59293.40 & +0.70 & $o$ & 18.05 & 0.03 & ATLAS \\
59297.36 & +4.66 & $o$ & 18.08 & 0.04 & ATLAS \\
59303.45 & +10.75 & $o$ & 17.24 & 0.05 & ATLAS \\
59304.39 & +11.69 & $o$ & 17.16 & 0.02 & ATLAS \\
59305.49 & +12.79 & $o$ & 17.13 & 0.02 & ATLAS \\
59306.39 & +13.69 & $o$ & 17.08 & 0.01 & ATLAS \\
59307.43 & +14.73 & $o$ & 17.08 & 0.03 & ATLAS \\
59319.41 & +26.71 & $o$ & 17.90 & 0.03 & ATLAS \\
59323.33 & +30.63 & $o$ & 18.18 & 0.03 & ATLAS \\
59324.44 & +31.74 & $o$ & 18.18 & 0.03 & ATLAS \\
59331.39 & +38.69 & $o$ & 18.66 & 0.14 & ATLAS \\
59332.43 & +39.73 & $o$ & 18.54 & 0.07 & ATLAS \\
59333.42 & +40.72 & $o$ & 18.55 & 0.06 & ATLAS \\
59334.39 & +41.69 & $o$ & 19.94 & 0.41 & ATLAS \\
59335.35 & +42.65 & $o$ & 18.58 & 0.05 & ATLAS \\
59337.47 & +44.77 & $o$ & 18.79 & 0.08 & ATLAS \\
59345.38 & +52.68 & $o$ & 18.98 & 0.07 & ATLAS \\
59349.33 & +56.63 & $o$ & 19.05 & 0.07 & ATLAS \\
59351.33 & +58.63 & $o$ & 19.11 & 0.08 & ATLAS \\
59353.31 & +60.61 & $o$ & 19.28 & 0.08 & ATLAS \\
59358.33 & +65.63 & $o$ & 19.24 & 0.19 & ATLAS \\
59359.30 & +66.60 & $o$ & 19.85 & 0.28 & ATLAS \\
59360.42 & +67.72 & $o$ & 19.46 & 0.17 & ATLAS \\
59361.36 & +68.66 & $o$ & 19.90 & 0.24 & ATLAS \\
59362.37 & +69.67 & $o$ & 19.68 & 0.23 & ATLAS \\
59363.34 & +70.64 & $o$ & 19.56 & 0.14 & ATLAS \\
59367.30 & +74.60 & $o$ & 19.60 & 0.12 & ATLAS \\
59371.34 & +78.64 & $o$ & 19.76 & 0.16 & ATLAS \\
59309.46 & +16.76 & $c$ & 17.51 & 0.02 & ATLAS \\
59313.44 & +20.74 & $c$ & 17.97 & 0.02 & ATLAS \\
59317.48 & +24.78 & $c$ & 18.45 & 0.04 & ATLAS \\
59321.43 & +28.73 & $c$ & 18.69 & 0.04 & ATLAS \\
59339.46 & +46.76 & $c$ & 19.55 & 0.34 & ATLAS \\
59343.33 & +50.63 & $c$ & 19.58 & 0.10 & ATLAS \\
59347.35 & +54.65 & $c$ & 20.04 & 0.14 & ATLAS \\
59369.32 & +76.62 & $c$ & 20.31 & 0.20 & ATLAS \\
59373.32 & +80.62 & $c$ & 20.87 & 0.29 & ATLAS \\
59377.28 & +84.58 & $c$ & 21.22 & 0.26 & ATLAS \\
\enddata
\tablenotetext{a}{Relative to explosion (MJD 59292.7 )}
\end{deluxetable}

\begin{deluxetable}{cccccc}[h!]
\tablecaption{Optical Photometry of SN~2021gno \label{tbl:phot_table_s}}
\tablecolumns{6}
\tablewidth{0.45\textwidth}
\tablehead{
\colhead{MJD} &
\colhead{Phase\tablenotemark{a}} &
\colhead{Filter} & \colhead{Magnitude} & \colhead{Uncertainty} & \colhead{Instrument}
}
\startdata
59295.18 & +2.48 & $g$ & 17.59 & 0.01 & DECam \\
59297.11 & +4.41 & $g$ & 18.27 & 0.02 & DECam \\
59297.16 & +4.46 & $g$ & 18.26 & 0.01 & DECam \\
59310.16 & +17.46 & $g$ & 17.88 & 0.01 & DECam \\
59313.14 & +20.44 & $g$ & 18.41 & 0.01 & DECam \\
59316.13 & +23.43 & $g$ & 18.81 & 0.02 & DECam \\
59319.15 & +26.45 & $g$ & 19.13 & 0.04 & DECam \\
59340.07 & +47.37 & $g$ & 20.02 & 0.02 & DECam \\
59343.10 & +50.40 & $g$ & 20.08 & 0.02 & DECam \\
59346.07 & +53.37 & $g$ & 20.22 & 0.02 & DECam \\
59349.06 & +56.36 & $g$ & 20.28 & 0.02 & DECam \\
59352.06 & +59.36 & $g$ & 20.48 & 0.02 & DECam \\
59367.06 & +74.36 & $g$ & 20.90 & 0.02 & DECam \\
59370.05 & +77.35 & $g$ & 21.14 & 0.02 & DECam \\
59373.04 & +80.34 & $g$ & 21.16 & 0.02 & DECam \\
59376.05 & +83.35 & $g$ & 21.21 & 0.02 & DECam \\
59579.34 & +286.64 & $g$ & 25.03 & 0.02 & DECam \\
59582.34 & +289.64 & $g$ & 23.46 & 0.02 & DECam \\
59592.34 & +299.64 & $g$ & 25.79 & 0.02 & DECam \\
59595.35 & +302.65 & $g$ & 24.26 & 0.02 & DECam \\
59297.11 & +4.41 & $r$ & 18.14 & 0.06 & DECam \\
59297.16 & +4.46 & $r$ & 18.12 & 0.03 & DECam \\
59304.18 & +11.48 & $r$ & 17.28 & 0.01 & DECam \\
59316.13 & +23.43 & $r$ & 17.81 & 0.01 & DECam \\
59337.08 & +44.38 & $r$ & 19.03 & 0.02 & DECam \\
59340.07 & +47.37 & $r$ & 19.12 & 0.02 & DECam \\
59346.07 & +53.37 & $r$ & 19.45 & 0.02 & DECam \\
59352.06 & +59.36 & $r$ & 19.75 & 0.02 & DECam \\
59358.07 & +65.37 & $r$ & 19.97 & 0.02 & DECam \\
59364.05 & +71.35 & $r$ & 20.38 & 0.02 & DECam \\
59367.06 & +74.36 & $r$ & 20.54 & 0.02 & DECam \\
59373.04 & +80.34 & $r$ & 20.98 & 0.02 & DECam \\
59295.19 & +2.49 & $i$ & 17.93 & 0.01 & DECam \\
59297.11 & +4.41 & $i$ & 18.13 & 0.01 & DECam \\
59304.18 & +11.48 & $i$ & 17.29 & 0.01 & DECam \\
59307.16 & +14.46 & $i$ & 17.16 & 0.01 & DECam \\
59310.16 & +17.46 & $i$ & 17.21 & 0.01 & DECam \\
59313.14 & +20.44 & $i$ & 17.38 & 0.01 & DECam \\
59319.16 & +26.46 & $i$ & 17.87 & 0.01 & DECam \\
59331.10 & +38.40 & $i$ & 18.37 & 0.01 & DECam \\
59333.09 & +40.39 & $i$ & 18.43 & 0.01 & DECam \\
59337.08 & +44.38 & $i$ & 18.54 & 0.01 & DECam \\
59343.10 & +50.40 & $i$ & 18.67 & 0.01 & DECam \\
59349.06 & +56.36 & $i$ & 18.82 & 0.01 & DECam \\
59358.07 & +65.37 & $i$ & 19.00 & 0.01 & DECam \\
59361.03 & +68.33 & $i$ & 19.09 & 0.01 & DECam \\
59364.05 & +71.35 & $i$ & 19.17 & 0.01 & DECam \\
59370.04 & +77.34 & $i$ & 19.18 & 0.03 & DECam \\
\enddata
\tablenotetext{a}{Relative to explosion (MJD 59292.7 )}
\end{deluxetable}

\begin{deluxetable}{cccccc}[h!]
\tablecaption{Optical Photometry of SN~2021gno \label{tbl:phot_table_s}}
\tablecolumns{6}
\tablewidth{0.45\textwidth}
\tablehead{
\colhead{MJD} &
\colhead{Phase\tablenotemark{a}} &
\colhead{Filter} & \colhead{Magnitude} & \colhead{Uncertainty} & \colhead{Instrument}
}
\startdata
59376.05 & +83.35 & $i$ & 19.39 & 0.01 & DECam \\
59571.34 & +278.64 & $i$ & 22.91 & 0.36 & DECam \\
59582.34 & +289.64 & $i$ & 22.78 & 0.20 & DECam \\
59588.35 & +295.65 & $i$ & 22.80 & 0.34 & DECam \\
59297.11 & +4.41 & $z$ & 18.23 & 0.05 & DECam \\
59307.16 & +14.46 & $z$ & 17.12 & 0.02 & DECam \\
59331.10 & +38.40 & $z$ & 17.90 & 0.02 & DECam \\
59333.09 & +40.39 & $z$ & 17.95 & 0.02 & DECam \\
59293.55 & +0.85 & $v$ & 17.33 & 0.22 & \emph{Swift} \\
59293.88 & +1.18 & $v$ & 17.50 & 0.25 & \emph{Swift} \\
59294.44 & +1.74 & $v$ & 17.48 & 0.25 & \emph{Swift} \\
59295.51 & +2.81 & $v$ & 17.26 & 0.21 & \emph{Swift} \\
59296.60 & +3.90 & $v$ & $>$17.85 & -- & \emph{Swift} \\
59298.29 & +5.59 & $v$ & 17.76 & 0.34 & \emph{Swift} \\
59302.67 & +9.97 & $v$ & 17.74 & 0.34 & \emph{Swift} \\
59303.15 & +10.45 & $v$ & 16.91 & 0.17 & \emph{Swift} \\
59308.72 & +16.02 & $v$ & 17.23 & 0.22 & \emph{Swift} \\
59313.30 & +20.60 & $v$ & $>$17.70 & -- & \emph{Swift} \\
59317.87 & +25.17 & $v$ & $>$17.84 & -- & \emph{Swift} \\
59325.65 & +32.95 & $v$ & $>$17.77 & -- & \emph{Swift} \\
59414.18 & +121.48 & $v$ & $>$17.62 & -- & \emph{Swift} \\
59293.54 & +0.84 & $b$ & 17.42 & 0.14 & \emph{Swift} \\
59293.88 & +1.18 & $b$ & 17.29 & 0.12 & \emph{Swift} \\
59294.44 & +1.74 & $b$ & 17.34 & 0.13 & \emph{Swift} \\
59295.51 & +2.81 & $b$ & 17.73 & 0.18 & \emph{Swift} \\
59296.60 & +3.90 & $b$ & 18.21 & 0.28 & \emph{Swift} \\
59298.28 & +5.58 & $b$ & 18.37 & 0.33 & \emph{Swift} \\
59302.66 & +9.96 & $b$ & 17.61 & 0.17 & \emph{Swift} \\
59303.14 & +10.44 & $b$ & 17.90 & 0.22 & \emph{Swift} \\
59308.71 & +16.01 & $b$ & 17.97 & 0.23 & \emph{Swift} \\
59313.30 & +20.60 & $b$ & $>$18.44 & -- & \emph{Swift} \\
59317.87 & +25.17 & $b$ & $>$18.52 & -- & \emph{Swift} \\
59325.64 & +32.94 & $b$ & $>$18.46 & -- & \emph{Swift} \\
59414.17 & +121.47 & $b$ & $>$18.21 & -- & \emph{Swift} \\
59293.54 & +0.84 & $u$ & 17.01 & 0.06 & \emph{Swift} \\
59293.88 & +1.18 & $u$ & 17.05 & 0.06 & \emph{Swift} \\
59294.44 & +1.74 & $u$ & 17.24 & 0.07 & \emph{Swift} \\
59295.51 & +2.81 & $u$ & 17.71 & 0.09 & \emph{Swift} \\
59296.60 & +3.90 & $u$ & 18.55 & 0.18 & \emph{Swift} \\
59298.28 & +5.58 & $u$ & 18.86 & 0.23 & \emph{Swift} \\
59302.66 & +9.96 & $u$ & 18.34 & 0.16 & \emph{Swift} \\
59303.14 & +10.44 & $u$ & 18.50 & 0.18 & \emph{Swift} \\
59308.71 & +16.01 & $u$ & 18.93 & 0.25 & \emph{Swift} \\
59313.30 & +20.60 & $u$ & $>$19.28 & -- & \emph{Swift} \\
59317.87 & +25.17 & $u$ & $>$19.38 & -- & \emph{Swift} \\
59325.64 & +32.94 & $u$ & $>$19.35 & -- & \emph{Swift} \\
59414.17 & +121.47 & $u$ & $>$19.10 & -- & \emph{Swift} \\
59293.54 & +0.84 & $w1$ & 17.06 & 0.05 & \emph{Swift} \\
\enddata
\tablenotetext{a}{Relative to explosion (MJD 59292.7 )}
\end{deluxetable}

\begin{deluxetable}{cccccc}[h!]
\tablecaption{Optical Photometry of SN~2021gno \label{tbl:phot_table_s}}
\tablecolumns{6}
\tablewidth{0.45\textwidth}
\tablehead{
\colhead{MJD} &
\colhead{Phase\tablenotemark{a}} &
\colhead{Filter} & \colhead{Magnitude} & \colhead{Uncertainty} & \colhead{Instrument}
}
\startdata
59293.87 & +1.17 & $w1$ & 17.13 & 0.05 & \emph{Swift} \\
59294.44 & +1.74 & $w1$ & 17.53 & 0.05 & \emph{Swift} \\
59295.50 & +2.80 & $w1$ & 18.43 & 0.08 & \emph{Swift} \\
59296.59 & +3.89 & $w1$ & 19.72 & 0.20 & \emph{Swift} \\
59298.28 & +5.58 & $w1$ & $>$20.44 & -- & \emph{Swift} \\
59302.66 & +9.96 & $w1$ & 20.23 & 0.31 & \emph{Swift} \\
59303.14 & +10.44 & $w1$ & 20.36 & 0.35 & \emph{Swift} \\
59308.71 & +16.01 & $w1$ & $>$20.51 & -- & \emph{Swift} \\
59313.30 & +20.60 & $w1$ & $>$20.33 & -- & \emph{Swift} \\
59317.86 & +25.16 & $w1$ & $>$20.51 & -- & \emph{Swift} \\
59325.64 & +32.94 & $w1$ & $>$20.47 & -- & \emph{Swift} \\
59414.17 & +121.47 & $w1$ & $>$20.27 & -- & \emph{Swift} \\
59293.55 & +0.85 & $m2$ & 16.96 & 0.04 & \emph{Swift} \\
59293.88 & +1.18 & $m2$ & 17.13 & 0.04 & \emph{Swift} \\
59294.44 & +1.74 & $m2$ & 17.62 & 0.04 & \emph{Swift} \\
59295.51 & +2.81 & $m2$ & 18.72 & 0.06 & \emph{Swift} \\
59296.60 & +3.90 & $m2$ & 20.43 & 0.17 & \emph{Swift} \\
59298.29 & +5.59 & $m2$ & 21.20 & 0.30 & \emph{Swift} \\
59302.67 & +9.97 & $m2$ & $>$21.42 & -- & \emph{Swift} \\
59303.15 & +10.45 & $m2$ & $>$21.39 & -- & \emph{Swift} \\
59308.72 & +16.02 & $m2$ & $>$21.47 & -- & \emph{Swift} \\
59313.30 & +20.60 & $m2$ & $>$21.24 & -- & \emph{Swift} \\
59317.88 & +25.18 & $m2$ & $>$21.50 & -- & \emph{Swift} \\
59325.65 & +32.95 & $m2$ & $>$21.39 & -- & \emph{Swift} \\
59414.18 & +121.48 & $m2$ & $>$21.31 & -- & \emph{Swift} \\
59293.54 & +0.84 & $w2$ & 16.90 & 0.04 & \emph{Swift} \\
59293.88 & +1.18 & $w2$ & 17.17 & 0.04 & \emph{Swift} \\
59294.44 & +1.74 & $w2$ & 17.72 & 0.04 & \emph{Swift} \\
59295.51 & +2.81 & $w2$ & 18.94 & 0.06 & \emph{Swift} \\
59296.60 & +3.90 & $w2$ & 20.09 & 0.11 & \emph{Swift} \\
59298.28 & +5.58 & $w2$ & 20.80 & 0.20 & \emph{Swift} \\
59302.67 & +9.97 & $w2$ & 20.95 & 0.22 & \emph{Swift} \\
59303.14 & +10.44 & $w2$ & 20.99 & 0.23 & \emph{Swift} \\
59308.72 & +16.02 & $w2$ & 21.12 & 0.25 & \emph{Swift} \\
59313.30 & +20.60 & $w2$ & $>$21.40 & -- & \emph{Swift} \\
59317.87 & +25.17 & $w2$ & $>$21.64 & -- & \emph{Swift} \\
59325.64 & +32.94 & $w2$ & 21.15 & 0.29 & \emph{Swift} \\
59414.18 & +121.48 & $w2$ & $>$21.48 & -- & \emph{Swift} \\
\enddata
\tablenotetext{a}{Relative to explosion (MJD 59292.7 )}
\end{deluxetable}

\begin{deluxetable}{cccccc}[h!]
\tablecaption{Optical Photometry of SN~2021gno \label{tbl:phot_table_s}}
\tablecolumns{6}
\tablewidth{0.45\textwidth}
\tablehead{
\colhead{MJD} &
\colhead{Phase\tablenotemark{a}} &
\colhead{Filter} & \colhead{Magnitude} & \colhead{Uncertainty} & \colhead{Instrument}
}
\startdata
59295.30 & +2.60 & $V$ & 17.59 & 0.14 & Nickel \\
59295.42 & +2.72 & $V$ & 17.60 & 0.17 & Nickel \\
59295.48 & +2.78 & $V$ & 17.56 & 0.23 & Nickel \\
59295.30 & +2.60 & $r$ & 17.77 & 0.11 & Nickel \\
59295.42 & +2.72 & $r$ & 17.79 & 0.15 & Nickel \\
59295.49 & +2.79 & $r$ & 17.78 & 0.22 & Nickel \\
59306.47 & +13.77 & $r$ & 17.28 & 0.15 & Nickel \\
59309.25 & +16.55 & $r$ & 17.32 & 0.11 & Nickel \\
59321.36 & +28.66 & $r$ & 18.39 & 0.22 & Nickel \\
59324.42 & +31.72 & $r$ & 18.47 & 0.28 & Nickel \\
59295.31 & +2.61 & $i$ & 17.92 & 0.13 & Nickel \\
59295.43 & +2.73 & $i$ & 17.91 & 0.12 & Nickel \\
59295.49 & +2.79 & $i$ & 17.91 & 0.24 & Nickel \\
59306.48 & +13.78 & $i$ & 17.14 & 0.11 & Nickel \\
59309.26 & +16.56 & $i$ & 17.16 & 0.12 & Nickel \\
59312.32 & +19.62 & $i$ & 17.38 & 0.10 & Nickel \\
59315.42 & +22.72 & $i$ & 17.71 & 0.12 & Nickel \\
59318.24 & +25.54 & $i$ & 17.82 & 0.14 & Nickel \\
59321.37 & +28.67 & $i$ & 18.00 & 0.17 & Nickel \\
59324.43 & +31.73 & $i$ & 18.09 & 0.25 & Nickel \\
\enddata
\tablenotetext{a}{Relative to explosion (MJD 59292.7 )}
\end{deluxetable}

\end{document}